\documentclass[preprint, superscriptaddress, preprintnumbers,amsmath,amssymb]{revtex4}

\usepackage{booktabs}
\usepackage{color,graphicx}
\usepackage{amsmath}
\usepackage{dcolumn}
\usepackage{bm}                  % bold math
\usepackage{soul}
\usepackage{enumerate}
\usepackage{multirow}
\usepackage{mathrsfs}
\usepackage{braket}
\usepackage{longtable}
\usepackage{pdflscape}

\usepackage[dvipdfm,
            pdfstartview=FitH,
            CJKbookmarks=true,
            bookmarksnumbered=true,
            bookmarksopen=true,
            colorlinks,
            pdfborder=001,
            linkcolor=blue,
            anchorcolor=blue,
            citecolor=blue
            ]{hyperref}

\usepackage{float}

\begin{document}
%\begin{CJK*}{GBK}{song}

\title{Deformed relativistic Hartree-Bogoliubov theory in continuum with point coupling functional: examples of even-even Nd isotopes}

\author{Kaiyuan Zhang}
\affiliation{State Key Laboratory of Nuclear Physics and Technology,
School of Physics, Peking University, Beijing 100871, China}

\author{Myung-Ki Cheoun}
\affiliation{Department of Physics and Origin of Matter and Evolution of Galaxy (OMEG) Institute, Soongsil University, Seoul 156-743, Korea}

\author{Yong-Beom Choi}
\affiliation{Department of Physics, Pusan National University, Busan 46241, Korea}

\author{Pooi Seong Chong}
\affiliation{Department of Physics, The University of Hong Kong, Pokfulam, 999077, Hong Kong}

\author{Jianmin Dong}
\affiliation{Institute of Modern Physics, Chinese Academy of Sciences, Lanzhou 730000, China}
\affiliation{School of Physics, University of Chinese Academy of Sciences, Beijing 100049, China}

\author{Lisheng Geng}
\affiliation{School of Physics, Beihang University, Beijing 102206, China}

\author{Eunja Ha}
\affiliation{Department of Physics and Origin of Matter and Evolution of Galaxy (OMEG) Institute, Soongsil University, Seoul 156-743, Korea}

\author{Xiaotao He}
\affiliation{College of Material Science and Technology, Nanjing University of Aeronautics and Astronautics, Nanjing 210016, China}

\author{Chan Heo}
\affiliation{Department of Physics, The University of Hong Kong, Pokfulam, 999077, Hong Kong}

\author{Meng Chit Ho}
\affiliation{Department of Physics, The University of Hong Kong, Pokfulam, 999077, Hong Kong}

\author{Eun Jin In}
\affiliation{Department of Energy Science, Sungkyunkwan University, Suwon 16419, Korea }

\author{Seonghyun Kim}
\affiliation{Department of Physics and Origin of Matter and Evolution of Galaxy (OMEG) Institute, Soongsil University, Seoul 156-743, Korea}

\author{Youngman Kim}
\affiliation{Rare Isotope Science Project, Institute for Basic Science, Daejeon 305-811, Korea}

\author{Chang-Hwan Lee}
\affiliation{Department of Physics, Pusan National University, Busan 46241, Korea}

\author{Jenny Lee}
\affiliation{Department of Physics, The University of Hong Kong, Pokfulam, 999077, Hong Kong}

\author{Zhipan Li}
\affiliation{School of Physical Science and Technology, Southwest University, Chongqing 400715, China}

\author{Tianpeng Luo}
\affiliation{State Key Laboratory of Nuclear Physics and Technology, School of Physics, Peking University, Beijing 100871, China}

\author{Jie Meng} \email{mengj@pku.edu.cn}
\affiliation{State Key Laboratory of Nuclear Physics and Technology, School of Physics, Peking University, Beijing 100871, China}

\author{Myeong-Hwan Mun}
\affiliation{Korea Institute of Science and Technology Information, Daejeon 34141, Korea}

\author{Zhongming Niu}
\affiliation{School of Physics and Materials Science, Anhui University, Hefei 230601, China}
\affiliation{Institute of Physical Science and Information Technology, Anhui University, Hefei 230601, China}

\author{Cong Pan}
\affiliation{State Key Laboratory of Nuclear Physics and Technology,
School of Physics, Peking University, Beijing 100871, China}

\author{Panagiota Papakonstantinou}
\affiliation{Institute for Basic Science, Rare Isotope Science Project, Daejeon 34000, South Korea}

\author{Xinle Shang}
\affiliation{Institute of Modern Physics, Chinese Academy of Sciences, Lanzhou 730000, China}
\affiliation{School of Physics, University of Chinese Academy of Sciences, Beijing 100049, China}

\author{Caiwan Shen}
\affiliation{School of Science, Huzhou University, Huzhou 313000, China}

\author{Guofang Shen}
\affiliation{School of Physics, Beihang University, Beijing 102206, China}

\author{Wei Sun}
\affiliation{School of Physical Science and Technology, Southwest University, Chongqing 400715, China}

\author{Xiang-Xiang Sun}
\affiliation{CAS Key Laboratory of Theoretical Physics, Institute of Theoretical Physics, Chinese Academy of Sciences, Beijing 100190, China}
\affiliation{School of Physical Sciences, University of Chinese Academy of Sciences, Beijing 100049, China}

\author{Chi Kin Tam}
\affiliation{Department of Physics, The University of Hong Kong, Pokfulam, 999077, Hong Kong}

\author{Thaivayongnou}
\affiliation{School of Physics, Beihang University, Beijing 102206, China}

\author{Chen Wang}
\affiliation{College of Material Science and Technology, Nanjing University of Aeronautics and Astronautics, Nanjing 210016, China}

\author{Sau Hei Wong}
\affiliation{Department of Physics, The University of Hong Kong, Pokfulam, 999077, Hong Kong}

\author{Xuewei Xia}
\affiliation{School of Physics and Electronic Engineering, Center for Computational Sciences, Sichuan Normal University, Chengdu 610068, China}

\author{Yijun Yan}
\affiliation{Institute of Modern Physics, Chinese Academy of Sciences, Lanzhou 730000, China}
\affiliation{School of Physics, University of Chinese Academy of Sciences, Beijing 100049, China}

\author{Ryan Wai-Yen Yeung}
\affiliation{Department of Physics, The University of Hong Kong, Pokfulam, 999077, Hong Kong}

\author{To Chung Yiu}
\affiliation{Department of Physics, The University of Hong Kong, Pokfulam, 999077, Hong Kong}

\author{Shuangquan Zhang}
\affiliation{State Key Laboratory of Nuclear Physics and Technology,
School of Physics, Peking University, Beijing 100871, China}

\author{Wei Zhang}
\affiliation{School of Physics and Microelectronics, Zhengzhou University, Zhengzhou 450001, China}

\author{Shan-Gui Zhou}
\affiliation{CAS Key Laboratory of Theoretical Physics, Institute of Theoretical Physics, Chinese Academy of Sciences, Beijing 100190, China}
\affiliation{School of Physical Sciences, University of Chinese Academy of Sciences, Beijing 100049, China}
\affiliation{Center of Theoretical Nuclear Physics, National Laboratory of Heavy Ion Accelerator, Lanzhou 730000, China}
\affiliation{Synergetic Innovation Center for Quantum Effects and Application, Hunan Normal University, Changsha, 410081, China}

\collaboration{DRHBc mass table collaboration}

\begin{abstract}

\begin{itemize}
\item[] \textbf{Background:}
The study of exotic nuclei far from the $\beta$ stability line is stimulated by the development of radioactive ion beam facilities worldwide and brings opportunities and challenges to existing nuclear theories. Including self-consistently the nuclear superfluidity, deformation, and continuum effects, the deformed relativistic Hartree-Bogoliubov theory in continuum (DRHBc) has turned out to be successful in describing both stable and exotic nuclei. Due to several challenges, however, the DRHBc theory has only been applied to study light nuclei so far.
\item[] \textbf{Purpose:}
The aim of this work is to develop the DRHBc theory based on the point-coupling density functional and examine its possible application for all even-even nuclei in the nuclear chart by taking Nd isotopes as examples.
\item[]\textbf{Method:}
The nuclear superfluidity is taken into account via Bogoliubov transformation. Densities and potentials are expanded in terms of Legendre polynomials to include the axial deformation degrees of freedom. Sophisticated relativistic Hartree-Bogoliubov equations in coordinate space are solved in a Dirac Woods-Saxon basis to consider the continuum effects.
\item[]\textbf{Results:}
Numerical convergence for energy cutoff, angular momentum cutoff, Legendre expansion, pairing strength, and (un)constrained calculations are confirmed for DRHBc from light nuclei to heavy nuclei. The ground-state properties of even-even Nd isotopes are calculated with the successful density functional PC-PK1 and compared with the spherical nuclear mass table based on the relativistic continuum Hartree-Bogoliubov (RCHB) theory as well as the data available. The calculated binding energies are in very good agreement with the existing experimental values with a rms deviation of $0.958$ MeV, which is remarkably smaller than $8.301$ MeV in the spherical case. The predicted proton and neutron drip-line nuclei for Nd isotopes are respectively $^{120}$Nd and $^{214}$Nd, in contrast with $^{126}$Nd and $^{228}$Nd in the RCHB theory. The experimental quadrupole deformations and charge radii are reproduced well. An interesting decoupling between the oblate shape $\beta_2=-0.273$ contributed by bound states and the nearly spherical one $\beta_2=0.047$ contributed by continuum is found in $^{214}$Nd. Contributions of different single-particle states to the total neutron density are investigated and an exotic neutron skin phenomenon is suggested for $^{214}$Nd. The proton radioactivity beyond the proton drip line is discussed and $^{114}$Nd, $^{116}$Nd, and $^{118}$Nd are predicted to be candidates for two-proton or even multi-proton radioactivity.
\item[]\textbf{Conclusions:}
The DRHBc theory based on the point-coupling density functional is developed and detailed numerical checks are performed. The techniques to construct the DRHBc mass table for even-even nuclei are explored and extended for all even-even nuclei in the nuclear chart by taking Nd isotopes as examples. The experimental data available are reproduced well. The deformation and continuum effects on drip-line nuclei, exotic neutron skin, and proton radioactivity are presented.
\end{itemize}

\end{abstract}

\date{\today}

\maketitle

%%%%%%%%%%%%%%%%%%%%%%%%%%%%%%%%%%%%%%%%%%%%%%%%%%%%%%%%%%
%                    begin  introduction
%%%%%%%%%%%%%%%%%%%%%%%%%%%%%%%%%%%%%%%%%%%%%%%%%%%%%%%%%%

%----------------------------------------------------------------------------------------
\section{Introduction}
%----------------------------------------------------------------------------------------

In nuclear physics, the study of the properties of exotic nuclei¡ªnuclei with extreme
numbers of protons or neutrons¡ªis one of the top priorities, as it can lead to new insights into
the origins of the chemical elements in stars and star explosions~\cite{Meng2016Book}. Although radioactive ion
beams (RIB) have extended our knowledge of nuclear physics from stable nuclei to exotic
ones far away from the valley of stability, it is still a dream to reach the neutron drip line up to
mass number $A\approx 100$ with the new generation of RIB facilities developed around the world,
including the Cooler Storage Ring (CSR) at the Heavy Ion Research Facility in Lanzhou (HIRFL) in China~\cite{ZHAN2010694c},the RIKEN Radioactive Ion Beam Factory (RIBF) in Japan~\cite{MOTOBAYASHI2010707c}, the Rare Isotope Science Project (RISP) in Korea~\cite{TSHOO2013242}, the Facility for
Antiproton and Ion Research (FAIR) in Germany~\cite{STURM2010682c}, the Second Generation System On-Line
Production of Radioactive Ions (SPIRAL2) at GANIL in France~\cite{GALES2010717c}, the Facility for Rare
Isotope Beams (FRIB) in the USA~\cite{THOENNESSEN2010688c}, etc.

The nuclear mass or binding energy is of crucial importance not only in nuclear physics, but also in other fields, such as astrophysics~\cite{Lunney20031021,BLAUM20061}. It has been always a priority in nuclear physics to explore the limit of nuclear binding~\cite{Erler2012,Thoennessen2013,Xia2018ADNDT}. Experimentally, the existence of about $3200$ isotopes has been confirmed~\cite{NNDC} and the masses of about $2500$ nuclides have been measured~\cite{AME20161,AME2016}. The proton drip line has been determined up to neptunium~\cite{PhysRevLett.122.192503}, but the neutron drip line is known only up to neon~\cite{NNDC}. In the foreseeable future, most of neutron-rich nuclei far from the valley of stability seem still beyond the experimental capability. Therefore, it is urgent to develop a theoretical nuclear mass table with predictive power to grasp a complete understanding of the nature.

Theoretically, a lot of efforts have been made to predict nuclear masses and to explore the great unknowns of the nuclear landscape~\cite{Moller2016ADNDT,Aboussir1995ADNDT,Wang2014PLB,Zhang2014NPA,Samyn2002NPA,Stoitsov2003PRC,Goriely2009PRLSkyrme,Goriely2013PRC,Erler2012,Hilaire2007EPJA,Goriely2009PRLGogny,Delaroche2010PRC,Lalazissis1999ADNDT,Geng2005Prog.Theor.Phys,Meng2013Front,Zhang2014Front.Phys.529,Agbemava2014PRC,Afanasjev2015PRC,Lu2015Phys.Rev.C27304,Pena-Arteaga2016EPJA,Xia2018ADNDT}. Precise descriptions of nuclear masses have been achieved with various macroscopic-microscopic models~\cite{Moller2016ADNDT,Aboussir1995ADNDT,Wang2014PLB,Zhang2014NPA}. Several Skyrme~\cite{Samyn2002NPA,Stoitsov2003PRC,Goriely2009PRLSkyrme,Goriely2013PRC,Erler2012} or Gogny~\cite{Hilaire2007EPJA,Goriely2009PRLGogny,Delaroche2010PRC} Hartree-Fock-Bogoliubov mass-table-type calculations have been performed based on the non-relativistic density functional theory. On the relativistic side, many investigations have been done and significant progresses have been made based on the covariant density functional theory~\cite{Lalazissis1999ADNDT,Geng2005Prog.Theor.Phys,Meng2013Front,Zhang2014Front.Phys.529,Agbemava2014PRC,Afanasjev2015PRC,Lu2015Phys.Rev.C27304,Pena-Arteaga2016EPJA,Xia2018ADNDT}.

The covariant density functional theory (CDFT) has been proved to be a powerful theory in nuclear physics by its successful description of many nuclear phenomena~\cite{Ring1996,Vretenar2005PhysRep,Meng2006PPNP,Niksic2011PPNP,Meng2013Front,Meng2015JPhysG,Zhou2016PhysScr,Meng2016Book,Shen2019PPNP}. As a microscopic and covariant theory, the CDFT has attracted a lot of attention in recent years for many attractive advantages, such as the automatic inclusion of nucleonic spin degree of freedom, explaining naturally the pseudospin symmetry in the nucleon spectrum~\cite{PhysRevLett.78.436,Meng1998Phys.Rev.C628,Meng1999Phys.Rev.C154,Chen2003Chin.Phys.Lett.358,Ginocchio2005PhysRep,Liang2015Phys.Rept.1} and the spin symmetry in anti-nucleon spectrum~\cite{Zhou2003Phys.Rev.Lett.262501,He2006Eur.Phys.J.A265,Liang2015Phys.Rept.1}, and the natural inclusion of the nuclear magnetism~\cite{Koepf1989NPA}, which plays an important role in nuclear magnetic moments~\cite{Yao2006Phys.Rev.C24307,Arima2011,Li2011Sci.ChinaPhys.Mech.Astron.204,Li2011Prog.Theor.Phys.1185,Li2018Front.Phys.Beijing132109} and nuclear rotations~\cite{Meng2013Front,PhysRevLett.71.3079,Afanasjev2000NPA,PhysRevC.62.031302,PhysRevC.82.034329,Zhao2011Phys.Rev.Lett.122501,Zhao2011Phys.Lett.B181,Zhao2012Phys.Rev.C54310,Zhao2015PRL,Wang2017Phys.Rev.C54324,Wang2018Phys.Rev.64321,Ren2018Sci}.

Based on the CDFT, assuming the spherical symmetry and taking into account both bound states and continuum via the Bogoliubov transformation in a microscopic and self-consistent way, the relativistic continuum Hartree-Bogoliubov (RCHB) theory was developed in Refs.~\cite{Meng1996PRL,Meng1998NPA} with the relativistic Hartree-Bogoliubov equations solved in the coordinate space. With the pairing correlation and the coupling to the continuum considered, the RCHB theory has achieved great success in reproducing and interpreting the halo in $^{11}$Li~\cite{Meng1996PRL}, predicting the giant halo phenomena~\cite{Meng1998PRL,Meng2002Phys.Rev.C41302,Zhang2002Chin.Phys.Lett.312}, reproducing the interaction cross section and charge-changing cross sections in sodium isotopes~\cite{Meng1998Phys.Lett.B1} and other light exotic nuclei~\cite{Meng2002Phys.Lett.B209}, interpreting the pseudospin symmetry in exotic nuclei~\cite{Meng1998Phys.Rev.C628,Meng1999Phys.Rev.C154}, and making predictions of new magic numbers in superheavy nuclei~\cite{Zhang2005Nucl.Phys.A106} and neutron halos in hypernuclei~\cite{Lu2003Eur.Phys.J.A19}. Recently, based on the RCHB theory with point-coupling density functional PC-PK1~\cite{Zhao2010Phys.Rev.C54319}, the first nuclear mass table including continuum effects has been constructed and the continuum effects on the limits of the nuclear landscape have been studied~\cite{Xia2018ADNDT}. It is demonstrated that the continuum effects are crucial for drip-line locations and there are totally $9035$ nuclei with $8\le Z\le 120$ predicted to be bound, which remarkably extends the existing nuclear landscapes. The RCHB mass table has been applied to investigate $\alpha$-decay energies~\cite{Zhang2016CPC} and proton radioactivity~\cite{Lim2016PRC}.

Except for doubly-magic nuclei, most nuclei in the nuclear chart deviate from the spherical shape. Solving the deformed relativistic Hartree-Bogoliubov equations in the coordinate space is extremely difficult if not impossible~\cite{Zhou2000CPL}. To provide a proper description of deformed exotic nuclei, the deformed relativistic Hartree-Bogoliubov theory in continuum (DRHBc) based on the meson-exchange density functional was developed in Refs.~\cite{Zhou2010PRC,Li2012PRC}, with the deformed relativistic Hartree-Bogoliubov equations solved in a Dirac Woods-Saxon basis~\cite{Zhou2003PRC}. Inheriting the advantages of RCHB theory and including the deformation degree of freedom, the DRHBc theory was applied to study magnesium isotopes and an interesting shape decoupling between the core and the halo was predicted in $^{44}$Mg and $^{42}$Mg~\cite{Zhou2010PRC,Li2012PRC}. The DRHBc theory has been extended to the version with density-dependent meson-nucleon couplings~\cite{Chen2012Phys.Rev.C67301}, and to incorporate the blocking effect~\cite{Li2012CPL}. The success of DRHBc theory has been demonstrated in resolving the puzzles concerning the radius and configuration of valence neutrons in $^{22}$C~\cite{Sun2018PLB}, and studying particles in the classically forbidden regions for magnesium isotopes~\cite{Zhang2019PRC}.

The deformation plays an important role in the description of nuclear masses and affects the location of neutron drip line~\cite{Xia2018ADNDT}. It is therefore necessary to construct an upgraded mass table including simultaneously the deformation and continuum effects using the DRHBc theory.

It is quite challenging to include both the deformation and continuum effects in coordinate space. In the DRHBc theory, the coupled relativistic Hartree-Bogoliubov equations are solved by the expansion on on a spherical Dirac Woods-Saxon basis~\cite{Zhou2003PRC}, with the Woods-Saxon parameters taken from Ref.~\cite{Koepf1991}. It is numerically much more complicated than the RCHB theory. So far, the DRHBc theory has been applied to light nuclei only~\cite{Zhou2010PRC,Li2012PRC,Li2012CPL,Chen2012Phys.Rev.C67301,Sun2018PLB,Zhang2019PRC}. In order to provide a unified description for all nuclei in the nuclear chart with the DRHBc theory, the difficulties include justifying a unified numerical setting, locating the ground-state deformation, and blocking the correct orbit(s) for odd-$A$ and odd-odd nuclei. Blocking the correct orbit for an odd-$A$ nucleus means that calculation should be performed by blocking several orbits near the Fermi level of its neighboring even-even nucleus independently and the one with the lowest energy should be identified~\cite{PeterBook,Li2012CPL,Xia2018ADNDT}. The blocking procedure for an odd-odd nucleus is similar to the odd-$A$ nuclei, but requires blocking for both the proton and neutron levels at the same time. Last but not the least, so far the DRHBc theory is based on the meson-exchange density functionals. It is necessary to develop the DRHBc theory with point-coupling density functionals to adopt the successful PC-PK1.

In this work, the DRHBc theory based on the point-coupling density functionals is developed and its application for even-even nuclei is discussed in detail. The formulism is presented in Sec.~\ref{theory}. Numerical checks are performed from light nuclei to heavy nuclei, and the details to construct a DRHBc mass table for even-even nuclei are suggested in Sec.~\ref{numerical}. As examples, the DRHBc calculated results for neodymium isotopes are compared with the RCHB mass table~\cite{Xia2018ADNDT} and the data available~\cite{AME2016,2016ADNDT,Angeli2013ADNDT} in Sec.~\ref{results}. A summary is given in Sec.~\ref{summary}.

%%%%%%%%%%%%%%%%%%%%%%%%%%%%%%%%%%%%%%%%%%%%%%%%%%%%%%%%%%
%                    begin  theoretical framework
%%%%%%%%%%%%%%%%%%%%%%%%%%%%%%%%%%%%%%%%%%%%%%%%%%%%%%%%%%

%----------------------------------------------------------------------------------------
\section{Theoretical framework} \label{theory}
%----------------------------------------------------------------------------------------

The DRHBc theory based on the meson-exchange density functionals has been developed~\cite{Zhou2010PRC} and the details can be found in Ref.~\cite{Li2012PRC}. In this paper the DRHBc theory with point-coupling density functionals is developed and its formulism is presented in the following in brief.

The point-coupling density functional starts from the following Lagrangian density~\cite{Meng2016Book}:
\begin{equation}\label{lagrangian}
\begin{aligned}
\mathcal L = & \bar\psi(i\gamma_\mu \partial^\mu -M)\psi-\frac{1}{2}\alpha_S(\bar\psi\psi)(\bar\psi\psi)\\
&-\frac{1}{2}\alpha_V(\bar\psi\gamma_\mu\psi)(\bar\psi\gamma^\mu\psi)-\frac{1}{2}\alpha_{TV}(\bar\psi\vec\tau\gamma_\mu\psi)(\bar\psi\vec\tau\gamma^\mu\psi)\\
&-\frac{1}{2}\alpha_{TS}(\bar\psi\vec\tau\psi)(\bar\psi\vec\tau\psi)-\frac{1}{3}\beta_S(\bar\psi \psi)^3\\
&-\frac{1}{4}\gamma_S(\bar \psi \psi)^4-\frac{1}{4}\gamma_V[(\bar\psi\gamma_\mu\psi)(\bar\psi\gamma^\mu\psi)]^2\\
&-\frac{1}{2}\delta_S\partial_\nu(\bar\psi\psi)\partial^\nu(\bar\psi\psi)
-\frac{1}{2}\delta_V\partial_\nu(\bar\psi\gamma_\mu\psi)\partial^\nu(\bar\psi\gamma^\mu\psi)\\
&-\frac{1}{2}\delta_{TV}\partial_\nu(\bar\psi\vec\tau\gamma_\mu\psi)\partial^\nu(\bar\psi\vec\tau\gamma^\mu\psi)\\
&-\frac{1}{2}\delta_{TS}\partial_\nu(\bar\psi\vec\tau\psi)\partial^\nu(\bar\psi\vec\tau\psi)\\
&-\frac{1}{4}F^{\mu\nu}F_{\mu\nu}-e\bar\psi\gamma^\mu \frac{1-\tau_3}{2}A_\mu \psi,
\end{aligned}
\end{equation}
where $M$ is the nucleon mass, $e$ is the charge unit, and $A_\mu$ and $F_{\mu\nu}$ are the four-vector potential and field strength tensor of the electromagnetic field, respectively. Here $\alpha_S, \alpha_V, \alpha_{TS},$ and $\alpha_{TV}$ represent the coupling constants for four-fermion terms, $\beta_S, \gamma_S,$ and $\gamma_V$ are those for the higher-order terms which are responsible for the medium effects, and $\delta_S, \delta_V, \delta_{TS},$ and $\delta_{TV}$ refer to those for the gradient terms which are included to simulate the finite-range effects. The subscripts $S, V,$ and $T$ stand for scalar, vector, and isovector, respectively. The isovector-scalar channel including the terms $\alpha_{TS}$ and $\delta_{TS}$ in Eq.~(\ref{lagrangian}) are neglected since including the isovector-scalar interaction does not improve the description of nuclear ground-state properties~\cite{PhysRevC.65.044308}.

From the Lagrangian density of Eq.~(\ref{lagrangian}), the energy density functional for the nuclear system can be constructed under the mean-field and no-sea approximations. By minimizing the energy density functional with respect to the densities, one obtains the Dirac equation for nucleons within the relativistic mean-field framework~\cite{Meng2016Book}. The pairing correlation is crucial in the description of open-shell nuclei. The conventional BCS theory used extensively in describing the pairing correlation turns out to be an insufficient approach for exotic nuclei~\cite{DOBACZEWSKI1984103}. The relativistic Hartree-Bogoliubov (RHB) theory can provide a unified and self-consistent treatment of both the mean field and the pairing correlation~\cite{Kucharek1991,GONZALEZLLARENA1996PLB,Meng1998NPA,Serra2002PRC}, and can describe the exotic nuclei properly in the coordinate space~\cite{Meng1998NPA} or a Dirac Woods-Saxon basis~\cite{Zhou2010PRC}.

The RHB equation reads
\begin{equation}\label{RHB}
\left(\begin{matrix}
\hat h_D-\lambda_\tau & \hat \Delta \\
-\hat \Delta^* &-\hat h_D^*+\lambda_\tau
\end{matrix}\right)\left(\begin{matrix}
U_k\\
V_k
\end{matrix}\right)=E_k\left(\begin{matrix}
U_k\\
V_k
\end{matrix}\right),
\end{equation}
where $\hat h_D$ is the Dirac Hamiltonian, $\hat\Delta$ is the pairing field, $\lambda_\tau$ is the Fermi energy for neutron or proton ($\tau=\mathrm{n,p}$), $E_k$ is the quasiparticle energy, and $U_k$ and $V_k$ are the quasiparticle wave functions.

The Dirac Hamiltonian in the coordinate space is
\begin{equation}
h_D(\bm{r})=\bm{\alpha}\cdot\bm{p}+V(\bm{r})+\beta[M+S(\bm{r})],
\end{equation}
with the scalar and vector potentials
\begin{align}
S(\bm r) & =  \alpha_S \rho_S + \beta_S \rho^2_S + \gamma_S \rho^3_S + \delta_S \Delta\rho_S,\label{S} \\
V(\bm r) & = \alpha_V \rho_V + \gamma_V \rho^3_V + \delta_V \Delta\rho_V + e A^0 + \alpha_{TV}\tau_3\rho_3 +\delta_{TV}\tau_3\Delta\rho_3,\label{V}
\end{align}
constructed by various densities
\begin{equation}\label{densities}
\begin{split}
&\rho_S(\bm r)=\sum_{k>0} V_k^\dag (\bm r)\gamma_0 V_k (\bm r),\\
&\rho_V(\bm r)=\sum_{k>0} V_k^\dag (\bm r) V_k (\bm r),\\
&\rho_3(\bm r)=\sum_{k>0} V_k^\dag (\bm r)\tau_3 V_k (\bm r).
\end{split}
\end{equation}
According to the no-sea approximation, the summations in above equations are performed over the quasiparticle states with positive energies in the Fermi sea.

The pairing potential is
\begin{equation}
\Delta(\bm{r}_1 s_1 p_1,\bm{r}_2 s_2 p_2)=\sum_{s_1'p_1'}^{s_2'p_2'}V^{pp}(\bm{r}_1,\bm{r}_2;s_1p_1,s_2p_2,s_1'p_1',s_2'p_2')\times
\kappa(\bm{r}_1 s_1' p_1',\bm{r}_2 s_2' p_2'),
\end{equation}
where $s$ represents the spin degree of freedom, $p$ represents the upper or lower component of the Dirac spinors, $\kappa(\bm{r}_1 s_1' p_1',\bm{r}_2 s_2' p_2')$ is the pairing tensor~\cite{PeterBook}, and $V^{pp}$ is the pairing interaction in the particle-particle channel. Here a density-dependent zero-range pairing force is adopted,
\begin{equation}\label{pair}
V^{\mathrm{pp}}(\bm r_1,\bm r_2)= V_0 \frac{1}{2}(1-P^\sigma)\delta(\bm r_1-\bm r_2)\left(1-\frac{\rho(\bm r_1)}{\rho_{\mathrm{sat}}}\right),
\end{equation}
with $V_0$ the pairing strength, $\rho_{\mathrm{sat}}$ the saturation density of nuclear matter, and $\frac{1}{2}(1-P^\sigma)$ projector for the spin $S=0$ component in the pairing channel. Details of the calculations of pairing tensor and pairing potential can be found in Ref.~\cite{Li2012PRC}.

%written as
%\begin{equation}\label{kappa}
%\kappa(\bm{r}_1 s_1' p_1',\bm{r}_2 s_2' p_2')=\sum_{k>0}  V_k^\dag(\bm{r}_1 s_1' p_1')  U_k (\bm{r}_2 s_2' p_2').
%\end{equation}

For axially deformed nuclei, the potentials in Eqs.~(\ref{S}) and (\ref{V}) together with densities in Eq.~(\ref{densities}) are expanded in terms of the Legendre polynomials~\cite{PhysRevC.36.354},
\begin{equation}\label{legendre}
f(\bm r)=\sum_\lambda f_\lambda(r)P_\lambda(\cos\theta),~~\lambda=0,2,4,\cdots,
\end{equation}
with
\begin{equation}
f_\lambda(r)= \frac{2\lambda+1}{4\pi}\int d\Omega f(\bm r) P_\lambda(\Omega).
\end{equation}
Because of the spatial reflection symmetry, $\lambda$ is restricted to be even numbers.

For exotic nuclei with the Fermi energy very close to the continuum threshold, the pairing interaction can scatter nucleons from bound states to the resonant states in the continuum. The density could become more diffuse due to this coupling to continuum, and the position of the drip-line might be influenced, which is the so-called continuum effects. In order to take into account the continuum effects, the deformed RHB equations are solved in a spherical Dirac Woods-Saxon basis, in which the radial wave functions have a proper asymptotic behavior for large $r$~\cite{Zhou2003PRC}. For techniques to treat strictly the boundary condition for continuum, see Refs.~\cite{Grasso2001PRC,Michel2008PRC,Pei2011PRC,Zhang2012PRC}.

The Dirac Woods-Saxon basis is obtained by solving a Dirac equation with spherical Woods-Saxon scalar and vector potentials~\cite{Koepf1991,Zhou2003PRC}. The basis wave function reads
\begin{equation}
\varphi_{n\kappa m}(\bm{r} s)=\frac{1}{r}\left( \begin{matrix}
iG_{n\kappa}(r) Y_{jm}^l(\Omega s) \\
-F_{n\kappa}(r) Y_{jm}^{\tilde{l}}(\Omega s)
\end{matrix}\right),
\end{equation}
with $G_{n\kappa}(r)/r$ and $F_{n\kappa}(r)/r$ the radial wave functions for large and small components, and $Y_{jm}^{l(\tilde{l})}(\Omega s)$ the spin spherical harmonics, where $n$ is the radial quantum number, $\kappa=(-1)^{j+l+1/2}(j+1/2)$, and $\tilde{l}=l+(-1)^{j+l-1/2}$. For the completeness of basis, the solutions in the Dirac sea should also be included in the basis space~\cite{Zhou2003PRC}.

With a set of complete Dirac Woods-Saxon basis, solving the RHB equation~(\ref{RHB}) is equivalent to the diagonalization of RHB matrix. Symmetries can simplify the calculation considerably. For axially deformed nuclei with the spatial reflection symmetry, the parity $\pi$ and the projection of the angular momentum on the symmetry axis $m$ are good quantum numbers. Therefore, the RHB matrix can be decomposed into different $m^\pi$ blocks. Moreover, because of the time-reversal symmetry, one only needs to diagonalize the RHB matrix in each positive-$m$ block,
\begin{equation}
\left( \begin{matrix}
\mathcal{A}-\lambda_\tau & \mathcal{B} \\
\mathcal{B}^\dag & -\mathcal{A}^*+\lambda_\tau
\end{matrix}\right)\left( \begin{matrix}
\mathcal{U}_k \\
\mathcal{V}_k
\end{matrix}\right)=
E_k\left( \begin{matrix}
\mathcal{U}_k \\
\mathcal{V}_k
\end{matrix}\right),
\end{equation}
where the matrix elements are
\begin{align}
& \mathcal{A}=(h_{D(n\kappa)(n'\kappa')}^{(m)})=(\langle n\kappa m|h_D|n'\kappa'm\rangle),\\
& \mathcal{B}=(\Delta_{(n\kappa)(n'\kappa')}^{(m)})=(\langle n\kappa m|\Delta|\overline{n'\kappa'm}\rangle).
\end{align}
For odd systems, the equal filling approximation that conserves time-reversal symmetry is adopted~\cite{Li2012CPL}. The details of the calculation of RHB matrix elements can be found in Ref.~\cite{Li2012PRC}. The obtained eigenvectors correspond to the expansion coefficients of quasiparticle wave functions in the Dirac Woods-Saxon basis
\begin{equation}
\mathcal{U}_k=(u_{k,(n\kappa)}^{(m)}),~~\mathcal{V}_k=(v_{k,(n\kappa)}^{(m)}).
\end{equation}
From these quasiparticle wave functions, new densities and potentials can be obtained, which are iterated in the RHB equations until the convergence is achieved.

Finally, one can calculate the total energy of a nucleus by~\cite{Meng1998NPA,Meng2016Book}
\begin{equation}
\begin{split}
E_{\mathrm{RHB}}=& \sum_{k>0} (\lambda_\tau-E_k)v_k^2-E_{\mathrm{pair}} \\
 & - \int \mathrm{d}^3 \bm r \left(\frac{1}{2}\alpha_S \rho_S^2 + \frac{1}{2}\alpha_V \rho_V^2 + \frac{1}{2}\alpha_{TV}\rho_3^2 \right.\\
 & + \left.\frac{2}{3}\beta_S \rho^3_S + \frac{3}{4}\gamma_S \rho^4_S + \frac{3}{4}\gamma_V \rho^4_V + \frac{1}{2}\delta_S \rho_S \Delta\rho_S \right. \\
 & + \left.\frac{1}{2} \delta_V \rho_V \Delta\rho_V +\frac{1}{2}\delta_{TV}\rho_3 \Delta\rho_3 +\frac{1}{2} \rho_p e A^0\right) \\
 & + E_{\mathrm{c.m.}},
\end{split}
\end{equation}
where
\begin{equation}
v_k^2=\int d^3 \bm r V_k^\dag(\bm r) V_k(\bm r).
\end{equation}
For the zero-range pairing force, the pairing field $\Delta(\bm r)$ is local, and the pairing energy is calculated as
\begin{equation}
E_{\mathrm{pair}}=-\frac{1}{2}\int d^3 \bm r\kappa(\bm r)\Delta(\bm r).
\end{equation}
The center-of-mass (c.m.) correction energy is calculated microscopically,
\begin{equation}
E_{\mathrm{c.m.}}=-\frac{1}{2mA}\langle \hat {\bm P}^2 \rangle,
\end{equation}
with $A$ the mass number and $\hat {\bm P}=\sum_i^A \hat {\bm p}_i$ the total momentum in the c.m. frame. It has been shown that the microscopic c.m. correction provides more reasonable and reliable results than phenomenological ones~\cite{Bender2000,Long2004Phys.Rev.C34319,Zhao2009Chin.Phys.Lett.112102}.

For deformed nuclei, as the rotational symmetry is broken in the mean-field approximation, the rotational correction energy, i.e., the energy gained by the restoration of rotational symmetry, should also be included properly~\cite{Zhao2010Phys.Rev.C54319}. Here the rotational correction energy is obtained from the cranking approximation,
\begin{equation}\label{rot}
E_{\mathrm{rot}}= -\frac{1}{2\mathscr{I}}\langle\hat {\bm J}^2\rangle,
\end{equation}
where $\mathscr{I}$ is the moment of inertia calculated by the Inglis-Belyaev formula~\cite{PeterBook} and $\hat {\bm J}=\sum_i^A \hat {\bm j}_i$ is the total angular momentum.

The root-mean-square (rms) radius is calculated as
\begin{equation}\label{radius}
R_{\tau,\mathrm{rms}}= \langle r^2\rangle^{1/2}= \sqrt{ \frac{\int d^3 \bm r [r^2\rho_\tau(\bm r)]}{N_\tau}}.
\end{equation}
where $\tau$ represents the proton, the neutron or the nucleon, and $\rho_\tau$ is the corresponding vector density, and $N_\tau$ refers to the corresponding particles number. The rms charge radius is simply calculated as
\begin{equation}\label{rch}
R_{\mathrm{c}}= \sqrt{R_{p,\mathrm{rms}}^2+0.64~\mathrm{fm}^2}.
\end{equation}

The intrinsic quadrupole moment is calculated by
\begin{equation}
Q_{\tau,2}= \sqrt{\frac{16\pi}{5}}\langle r^2 Y_{20}(\theta,\varphi)\rangle.
\end{equation}
The quadrupole deformation parameter is obtained from the quadrupole moment by
\begin{equation}\label{def}
\beta_{\tau,2}= \frac{\sqrt{5\pi}Q_{\tau,2}}{3N_\tau\langle r_\tau^2\rangle}.
\end{equation}

The canonical basis $\ket{\psi_i}$ can be obtained by diagonalizing the density matrix $\hat{\rho}$~\cite{PeterBook},
\begin{equation}\label{canonical}
 \hat{\rho} \ket{\psi_i} = v_i^2 \ket{\psi_i},
\end{equation}
where the eigenvalue $v_i^2$ is the corresponding occupation probability of $\ket{\psi_i}$.
It has to be emphasized that, in a diagonalization problem, the degeneration of eigenvalues will lead to an arbitrary mixture of the eigenvectors that satisfies the unitary transformation in the corresponding subspace. As a consequence, the canonical states are not uniquely defined when their occupation probabilities are degenerate. The problem can be solved by diagonalizing $\hat{h}$ in the subspace with degenerate occupation probabilities to determine the canonical single-particle states uniquely~\cite{Meng2016Book}.

%%%%%%%%%%%%%%%%%%%%%%%%%%%%%%%%%%%%%%%%%%%%%%%%%%%%%%%%%%
%                    begin  numerical details
%%%%%%%%%%%%%%%%%%%%%%%%%%%%%%%%%%%%%%%%%%%%%%%%%%%%%%%%%%

%----------------------------------------------------------------------------------------
\section{Numerical details}\label{numerical}
%----------------------------------------------------------------------------------------

Here we concentrate on the numerical details in the systematic calculations for even-even nuclei from the proton drip lines to the neutron drip lines in the nuclear chart with the DRHBc theory. For the particle-hole channel, the relativistic density functional PC-PK1~\cite{Zhao2010Phys.Rev.C54319}, which has turned out to be very successful in providing good descriptions of the isospin dependence of the binding energy along both the isotopic and the isotonic chain~\cite{Zhao2012Phys.Rev.C64324,Zhang2014Front.Phys.529,Lu2015Phys.Rev.C27304}, is adopted. For the particle-particle channel, the density-dependent zero-range pairing force in Eq.~(\ref{pair}) is used.

In the DRHBc theory, the relativistic Hartree-Bogoliubov equations are solved by the expansion on a spherical Dirac Woods-Saxon basis~\cite{Zhou2003PRC}, with the Woods-Saxon parameters taken from Ref.~\cite{Koepf1991}. Therefore, the box size $R_{\mathrm{box}}$ and the mesh size $\Delta r$ for the Dirac Woods-Saxon basis should be determined. Secondly, for the completeness of basis space, an angular momentum cutoff $J_{\mathrm{max}}$, an energy cutoff $E^+_{\mathrm{cut}}$ for the Woods-Saxon basis in the Fermi sea, and the number of states in the Dirac sea should be chosen properly. Thirdly, the convergence of Legendre expansion in Eq.~(\ref{legendre}) for the deformed densities and potentials should be guaranteed. Finally, the pairing strength in Eq.~(\ref{pair}) should be justified properly.

In Ref.~\cite{Zhou2003PRC}, the solutions of Dirac equations in the Dirac Woods-Saxon basis with $R_{\mathrm{box}}=20$ fm and $\Delta r=0.1$ fm reproduce accurately the results obtained by the shooting method. In the RCHB mass table~\cite{Xia2018ADNDT}, $R_{\mathrm{box}}=20$ fm and $\Delta r=0.1$ fm have been chosen.  Here we have further checked the convergence of DRHBc solutions with respect to $R_{\mathrm{box}}$ and $\Delta r$ for deformed nuclei $^{20}$Ne, $^{112}$Mo, and $^{300}$Th, and found that $R_{\mathrm{box}}=20$ fm and $\Delta r=0.1$ fm lead to a satisfactory accuracy of less than $0.01\%$ of the binding energies. Therefore, the box size $R_{\mathrm{box}}=20$ fm and the mesh size $\Delta r=0.1$ fm are used in the present DRHBc calculations.

In the following, numerical checks for the energy cutoff $E^+_{\mathrm{cut}}$ and the angular momentum cutoff $J_{\mathrm{max}}$ will be performed. The number of states in the Dirac sea is taken to be the same as that in the Fermi sea~\cite{Zhou2003PRC,Zhou2010PRC,Li2012PRC}. Convergence check for the Legendre expansion will also be performed. In addition, the pairing strength will be determined by reproducing experimental odd-even mass differences, and the strategy to determine ground states in the DRHBc calculations will be suggested according to the self-consistency between unconstrained and constrained calculations.

\subsection{Energy cutoff for Woods-Saxon basis}

%----------------------------------------------------------------------------------------
\begin{figure}[htbp]
  \centering
  \includegraphics[scale=0.55,angle=0]{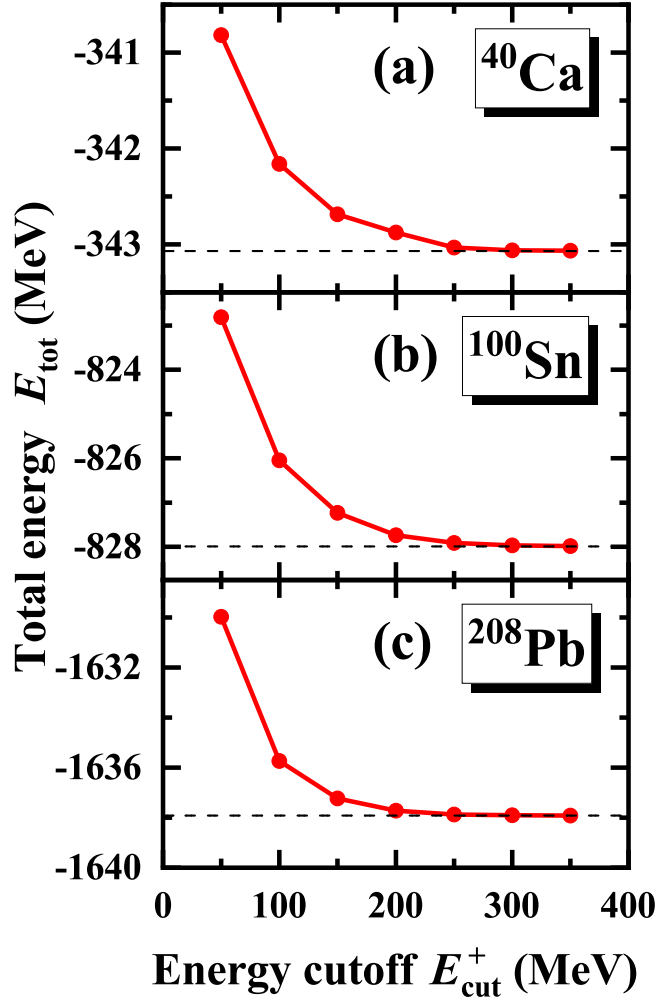}
  \caption{Total energy as a function of the energy cutoff $E^+_{\mathrm{cut}}$ for doubly-magic nuclei $^{40}$Ca (a), $^{100}$Sn (b), and $^{208}$Pb (c) calculated by the DRHBc theory. Dashed lines show total energies of these three nuclei in the RCHB mass table~\cite{Xia2018ADNDT}. Same as the RCHB calculations~\cite{Xia2018ADNDT}, the angular momentum cutoff $J_{\mathrm{max}}=19/2~\hbar$ is used here.}
\label{fig1}
\end{figure}
%----------------------------------------------------------------------------------------

In Ref.~\cite{Zhou2003PRC}, it is found that the results of the calculations with the Dirac Woods-Saxon basis converge to the exact ones with the energy cutoff $E^+_{\mathrm{cut}}\approx 300$ MeV. Here we perform the fully self-consistent calculations to examine the convergence of total energy with the energy cutoff $E^+_{\mathrm{cut}}$, as seen in Fig.~\ref{fig1}, for doubly-magic nuclei $^{40}$Ca, $^{100}$Sn, and $^{208}$Pb. The results from the RCHB mass table~\cite{Xia2018ADNDT} are also shown for comparison. The total energy of each nucleus converges gradually to the corresponding RCHB result with the increasing $E^+_{\mathrm{cut}}$. When $E^+_{\mathrm{cut}}=300$ MeV, the total energy differences between the DRHBc and RCHB calculations for $^{40}$Ca, $^{100}$Sn, and $^{208}$Pb are $0.0097$ MeV, $0.0193$ MeV, and $0.0179$ MeV, respectively. Changing $E^+_{\mathrm{cut}}$ from $300$ MeV to $350$ MeV, the total energy varies by $0.0037$ MeV, $0.0090$ MeV, and $0.0093$ MeV for $^{40}$Ca, $^{100}$Sn, and $^{208}$Pb, respectively. Therefore, consistent with the conclusion in Ref.~\cite{Zhou2003PRC} and the RCHB mass table~\cite{Xia2018ADNDT}, $E^+_{\mathrm{cut}}=300$ MeV is a reasonable choice for the DRHBc mass table calculations.

\subsection{Angular momentum cutoff}

%----------------------------------------------------------------------------------------
\begin{figure}[htbp]
  \centering
  \includegraphics[scale=0.55,angle=0]{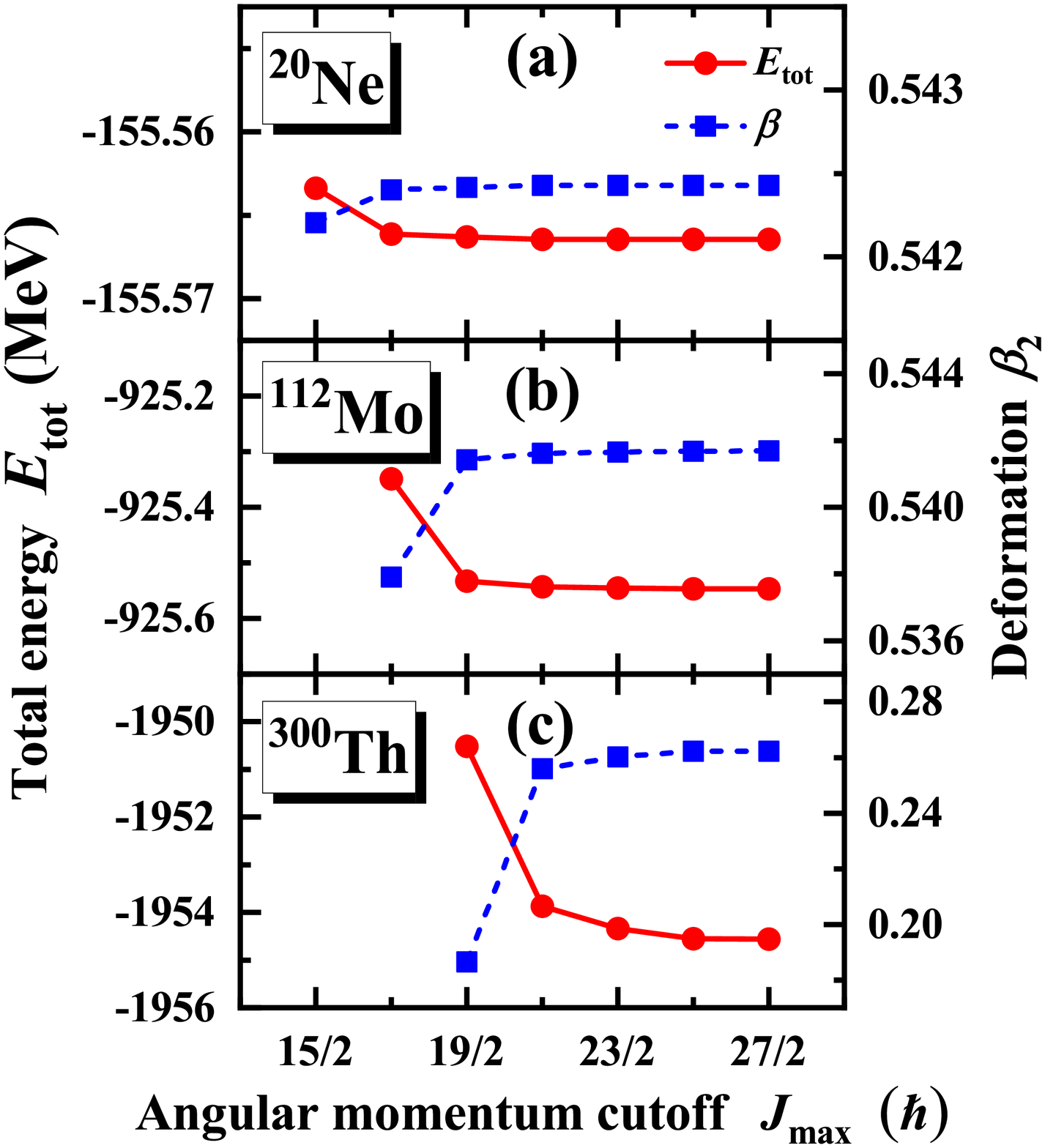}
  \caption{Total energy and deformation versus the angular momentum cutoff $J_{\mathrm{max}}$ for deformed nuclei $^{20}$Ne (a), $^{112}$Mo (b), and $^{300}$Th (c), with the energy cutoff $E^+_{\mathrm{cut}}=300$ MeV. Here the pairing correlation is neglected.}
\label{fig2}
\end{figure}
%----------------------------------------------------------------------------------------

In the RCHB mass table calculations, the convergence has been confirmed for the angular momentum cutoff $J_{\mathrm{max}}=19/2~\hbar$~\cite{Xia2018ADNDT}. With deformation effects included in DRHBc calculations, further numerical checks for $J_{\mathrm{max}}$ are necessary.

Figure~\ref{fig2} shows the total energy and deformation versus the angular momentum cutoff $J_{\mathrm{max}}$ for deformed nuclei $^{20}$Ne, $^{112}$Mo, and $^{300}$Th, where $E^+_{\mathrm{cut}}=300$ MeV and the pairing is neglected. The stable light nucleus $^{20}$Ne, short-lived medium-heavy nucleus $^{112}$Mo, and neutron-rich heavy nucleus $^{300}$Th are chosen in order to determine a universal angular momentum cutoff. It is found that $J_{\mathrm{max}}=19/2~\hbar$ is enough for light nuclei like $^{20}$Ne and medium-heavy nuclei like $^{112}$Mo. For heavy nucleus $^{300}$Th, changing $J_{\mathrm{max}}$ from $19/2~\hbar$ to $27/2~\hbar$, the deformation varies by about $0.08$ and the total energy varies by $4.0406$ MeV. Changing $J_{\mathrm{max}}$ from $23/2~\hbar$ to $27/2~\hbar$, the deformation varies by about $0.002$ and the total energy varies by $0.2180$ MeV, which is about $0.01\%$ of its total energy. Therefore, a unified angular momentum cutoff $J_{\mathrm{max}}=23/2~\hbar$ is suggested in the DRHBc calculations in order to achieve a satisfactory accuracy for the entire nuclear landscape.

\subsection{Legendre expansion}

%----------------------------------------------------------------------------------------
\begin{figure}[htbp]
  \centering
  \includegraphics[scale=0.55,angle=0]{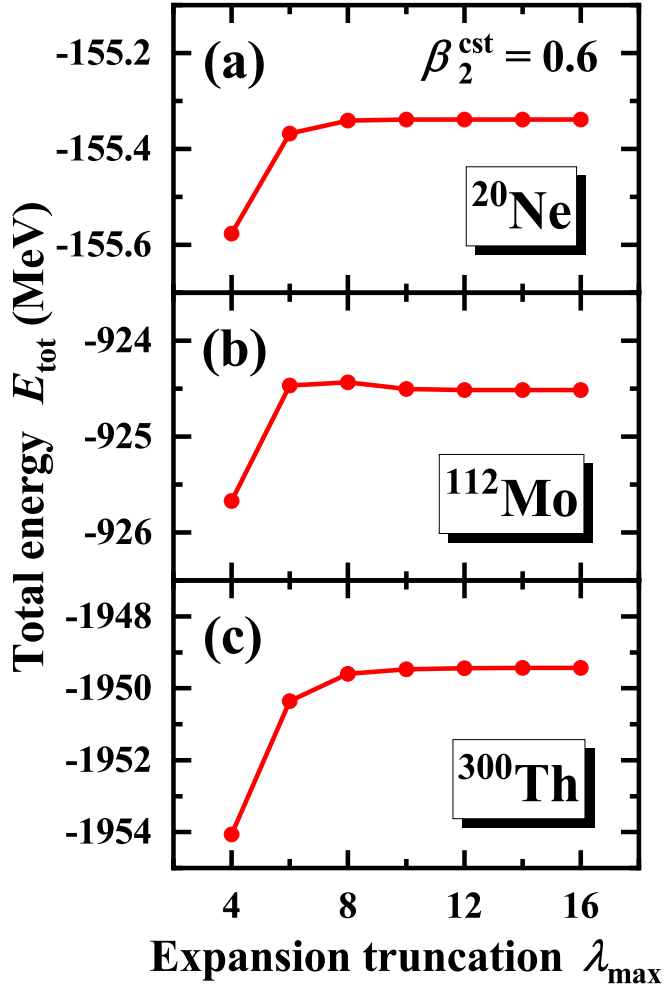}
  \caption{Total energy as a function of the Legendre expansion truncation $\lambda_{\max}$ for nuclei $^{20}$Ne (a), $^{112}$Mo (b), and $^{300}$Th (c) from constrained DRHBc calculations at the quadrupole deformation $\beta_2=0.6$, with the energy cutoff $E^+_{\mathrm{cut}}=300$ MeV and the angular momentum cutoff $J_{\max}=23/2~\hbar$. Here the pairing correlation is neglected.}
  \label{fig3}
\end{figure}
%----------------------------------------------------------------------------------------

In the DRHBc theory, the deformed densities and potentials are expanded in terms of the Legendre polynomials as in Eq.~(\ref{legendre})~\cite{Zhou2010PRC}. Since a nucleus with a large deformation may need higher orders in the Legendre expansion, the convergence of the expansion truncation $\lambda_{\max}$ is checked for nuclei $^{20}$Ne, $^{112}$Mo, and $^{300}$Th at the constrained deformation $\beta_2=0.6$. Figure~\ref{fig3} shows the total energies as a function of the Legendre expansion truncation $\lambda_{\max}$ for $^{20}$Ne, $^{112}$Mo, and $^{300}$Th. Changing $\lambda_{\max}$ from $6$ to $16$, the total energy varies by 0.03 MeV for $^{20}$Ne and $0.05$ MeV for $^{112}$Mo, i.e., less than $0.03\%$ of their total energies. Changing $\lambda_{\max}$ from $8$ to $16$, the total energy of $^{300}$Th varies by $0.16$ MeV, i.e., less than $0.01\%$ of its total energy. Therefore, in the mass table calculations, for light nuclei like $^{20}$Ne and medium-heavy nuclei like $^{112}$Mo, $\lambda_{\max}=6$ can provide converged results. For heavy nuclei like $^{300}$Th, $\lambda_{\max}=8$ is necessary in order to achieve convergence. Although the pairing correlation is neglected, the conclusion is also valid after the inclusion of the pairing correlation~\cite{Pan2019IJMPE}.

%----------------------------------------------------------------------------------------
\begin{figure}[htbp]
  \centering
  \includegraphics[scale=0.5,angle=0]{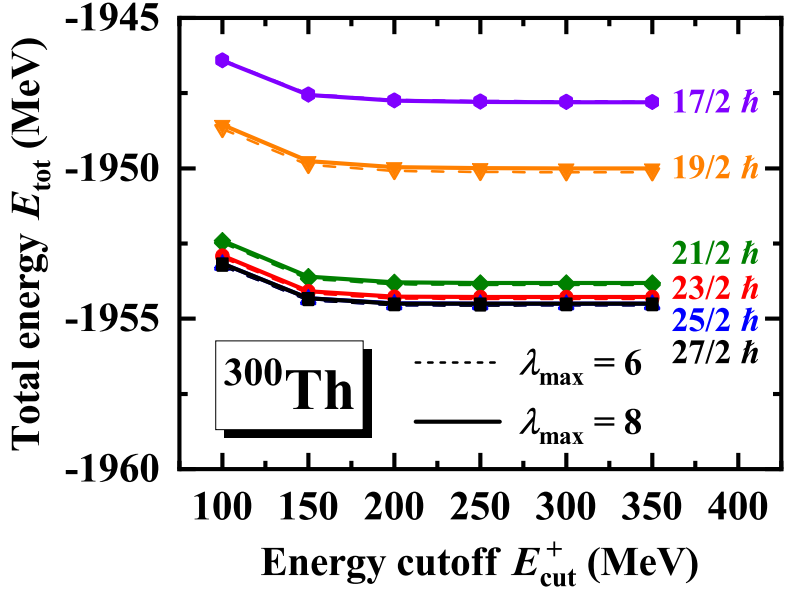}
  \caption{The convergence of the total energy in $^{300}$Th with the energy cutoff for the angular momentum cutoff from $17/2~\hbar$ to $27/2~\hbar$, in which the Legendre expansion truncation $\lambda_{\max} = 6$ and 8, labeled by dashed and solid lines, respectively. Here the pairing correlation is neglected.}
  \label{fig4}
\end{figure}
%----------------------------------------------------------------------------------------

In order to confirm the above-obtained numerical settings, Fig.~\ref{fig4} shows the convergence of the total energy in $^{300}$Th with respect to the energy cutoff, the angular momentum cutoff, and the Legendre expansion truncation simultaneously. For given angular momentum cutoff and Legendre expansion truncation, the energy difference is less than $0.01$ MeV between $E_{\mathrm{cut}}^+ = 300$ MeV and $350$ MeV. For given energy cutoff and Legendre expansion truncation, the energy difference is less than $0.27$ MeV between $J_{\max} = 23/2~\hbar$ and $27/2~\hbar$. For given energy cutoff and angular momentum cutoff, the energy difference is less than $0.13$ MeV between $\lambda_{\max} = 6$ and 8. Therefore, the suggested numerical settings obtained by the independent convergence check of each parameter are confirmed by varying all three parameters at the same time.

\subsection{Pairing strength}

\begin{figure}[htbp]
  \centering
  \includegraphics[scale=0.6,angle=0]{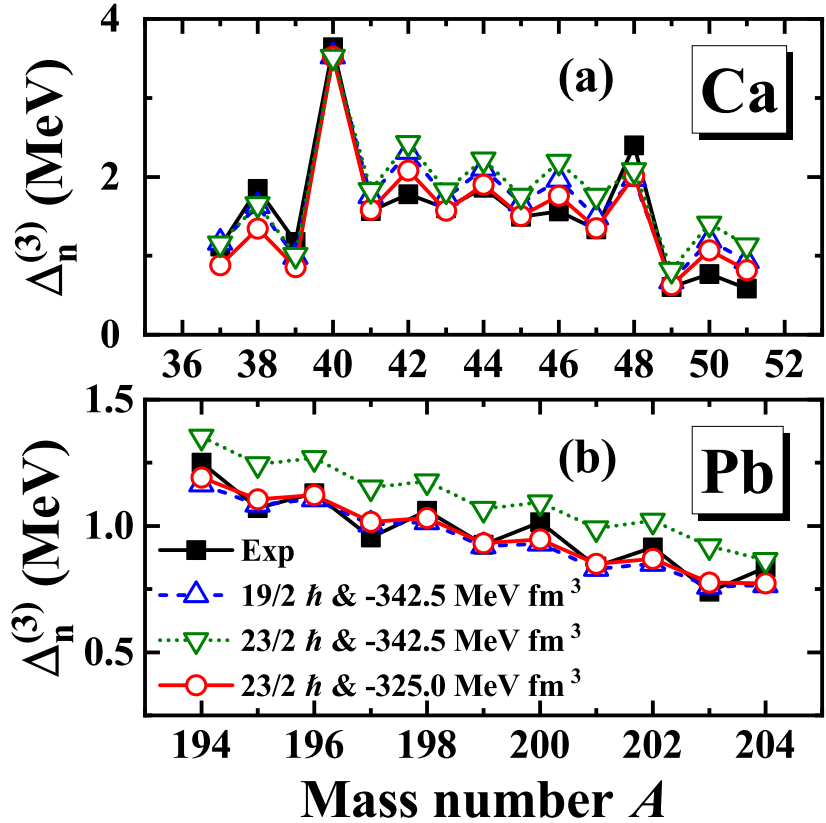}
  \caption{Odd-even mass differences of Ca (a) and Pb (b) isotopic chains in the DRHBc calculations versus the mass number, for $V_0=-342.5~\mathrm{MeV~fm}^3$ with $J_{\mathrm{max}}=23/2~\hbar$ (inverted triangle) and for $V_0=-325.0~\mathrm{MeV~fm}^3$ with $J_{\mathrm{max}}=23/2~\hbar$ (circle). The corresponding experimental data~\cite{AME2016} (square) and the results in the RCHB mass table~\cite{Xia2018ADNDT} (triangle) are shown for comparison.}
\label{fig5}
\end{figure}

In the present DRHBc calculations, the saturation density $\rho_{\mathrm{sat}}=0.152~\mathrm{fm}^{-3}$ in Eq.~(\ref{pair}) is used. Same as the calculations for the RCHB mass table~\cite{Xia2018ADNDT}, a cutoff energy $100$ MeV in the quasiparticle space is used for the pairing window. With the angular momentum cutoff $J_{\mathrm{max}}=23/2~\hbar$, the pairing strength is chosen to reproduce the experimental odd-even mass differences,
\begin{equation}
\Delta_{\mathrm{n}}^{(3)}=\frac{(-1)^N}{2}[E_{\mathrm{b}}(Z,N+1)-2E_{\mathrm{b}}(Z,N)+E_{\mathrm{b}}(Z,N-1)].
\end{equation}
The odd-even mass differences in Ca and Pb isotopic chains are used to fix the pairing strength. As Ca and Pb isotopes are proton magic nuclei, the spherical symmetry is assumed in the calculation, i.e., the Legendre expansion truncation is taken as $\lambda_{\max} = 0$. We realize that some odd-mass isotopes~\cite{Hofmann1988PLB,Rutz1998NPA,Rutz1999PLB} and some mid-shell neutron-deficient Pb isotopes~\cite{Agbemava2014PRC} might not be spherical. As shown in Fig.~\ref{fig5}, the experimental odd-even mass differences can be nicely reproduced. The pairing strength thus obtained will be used to construct the mass table.

%----------------------------------------------------------------------------------------
%----------------------------------------------------------------------------------------

Figure~\ref{fig5} shows the DRHBc calculated odd-even mass differences for
Ca isotopes and Pb isotopes, the corresponding experimental data~\cite{AME2016}, as well as the results in the RCHB mass table~\cite{Xia2018ADNDT}. For $J_{\mathrm{max}}=23/2~\hbar$, if the pairing strength $V_0=-342.5~\mathrm{MeV~fm}^3$ in Ref.~\cite{Xia2018ADNDT} is adopted, the odd-even mass differences will be overestimated for most of Ca and Pb isotopes. In order to reproduce the experimental values, $V_0=-325.0~\mathrm{MeV~fm}^3$ should be adopted. Therefore, the pairing strength $V_0=-325.0~\mathrm{MeV~fm}^3$ will be used in the DRHBc mass table calculations.

\subsection{Constrained calculations}

In order to describe the shape of the atomic nucleus and understand the shape coexistence, it is crucial to obtain the potential energy surface (PES) of the nucleus as a function of the deformation~\cite{Zhou2016PhysScr}. In microscopic models, there are two different ways to obtain the PES, i.e., the adiabatic and configuration-fixed (diabatic) approaches~\cite{Meng2006Phys.Rev.C37303,Lu2007Eur.Phys.J.A273,Sun2008Chin.Phys.C882,Li2009Chin.Phys.C98,Zhang2009Chin.Phys.Lett.52101}. In the present DRHBc calculations, the adiabatic constrained calculation is adopted to obtain the potential energy curve (PEC) of the nucleus as a function of the quadrupole deformation and the augmented Lagrangian method~\cite{Staszczak2010} is used.

For each nucleus, in order to find the ground state, the DRHBc calculations are performed with initial deformations $\beta_2=-0.4,-0.2,0.0,0.2,0.4,$ and $0.6$. The solution with the lowest total energy corresponds to the ground state. Sometimes the constrained calculation is also necessary if the PEC is very soft, or several local minima are close to each other. In present calculations, the rotational correction is added to the mean-field minimum instead of the PEC. In general, adding the rotational correction to the PEC will lead to different ground state, in particular for the nucleus with a soft PEC or shape coexistence, which requires beyond-mean-field investigation and is out of the scope for the present study.

%----------------------------------------------------------------------------------------
\begin{figure}[htbp]
  \centering
  \includegraphics[scale=0.5,angle=0]{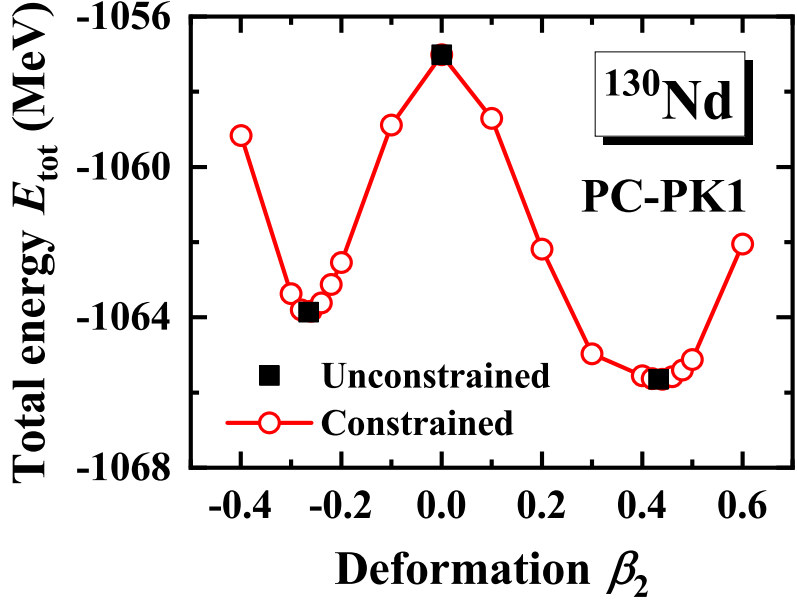}
  \caption{Potential energy curve (PEC) of $^{130}$Nd in constrained DRHBc calculations. The unconstrained results are also shown. Here the energy cutoff $E^+_{\mathrm{cut}}=300$ MeV, the angular momentum cutoff $J_{\max}=23/2~\hbar$, the Legendre expansion truncation $\lambda_{\max}=6$, and the pairing strength $V_0=-325.0~\mathrm{MeV~fm}^3$ are used.}
  \label{fig6}
\end{figure}
%----------------------------------------------------------------------------------------

Taking $^{130}$Nd as an example, the unconstrained calculations with different initial deformations respectively converge to $\beta_2=-0.27$, $0.00$, and $0.43$, as shown in Fig.~\ref{fig6}. The prolate solution has the lowest total energy and thus is considered to be the ground state. The constrained calculations are further performed for $^{130}$Nd and shown in Fig.~\ref{fig6}. Both the unconstrained prolate and oblate solutions correspond to the local minima in the PEC. Although the solution at $\beta_2=0.00$ is not a local minimum, the calculated total energy agrees with the constrained one. The self-consistency is therefore guaranteed and the strategy to find the ground state from unconstrained calculations is reasonable and practicable. Of course, whenever necessary, constrained calculations can be performed to build the PEC and confirm the ground state.

~~

Summarizing the above discussions, the numerical details for the DRHBc mass table calculations including the box size $R_{\mathrm{box}}=20$ fm, the mesh size $\Delta r=0.1$ fm, the energy cutoff $E^+_{\mathrm{cut}}=300$ MeV, the angular momentum cutoff $J_{\max}=23/2~\hbar$, the pairing strength $V_0=-325.0~\mathrm{MeV~fm}^3$, and the sharp pairing window of $100$ MeV are suggested. For Legendre expansion, the expansion truncation $\lambda_{\max}=6$ is suggested for $Z\le 80$, and $\lambda_{\max}=8$ is suggested for $Z>80$.

With the present numerical settings, the level of convergence up to $0.03\%$ for the total energy can be expected for all nuclei at ground state with deformation in the region $-0.4 \le \beta_2 \le 0.6$. Taking the possible heaviest nucleus $^{400}120$ as an example, the predicted ground-state binding energy at deformation $\beta_2 = 0.35$ is $2345.94$ MeV, with an accuracy of $550$ keV that is less than $0.03\%$ of its binding energy. For $^{230}$Hg, the predicted ground-state binding energy at deformation $\beta_2 = 0.30$ is $1707.18$ MeV, with an accuracy of $50$ keV that is less than $0.003\%$ of its binding energy.

%%%%%%%%%%%%%%%%%%%%%%%%%%%%%%%%%%%%%%%%%%%%%%%%%%%%%%%%%%
%                    begin  results and discussions
%%%%%%%%%%%%%%%%%%%%%%%%%%%%%%%%%%%%%%%%%%%%%%%%%%%%%%%%%%

%----------------------------------------------------------------------------------------
\section{Results and Discussion}\label{results}

Taking even-even Nd isotopes as examples, the DRHBc calculations with the suggested numerical details in Sec.~\ref{numerical} are performed and the ground-state properties, such as binding energy, two-neutron separation energy, Fermi energy, quadrupole deformation, rms radius, density distribution, as well as the single-particle levels are obtained. In this section, ground-state properties of Nd isotopes will be discussed and compared with those predicted by the RCHB theory~\cite{Xia2018ADNDT} and with data available~\cite{AME2016,2016ADNDT,Angeli2013ADNDT}. The ground-state properties of even-even Nd isotopes are also tabulated in Appendix~\ref{appendix}.

\subsection{Binding energy}

%----------------------------------------------------------------------------------------
\begin{figure}[htbp]
  \centering
  \includegraphics[scale=0.5,angle=0]{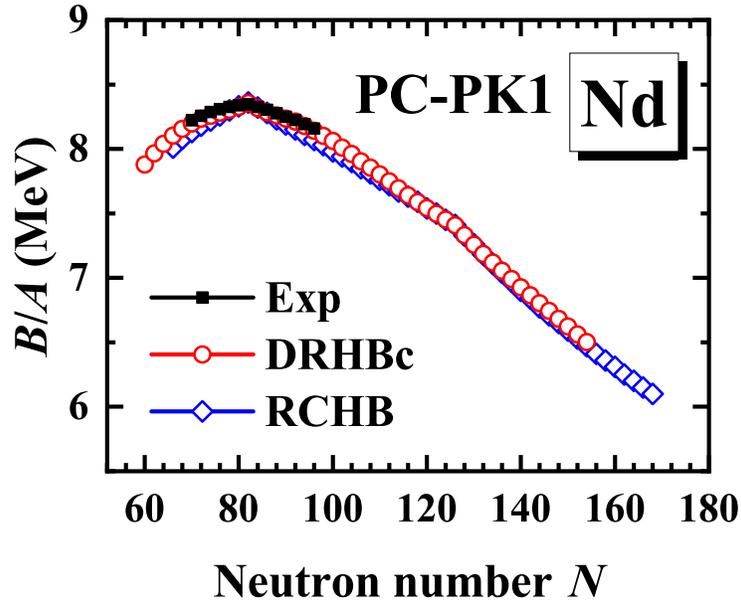}
  \caption{Binding energy per nucleon of Nd isotopes from the DRHBc calculations as a function of the neutron number. The results in the RCHB mass table~\cite{Xia2018ADNDT} and the experimental data from Ref.~\cite{AME2016} are shown for comparison.}
\label{fig7}
\end{figure}
%----------------------------------------------------------------------------------------

In Fig.~\ref{fig7}, the binding energies per nucleon for neodymium isotopes from the DRHBc calculations are shown versus the neutron number together with the RCHB results~\cite{Xia2018ADNDT} and data available~\cite{AME2016}. The most stable nucleus $^{142}$Nd in neodymium isotopes with the magic number $N=82$ is well reproduced by the DRHBc theory. Distinguishable differences between the DRHBc and the RCHB calculations can be seen. Away from the neutron shell closures $82$ and $126$, the deformation effects in the DRHBc calculations improve the RCHB results and reproduce better the data.

%----------------------------------------------------------------------------------------
\begin{figure}[htbp]
  \centering
  \includegraphics[scale=0.5,angle=0]{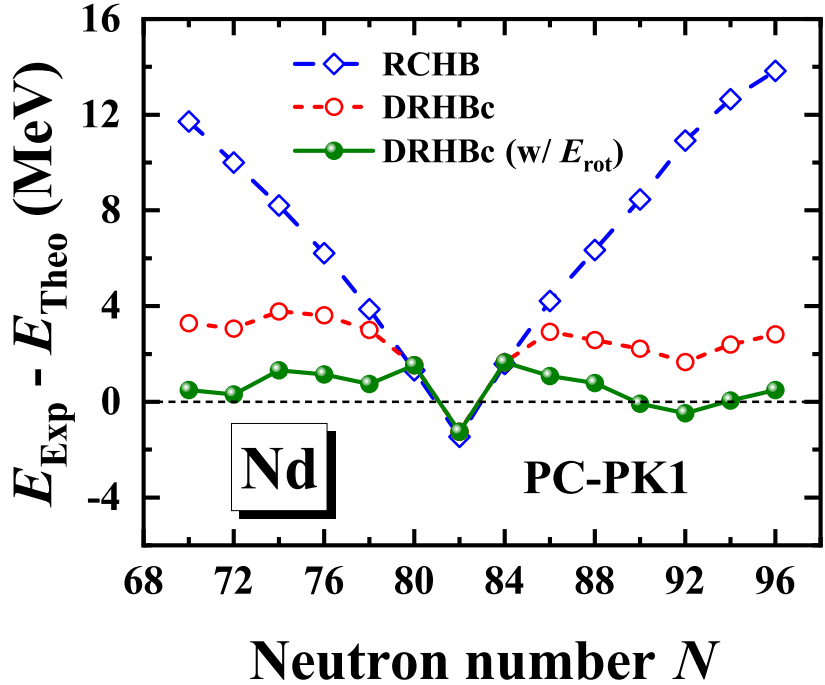}
  \caption{The difference between the experimental binding energy~\cite{AME2016} and the DRHBc calculations for Nd isotopes versus the neutron number. The results of the DRHBc calculations including rotational correction and the RCHB mass table~\cite{Xia2018ADNDT} are also shown for comparison.
  }
\label{fig8}
\end{figure}
%----------------------------------------------------------------------------------------

Figure~\ref{fig8} shows the differences between calculated binding energies and the data available~\cite{AME2016}. The deformation effects in the DRHBc calculations dramatically reduce the deviation between the RCHB calculations and the data from up to $13.8$ MeV to less than $3.8$ MeV. The rms deviation for the binding energy is reduced from $8.301$ MeV in the RCHB calculations to $2.668$ MeV in the DRHBc ones.

Following Ref.~\cite{Zhao2010Phys.Rev.C54319}, the differences after including rotational correction energies in Eq.~(\ref{rot}) in the DRHBc theory are also shown in Fig.~\ref{fig8}. The largest deviation becomes less than $1.7$ MeV and the rms deviation is reduced to $0.958$ MeV. It should be noted that one can use the Thouless-Valatin formula to better estimate the moment of inertia in the calculation of rotation correction energy~\cite{Delaroche2010PRC,Li2012Phys.Rev.C34334} and to examine if the rms deviation can be further reduced. Furthermore, the collective Hamiltonian method provides another method to better estimate the beyond-mean-field correlation energies as shown in Refs.~\cite{Libert1999PRC,Delaroche2010PRC,Lu2015Phys.Rev.C27304}.

The improved agreements between the calculated binding energies and the data by the deformation and beyond-mean-field correlation have been demonstrated by other density functional calculations as well. In Refs.~\cite{Bender2006PRC,Bender2008PRC}, based on the Hartree-Fock plus BCS theory using the Skyrme SLy4 interaction and a density-dependent zero-range pairing force together with the generator coordinate method, total energies obtained from spherical, deformed, and beyond-mean-field calculations have been compared for 605 even-even nuclei. The rms deviations from the experimental data are respectively 11.7 MeV, 5.3 MeV, and 4.4 MeV for spherical, deformed, and beyond-mean-field calculations. In Ref.~\cite{Delaroche2010PRC}, based on the constrained-Hartree-Fock-Bogoliubov theory using the Gogny D1S interaction together with a five-dimensional collective Hamiltonian, total energies obtained from spherical, deformed, and beyond-mean-field calculations have been compared for even-even nuclei with $10 \le Z \le 110$ and $N \le 200$.

\subsection{Two-neutron separation energy}

%----------------------------------------------------------------------------------------
\begin{figure}[htbp]
  \centering
  \includegraphics[scale=0.5,angle=0]{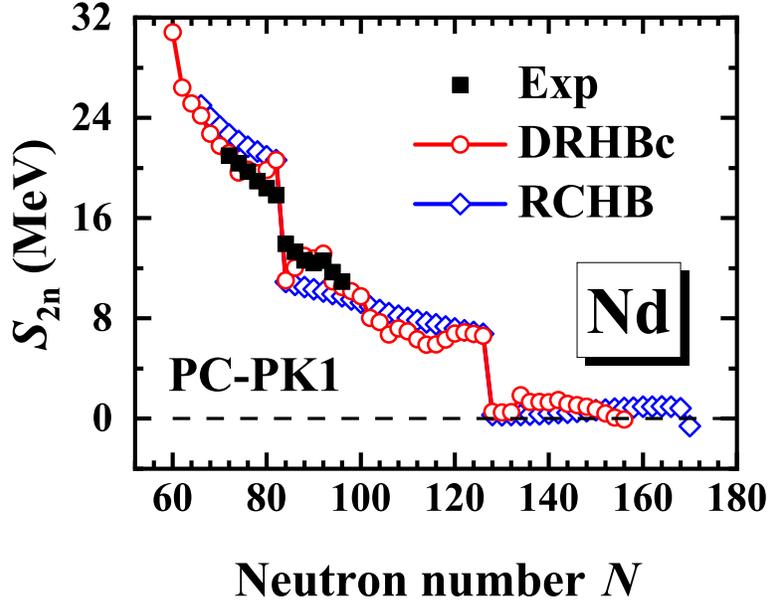}
  \caption{Two-neutron separation energy as a function of the neutron number for Nd isotopes in the DRHBc calculations. The RCHB results~\cite{Xia2018ADNDT} and the data available~\cite{AME2016} are also shown for comparison.}
\label{fig9}
\end{figure}
%----------------------------------------------------------------------------------------

From binding energies, the two-neutron separation energy can be calculated and the neutron drip line can be decided. Figure~\ref{fig9} shows the DRHBc and RCHB calculated two-neutron separation energies of neodymium isotopes, in comparison with the existing experimental data~\cite{AME2016}. The DRHBc results are consistent with the RCHB ones for spherical nuclei near the neutron magic numbers $N=82$ and $N=126$. From $^{132}$Nd to $^{140}$Nd and $^{146}$Nd to $^{156}$Nd, the DRHBc calculations including deformation effects reproduce better the experimental values.
From the DRHBc calculated two-neutron separation energies, the neutron drip-line (last bound) nucleus is predicted to be $^{214}$Nd, while it is $^{228}$Nd in the RCHB theory~\cite{Xia2018ADNDT}. Including the deformation degrees of freedom, the predicted neutron drip-line location varies by $14$ neutrons. It is an interesting topic to investigate the deformation effects on the verge of whole nuclear landscape.

\subsection{Fermi energy}

%----------------------------------------------------------------------------------------
\begin{figure}[htbp]
  \centering
  \includegraphics[scale=0.45,angle=0]{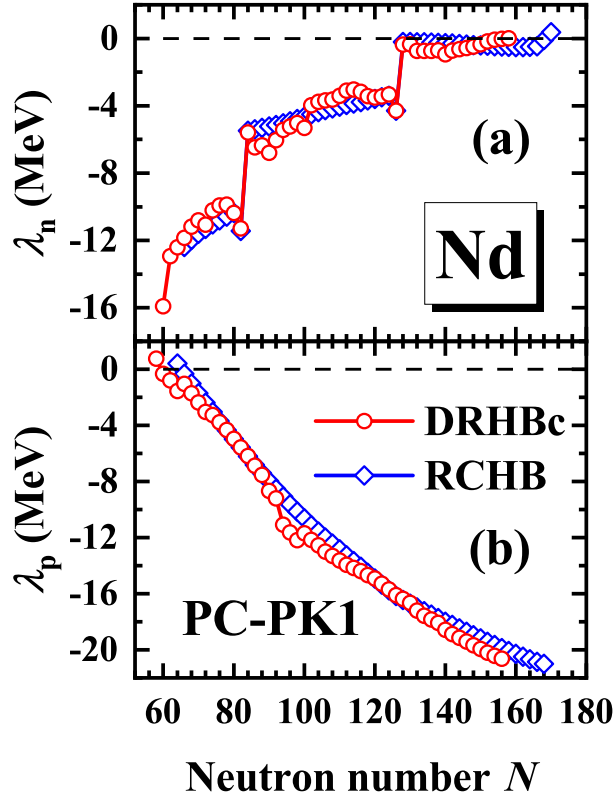}
  \caption{Neutron (a) and proton (b) Fermi energies for Nd isotopes in the DRHBc calculations versus the neutron number. The results from the RCHB mass table~\cite{Xia2018ADNDT} are shown for comparison.}
\label{fig10}
\end{figure}
%----------------------------------------------------------------------------------------

The Fermi energy represents the change of the total energy for a change in the particle number~\cite{PeterBook}. In addition to the two-neutron separation energy, the Fermi energy can also provide information about the nucleon drip line. Figure~\ref{fig10} shows the neutron and proton Fermi energies in the DRHBc calculations, in comparison with the RCHB results~\cite{Xia2018ADNDT}. If the pairing energy vanishes, the Fermi energy is chosen to be the energy of the last occupied single-particle state. In Fig.~\ref{fig10}(a), the neutron Fermi energy becomes positive at $^{218}$Nd and $^{230}$Nd in the RCHB ones. In the DRHBc calculations, although the neutron Fermi energy for $^{216}$Nd is negative with $\lambda_n=-0.025~\mathrm{MeV}$, it is unstable against neutron emission with $S_{2n}=-0.057~\mathrm{MeV}$ in Fig.~\ref{fig9}. In the RCHB calculations, the neutron drip line from the Fermi energy is consistent with the two-neutron separation energies. The sudden increases in the neutron Fermi energy reflect the shell closures at $N=82$ and $N=126$. In Fig.~\ref{fig10}(b), the proton Fermi energy becomes positive at $^{118}$Nd in the DRHBc calculations and $^{124}$Nd in the RCHB ones. Therefore, the deformation effects influence not only the neutron but also the proton drip line for neodymium isotopes.
Near $N=82$ and $126$, the Fermi energies in the DRHBc calculations agree more or less with the RCHB ones. Moving away from the shell closures, the smooth evolution of Fermi energy does not exist in the DRHBc calculations due to deformation effects.

%Interestingly, compared with the RCHB theory, the proton drip line is located on the proton-richer side in the DRHBc theory, whereas the neutron drip line is located on the neutron side.

\subsection{Quadrupole deformation}

%----------------------------------------------------------------------------------------
\begin{figure}[htbp]
  \centering
  \includegraphics[scale=0.5,angle=0]{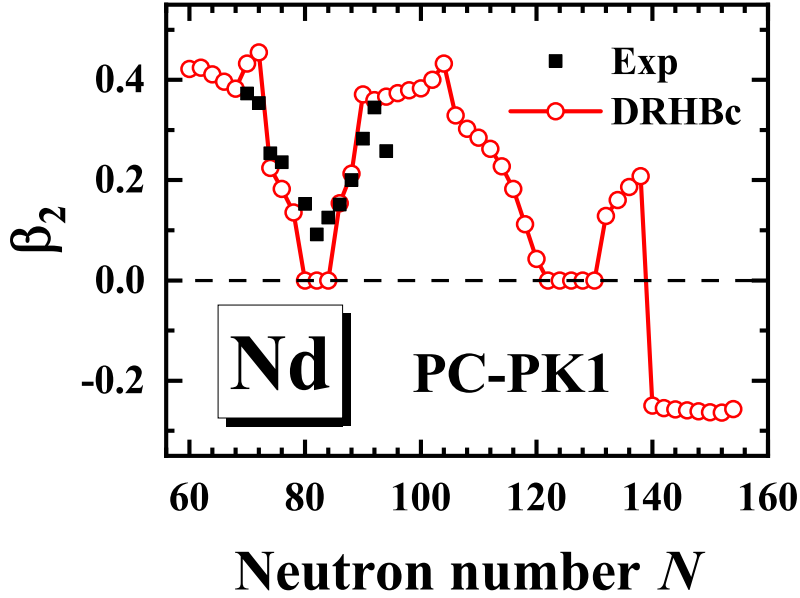}
  \caption{Quadrupole deformation as a function of the neutron number in the DRHBc calculations for Nd isotopes. The data available~\cite{2016ADNDT} are shown for comparison.}
\label{fig11}
\end{figure}
%----------------------------------------------------------------------------------------

The ground-state quadrupole deformation parameters in Eq.~(\ref{def}) in the DRHBc calculations for neodymium isotopes are shown in Fig.~\ref{fig11} and compared with the available data~\cite{2016ADNDT}. Generally, the DRHBc calculated ground-state quadrupole deformations reproduce well the data. There exists some difference between the calculated deformation and the data, which might be explained by the fact that the data are not directly observed. The deformation data in Ref.~\cite{2016ADNDT} are extracted from the observed $B(E2, 0_{1}^+ \rightarrow 2_{1}^+)$ with the assumption of the nucleus as a rigid rotor which might not be true for all nuclei. The nuclei near $N=82$ and $N=126$ exhibit the spherical shape due to the shell effects. For these nuclei, the bulk properties in the DRHBc calculations discussed above are consistent with the RCHB ones. The shape evolution is following, a) from the proton drip line to $N=80$, the shape changes from prolate to spherical; b) from $N=84$ to $N=122$, the shape changes from spherical to prolate and then back to spherical; c) from $N=130$ to $N=138$, the shape changes from spherical to prolate; d) from $N=140$ to the neutron drip line, the shape changes to oblate.

\subsection{Rms radii}

%----------------------------------------------------------------------------------------
\begin{figure}[htbp]
  \centering
  \includegraphics[scale=0.25,angle=0]{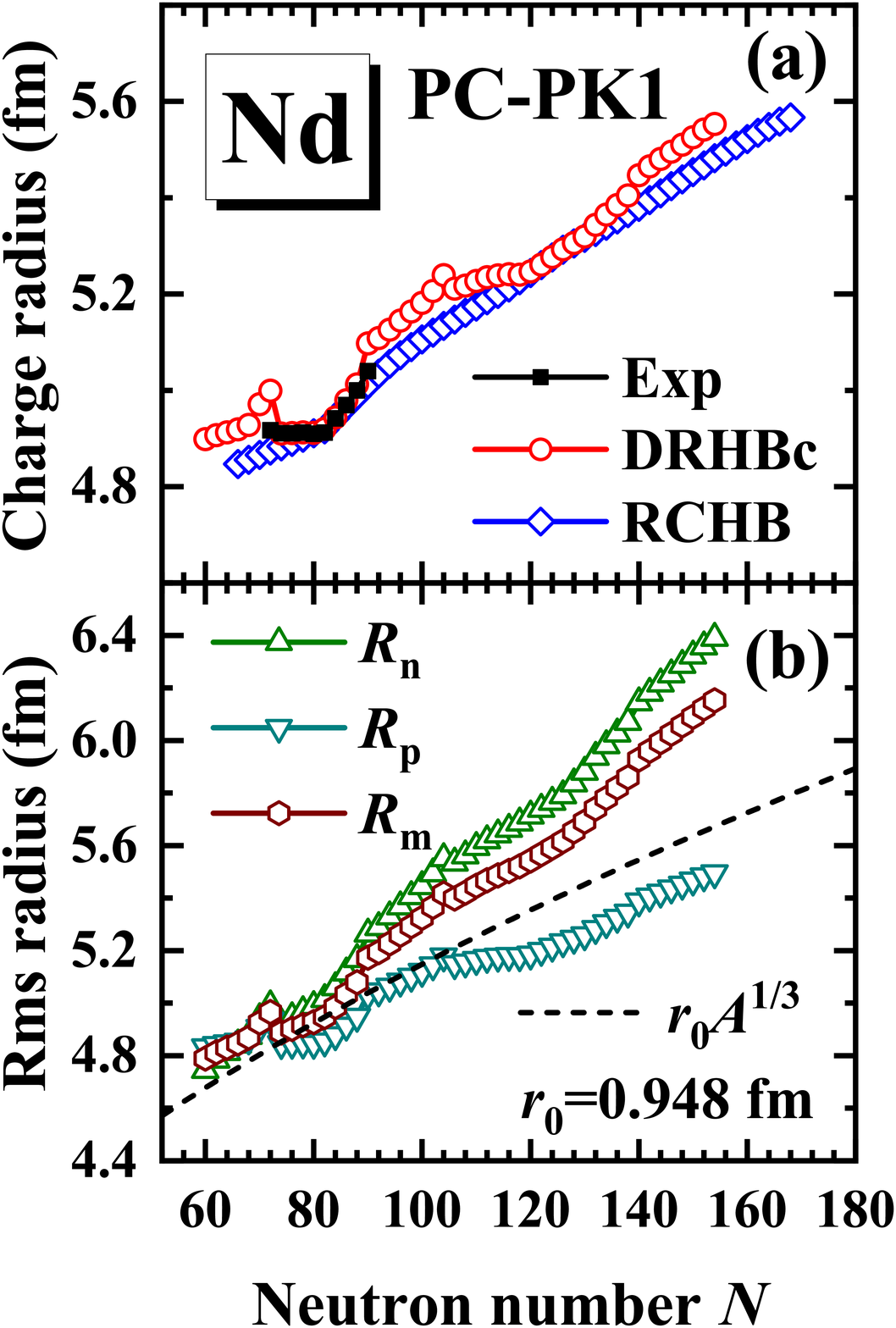}
  \caption{(a) Charge radius as a function of the neutron number in the DRHBc calculations for Nd isotopes. The results in the RCHB mass table~\cite{Xia2018ADNDT} and existing data from Ref.~\cite{Angeli2013ADNDT} are shown for comparison. (b) Rms neutron radius, proton radius, and matter radius as functions of the neutron number in the DRHBc calculations for Nd isotopes. The empirical matter radii $r_0 A^{1/3}$, in which $r_0=0.948$~fm determined by $^{142}$Nd, are also shown to guide the eye.}
\label{fig12}
\end{figure}
%----------------------------------------------------------------------------------------

In Fig.~\ref{fig12}(a), the charge radius as a function of the neutron number in the DRHBc calculations for neodymium isotopes are shown, together with the RCHB results~\cite{Xia2018ADNDT} and the available data~\cite{Angeli2013ADNDT}. In general, the data are reproduced well by both the RCHB and DRHBc calculations. In particular, the DRHBc calculations reproduce well not only the data  from $^{134}$Nd to $^{148}$Nd but also the kink at $^{142}$Nd, in which the deformation plays a crucial role. For $^{132}$Nd and $^{150}$Nd, the charge radii are underestimated by the RCHB calculations due to the neglect of deformation, and are slightly overestimated by the DRHBc ones due to the overestimated deformation in Fig.~\ref{fig11}. The overestimation of deformation for $^{132}$Nd and $^{150}$Nd might be due to their soft PESs shown in Refs.~\cite{Xiang2018PRC,Li2009PRC}.

In Fig.~\ref{fig12}(b), the rms neutron radii $R_n$, proton radii $R_p$, and matter radii $R_m$ in the DRHBc calculations for neodymium isotopes are shown. The empirical matter radii $r_0 A^{1/3}$ with $r_0$ determined by the most stable neodymium isotope $^{142}$Nd are shown to guide the eye. Starting from the proton drip line, $R_n$, $R_p$, and $R_m$ are close to each other and gradually increase with the neutron number. There is a sudden decrease from $^{132}$Nd to $^{134}$Nd because the quadrupole deformation parameter $\beta_2$ decreases from $0.46$ to $0.23$.
Beyond $^{134}$Nd, the proton radius increases gradually, the neutron radius increases more rapidly, and the matter radius is in between.
By scaling the empirical matter radius $r_0 A^{1/3}$ by the most stable nucleus $^{142}$Nd, for the nuclei far away from the stability line, the calculated radii are systematically larger than the empirical ones.
In particular, for nuclei with $N>126$, the ever increasing deviation from the empirical value may indicate some underlying exotic structure.

\subsection{Neutron density distribution}

%----------------------------------------------------------------------------------------
\begin{figure}[htbp]
  \centering
  \includegraphics[scale=0.35,angle=0]{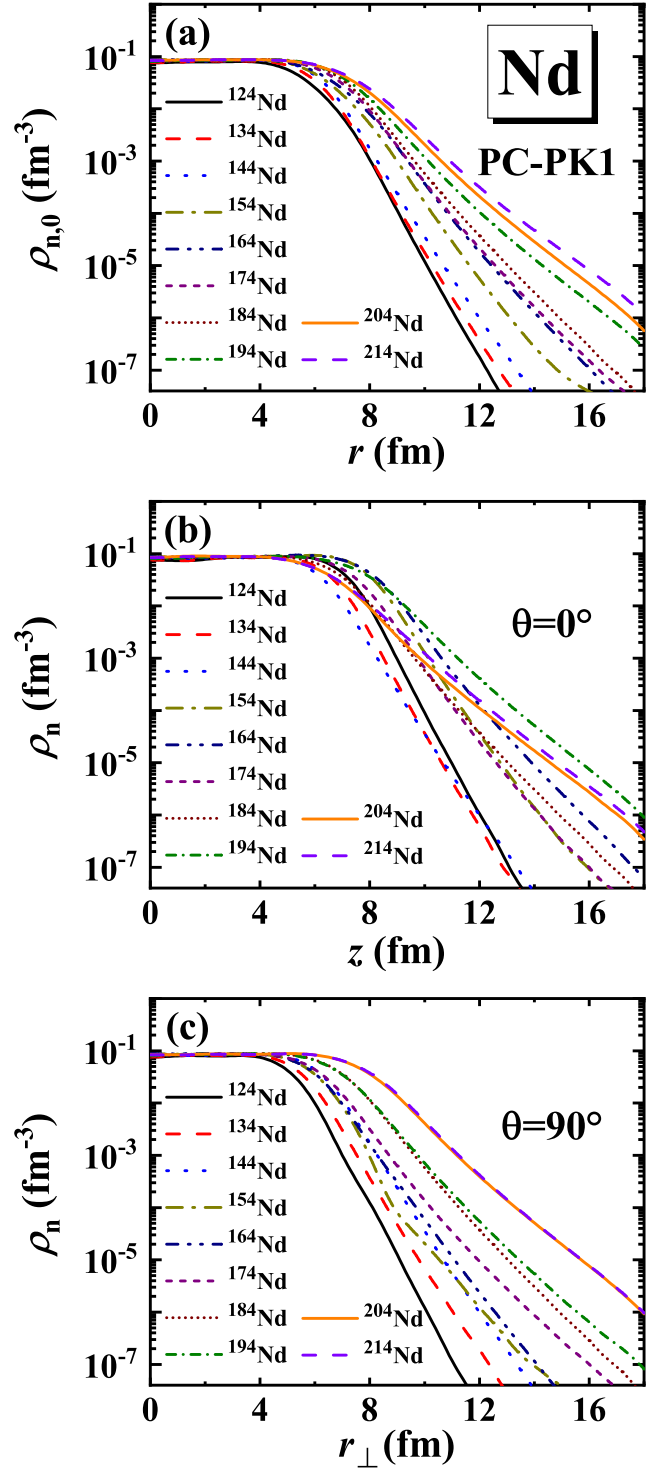}
  \caption{(a) Spherical component of the neutron density distribution, (b) the neutron density distribution along the symmetry axis $z$, and (c) the neutron density distribution perpendicular to the symmetry axis with $r_\perp=\sqrt{x^2+y^2}$, for selected even-even neodymium isotopes $^{124,134,\cdots,214}$Nd in the DRHBc calculations.}
\label{fig13}
\end{figure}
%----------------------------------------------------------------------------------------

Figure~\ref{fig13} shows neutron density profiles of selected even-even neodymium isotopes $^{124, 134, \cdots, 214}$Nd. In Fig.~\ref{fig13}(a), $\rho_{n,0}$ represents the spherical component of the neutron density distribution [cf.~Eq.~(\ref{legendre})]. In Figs.~\ref{fig13}(b) and \ref{fig13}(c), the total neutron density distributions along $(\theta=0^\circ)$ and perpendicular $(\theta=90^\circ)$ to the symmetry axis $z$ are shown, respectively. In Fig.~\ref{fig13}(a), for the spherical component, the neutron density distribution becomes more diffuse monotonically with the increasing mass number. In Figs.~\ref{fig13}(b) and \ref{fig13}(c), the neutron density distributions manifest not only the diffuseness with the increasing neutron number but also the deformation effects. In Fig.~\ref{fig13}(b), although $^{134}$Nd has ten more neutrons than $^{124}$Nd, its density along the symmetry axis is smaller than that of $^{124}$Nd for $z\gtrsim 6$~fm. This can be understood from the deformation $\beta_2=0.41$ for $^{124}$Nd and $\beta_2=0.23$ for $^{134}$Nd. Due to their oblate deformation, the densities perpendicular to the symmetry axis for $^{204}$Nd and $^{214}$Nd at $z \ge 8$ fm are less than $^{194}$Nd, as shown in Fig.~\ref{fig13}(b). However, their slopes of the density are still the smallest. As discussed in Ref.~\cite{Scamps2013PRC}, a reduction of the diffuseness along the main axis of deformation develops simultaneously with an increase of the diffuseness along the other axis. Therefore, the oblate deformed $^{204}$Nd and $^{214}$Nd are the most diffuse ones along the symmetry axis. In Fig.~\ref{fig13}(c), since the deformation for $^{124, 134,\cdots, 194}$Nd is either spherical or prolate, its density distribution perpendicular to the symmetry axis is equal to or smaller than that along the symmetry axis. The densities for oblate $^{204}$Nd and $^{214}$Nd are much more elongated perpendicular to the symmetry axis and lead to significantly larger density distributions along $r_\perp$.

\subsection{Single-neutron levels in canonical basis}

The canonical basis is obtained by diagonalizing the density matrix, with the eigenvalues corresponding to the occupation probabilities [cf.~Eq.~(\ref{canonical})]. The single-particle energies in the canonical basis are the diagonal matrix elements of the single-particle Hamiltonian in the canonical basis. The canonical basis is very useful to discuss the physics in exotic nuclei~\cite{DOBACZEWSKI1984103,Dobaczewski1996PRC,Meng1996PRL,Meng1998PRL,Zhou2010PRC,Li2012PRC}.

%----------------------------------------------------------------------------------------
\begin{figure}[htbp]
  \centering
  \includegraphics[scale=0.5,angle=0]{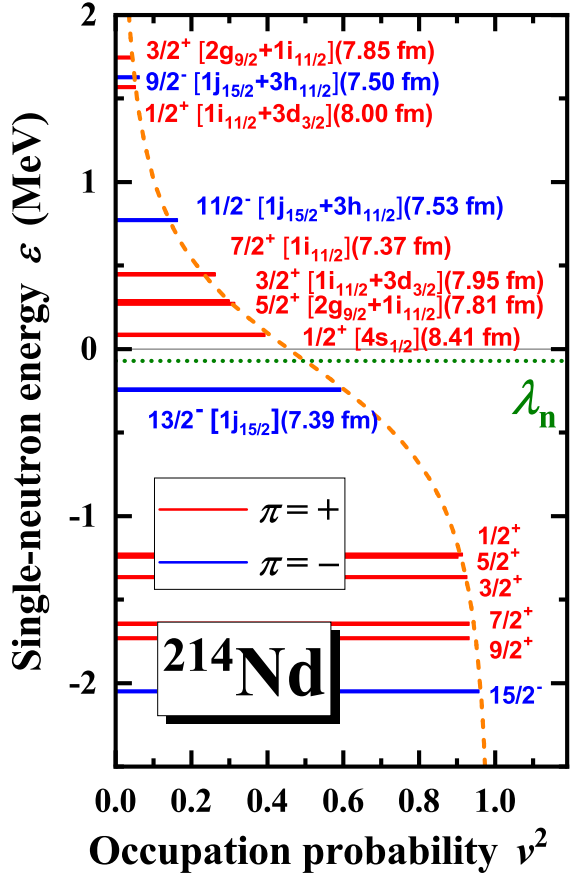}
  \caption{Single-neutron levels around the Fermi energy in the canonical basis for $^{214}$Nd versus the occupation probability $v^2$ in the DRHBc calculations. Each level is labeled by the quantum numbers $m^\pi$. The main components and rms radii for the levels with $\epsilon>-0.3~\mathrm{MeV}$ are also given. The neutron Fermi energy $\lambda_\mathrm{n}$ is shown with the dotted line. The occupation probability from the BCS formula with the average pairing gap is given by dashed line. The thin solid line represents the continuum threshold.}
\label{fig14}
\end{figure}
%----------------------------------------------------------------------------------------

The single-neutron spectrum around the neutron Fermi energy $\lambda_n$ in the canonical basis for $^{214}$Nd is shown in Fig.~\ref{fig14}. As the projection of the angular momentum on the symmetry axis $m$ and the parity $\pi$ are good quantum numbers in the axially deformed system with the spatial reflection symmetry, each state is labeled with $m^\pi$. The main components in the spherical Woods-Saxon basis and rms radii for the states with single-neutron energies higher than $-0.3$ MeV are also given. The lengths of horizontal lines represent the occupation probabilities $v^2$ in Fig.~\ref{fig14}. The occupation probabilities calculated by the BCS formula~\cite{PeterBook} with the average pairing gap and single-neutron energies in canonical basis is shown by the dashed line. The bound single-neutron levels are occupied with considerable probabilities, and those with single-neutron energies smaller than $-1$ MeV are almost fully occupied. As the neutron Fermi energy $\lambda_n=-0.07$ MeV and is close to the threshold, the states in continuum have noticeable occupation probabilities due to the pairing correlation. Since the neutron Fermi energy is negative, the single-neutron densities in continuum are localized~\cite{DOBACZEWSKI1984103} and the nucleus is still bound. The occupation probabilities of both bound states and continuum states are roughly consistent with those calculated by BCS formula. By summing the number of neutrons in the continuum, one obtains about $4$ neutrons in the continuum, which could be related to the possible neutron halo phenomenon~\cite{Meng1996PRL,Meng1998PRL,Zhang2002Chin.Phys.Lett.312,Meng2002Phys.Rev.C41302,Terasaki2006Phys.Rev.C54318,Zhou2010PRC,Li2012PRC}. The states whose main components are $s$ waves or $d$ waves with low centrifugal barriers have relatively larger rms radii, and are helpful in the formation of halos. It can also be seen in Fig.~\ref{fig13} that, $^{214}$Nd has significant density distributions in the region of large $r$, which could be an indicator of exotic structure such as the existence of the neutron halo. This could be also an interesting topic worth further studying and the strategy in Refs.~\cite{Zhou2010PRC,Li2012PRC} can be employed to investigate such exotic structure.

\subsection{Neutron skin and proton radioactivity}

In order to explore the possible exotic structures in Nd isotopes, the thickness of the neutron skin, the particles number in the continuum, contributions of different states to the total density, and the proton radioactivity are investigated and discussed in detail.

%----------------------------------------------------------------------------------------
\begin{figure}[htbp]
  \centering
  \includegraphics[scale=0.4,angle=0]{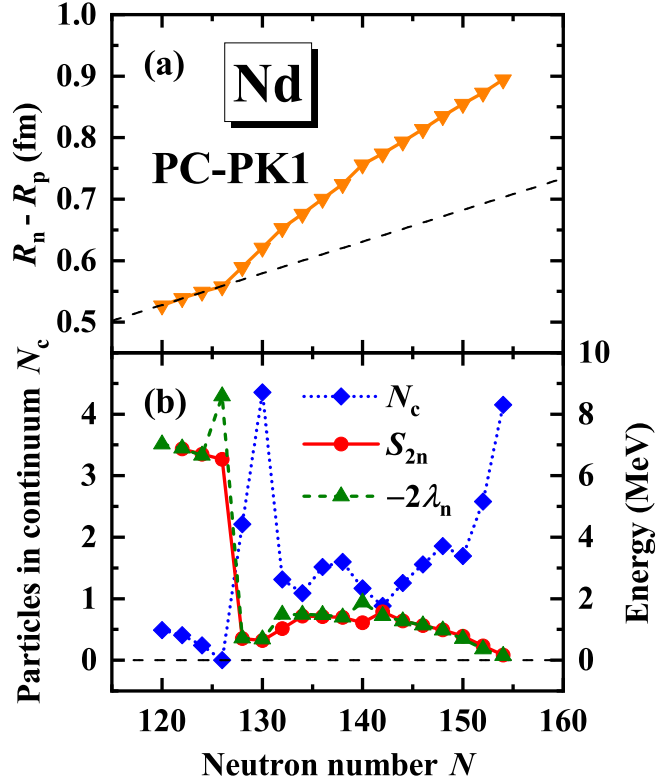}
  \caption{(a) Thickness of the neutron skin $R_\mathrm{n}-R_\mathrm{p}$ together with its increasing trend for $N\le126$ as a dashed line, and (b) the number of particles in the continuum $N_c$, two-neutron separation energy $S_{2\mathrm{n}}$, and two times the negative neutron Fermi energy $-2\lambda_\mathrm{n}$ as functions of the neutron number for neutron-rich neodymium isotopes with $N\ge120$ in the DRHBc calculations.}
\label{fig15}
\end{figure}
%----------------------------------------------------------------------------------------

Figure~\ref{fig15}(a) shows the thickness of the neutron skin $R_\mathrm{n}-R_\mathrm{p}$ for Nd isotopes with $N\ge120$. The thickness of the neutron skin increases gradually from $N=120$ to $N=126$ and significantly after the neutron shell closure $N=126$, and reaches the maximum at the neutron drip-line nucleus $^{214}$Nd.

In Fig.~\ref{fig15}(b), the number of particles in the continuum $N_{\mathrm{c}}$, the two-neutron separation energy $S_{2n}$, and two times the negative neutron Fermi energy $-2\lambda_n$ for Nd isotopes with $N\ge120$ are shown. The number of particles in the continuum is the sum of occupation probabilities over positive-energy states in the canonical basis. The relation $S_{2\mathrm{n}}\approx -2\lambda_{\mathrm{n}}$ is reproduced except for $^{186}$Nd due to the pairing collapse and $^{192,200}$Nd due to the change of deformation (configuration). For the nuclei with $N>126$, the neutron Fermi energy is close to the continuum threshold ($\lambda_n>-1~\mathrm{MeV}$), as a result neutrons can be scattered into the continuum due to the pairing correlation~\cite{DOBACZEWSKI1984103,Meng1998NPA}. The sudden increase of $N_{\mathrm{c}}$ and the sudden decrease of $S_{2\mathrm{n}}$ after $N=126$ in Fig.~\ref{fig15}(b) coincide with the abrupt change in the thickness of neutron skin $R_\mathrm{n}-R_\mathrm{p}$ in Fig.~\ref{fig15}(a). The nuclei with more than $2$ neutrons in the continuum, $^{188}$Nd, $^{190}$Nd, $^{212}$Nd, and $^{214}$Nd, have the smallest two-neutron separation energies.

Since for $^{214}$Nd, there are more than $4$ neutrons in the continuum, and the two-neutron separation energy is less than $0.1~\mathrm{MeV}$, and the thickness of the neutron skin $R_\mathrm{n}-R_\mathrm{p}$ is around $0.9~\mathrm{fm}$, it is encouraging to investigate its density distribution to explore the existence of possible neutron skin or neutron halo.

%----------------------------------------------------------------------------------------
\begin{figure}[htbp]
  \centering
  \includegraphics[scale=0.4,angle=0]{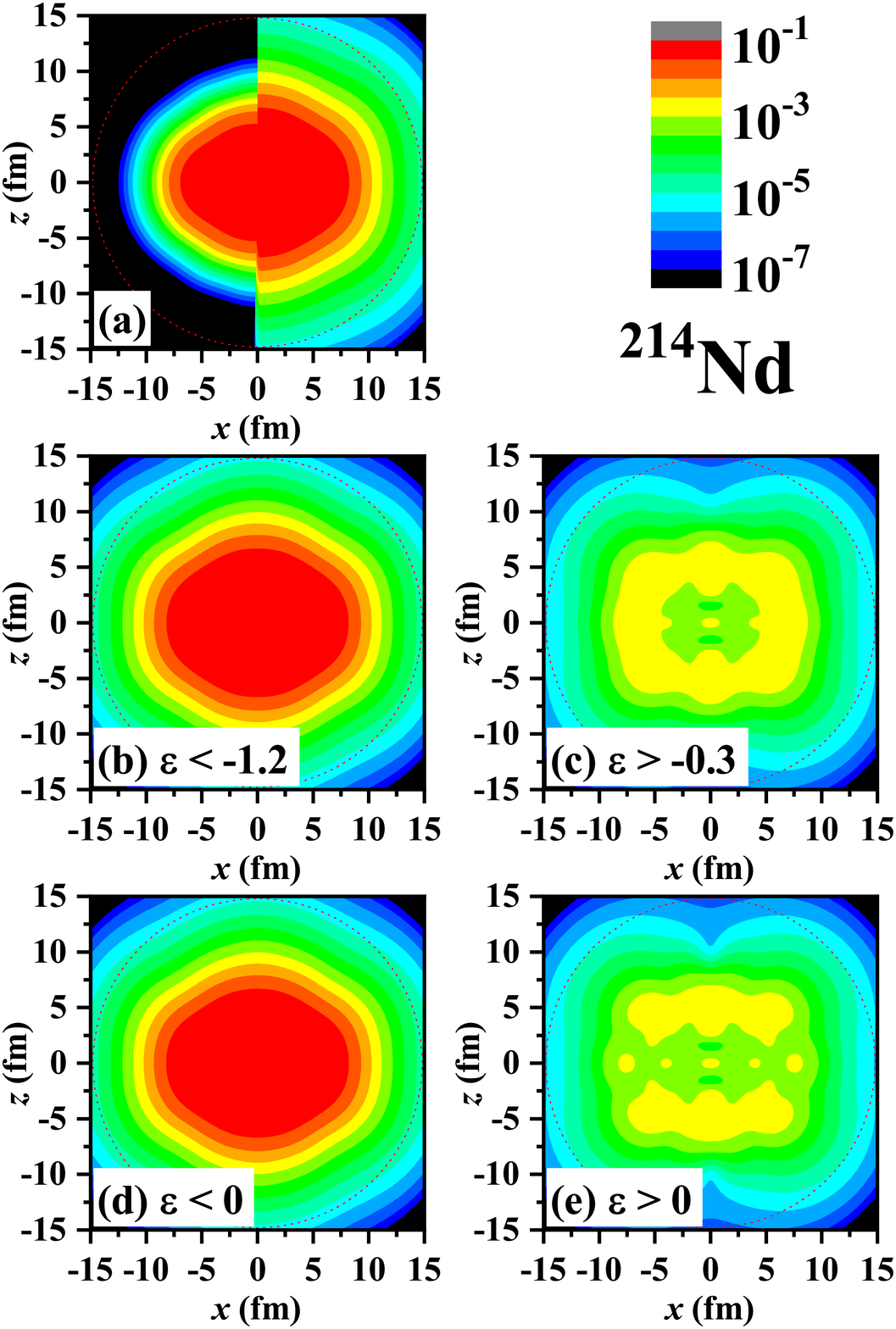}
  \caption{Density distributions with $z$ axis as the symmetry axis in $^{214}$Nd for (a) the proton (for $x<0$) and the neutron (for $x>0$), and the neutron with single-particle energy (b) $\epsilon<-1.2~\mathrm{MeV}$, (c) $\epsilon>-0.3~\mathrm{MeV}$, (d) $\epsilon<0~\mathrm{MeV}$, and (e) $\epsilon>0~\mathrm{MeV}$ in the canonical basis. In each plot, a dotted circle is drawn to guide the eye.}
\label{fig16}
\end{figure}
%----------------------------------------------------------------------------------------

In Fig.~\ref{fig16}(a), the neutron and proton density distributions for $^{214}$Nd are shown. Both the neutron and proton density distributions show oblate shapes, in consistent with Fig.~\ref{fig11}. Owing to the large neutron excess, the neutron density extends much farther than the proton.

According to the single-particle levels in Fig.~\ref{fig14}, there is a gap between the levels with $\epsilon<-1.2~\mathrm{MeV}$ and those with $\epsilon>-0.3~\mathrm{MeV}$. Following the strategy in Refs.~\cite{Zhou2010PRC,Li2012PRC}, the neutron density is decomposed into two parts as shown in Figs.~\ref{fig16}(b) for $\epsilon<-1.2~\mathrm{MeV}$ and \ref{fig16}(c) for $\epsilon>-0.3~\mathrm{MeV}$. The quadrupole deformations are respectively $\beta_2=-0.268$ for $\epsilon<-1.2~\mathrm{MeV}$ in Fig.~\ref{fig16}(b) and $\beta_2=-0.160$ for $\epsilon>-0.3~\mathrm{MeV}$ in Fig.~\ref{fig16}(c). While both are oblate, they are still slightly decoupled.
Although the density in Fig.~\ref{fig16}(c) is contributed by the weakly bound states and continuum, it is less diffuse than that in Fig.~\ref{fig16}(b) both along and perpendicular to the symmetry axis.

Similarly, the neutron density can be decomposed into the part for bound states with $\epsilon<0~\mathrm{MeV}$ in Fig.~\ref{fig16}(d), and the part for continuum with $\epsilon>0~\mathrm{MeV}$ in Fig.~\ref{fig16}(e). The difference between such decomposition and the previous one is the allocation of the weakly bound state $13/2^-$, which corresponds to an oblate shape and the main component $1j_{15/2}$ with a mixing of $|\mathrm{Y}_{76}(\theta,\varphi)|^2$ and $|\mathrm{Y}_{77}(\theta,\varphi)|^2$. This allocation hardly influences the density distribution in Fig.~\ref{fig16}(b) and the quadrupole deformation changes slightly to $\beta_2=-0.273$ in Fig.~\ref{fig16}(d). In contrast, the density distribution changes from oblate with $\beta_2=-0.160$ in Fig.~\ref{fig16}(c) to nearly spherical with $\beta_2=0.047$ in Fig.~\ref{fig16}(e). The decoupling between the oblate shape contributed by bound states and the nearly spherical one by continuum is remarkable. By comparing Figs.~\ref{fig16}(b) and \ref{fig16}(c) or \ref{fig16}(d) and \ref{fig16}(e), there is no clear clue for a halo structure in $^{214}$Nd.

Although the theoretical description of light halo nuclei is well under control, as discussed in Refs.~\cite{Rotival2009PRC1,Rotival2009PRC2,Meng2015JPhysG}, existing definitions and tools are often too qualitative and the associated observables are incomplete for heavier ones. There has been much effort put into quantifying halos by examining the separation energy, the density profiles, the particles in the classically forbidden region, and weakly-bound particles obtained from mean field calculations~\cite{Meng1998Phys.Lett.B1,Im2000PRC,Mizutori2000PRC,Rotival2009PRC1}. In this paper, the total neutron density for $^{214}$Nd is decomposed into the contributions of different states to examine quantitatively whether the density in the region of large $r$ is mainly contributed by the narrow bunch of weakly bound and positive-energy states to distinguish its halo character.

%----------------------------------------------------------------------------------------
\begin{figure}[htbp]
  \centering
  \includegraphics[scale=0.4,angle=0]{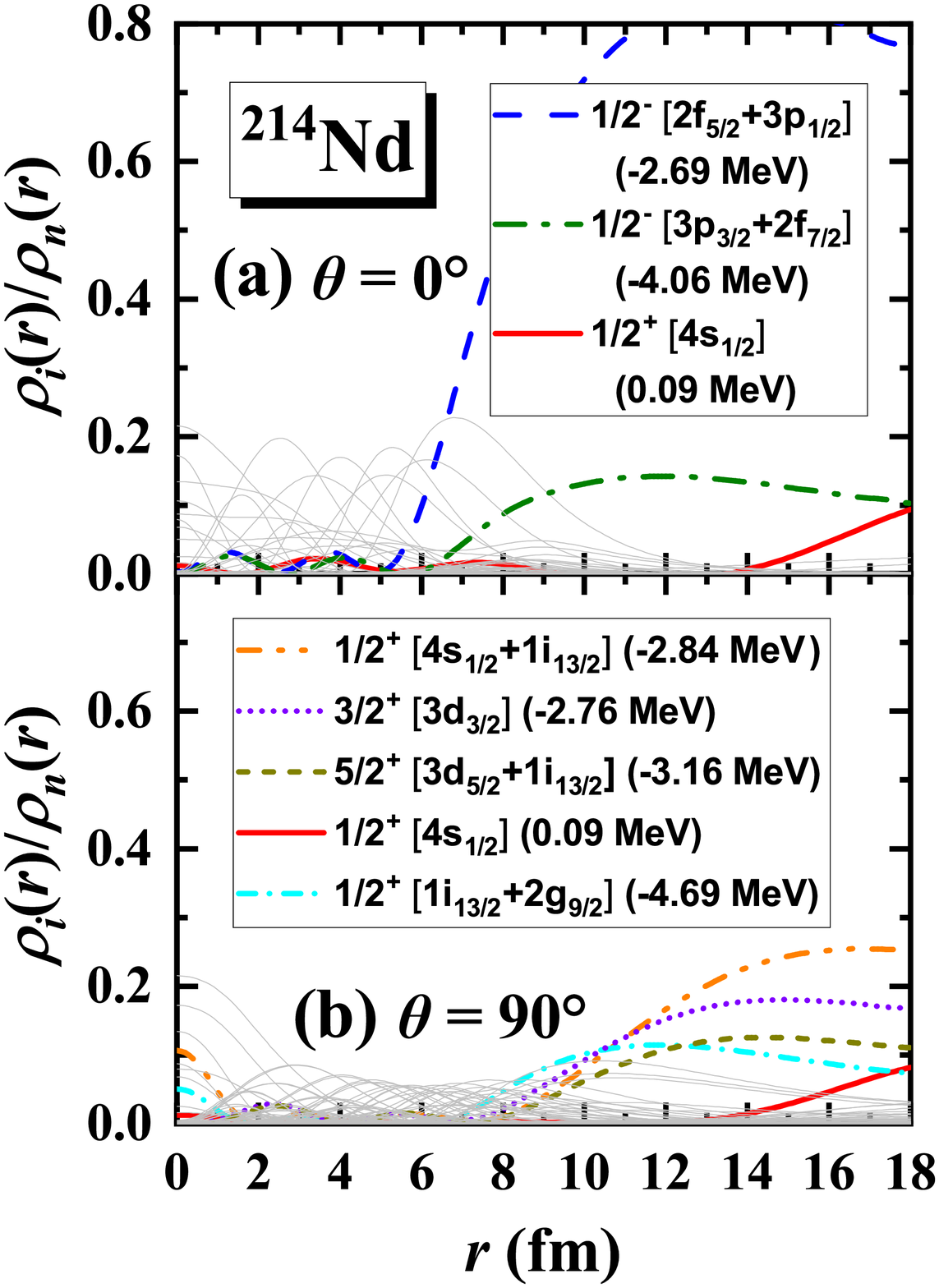}
  \caption{Contribution of each single-particle state in the canonical basis to the total neutron density at (a) $\theta=0^\circ$ (along the symmetry axis) and (b) $\theta=90^\circ$ (perpendicular to the symmetry axis) in $^{214}$Nd as a function of the radius. The states with significant contributions ($\gtrsim 0.1$) in the asymptotic area are highlighted and their energies and main components are given.}
\label{fig17}
\end{figure}
%----------------------------------------------------------------------------------------

In order to further identify the nature of neutron halo or neutron skin in $^{214}$Nd, the contribution of each single-particle state in the canonical basis to the total neutron density  is shown in Fig.~\ref{fig17}. Along the symmetry axis, the state $1/2^-$ with $\epsilon=-2.69~\mathrm{MeV}$ plays the dominant role for large $r$ as shown in Fig.~\ref{fig17}(a). Another $1/2^-$ state with $\epsilon=-4.06~\mathrm{MeV}$ also makes distinguishable contributions for large $r$. The contribution of the $1/2^+$ state embedded in the continuum becomes more and more important for $r\gtrsim 14$~fm because its main component $s$ wave is free from the centrifugal barrier. Perpendicular to the symmetry axis, several bound states with $\epsilon<-2.6~\mathrm{MeV}$ together with the $1/2^+$ state in the continuum contribute to the total neutron density for large $r$ as shown in Fig.~\ref{fig17}(b). The contribution of the $1/2^+$ state in the continuum evolves similarly at both $\theta=0^\circ$ and $90^\circ$ due to its nearly spherical density distribution. From Figs.~\ref{fig17}(a) and \ref{fig17}(b), it can be clearly seen that the density in the region of large $r$ is mainly contributed by the deeply bound low-$m$ states with $\epsilon<-2.6~\mathrm{MeV}$, and the contributions of continuum states except for the $1/2^+$ state are very small because of their suffered high centrifugal barriers as shown in Fig.~\ref{fig14}, explaining why the density distributions in Figs.~\ref{fig16}(b) and \ref{fig16}(d) are more diffuse than those in Figs.~\ref{fig16}(c) and \ref{fig16}(e). Therefore, the halo character in $^{214}$Nd can be excluded.

On the proton-rich side, possible exotic phenomena include the proton halo and the proton radioactivity. The interest in the proton radioactivity has been boosted significantly by the discoveries of one- and two-proton emission beyond the proton drip lines~\cite{Blank2008PPNP,Pfutzner2012RMP}. Comprehensive theoretical efforts have been made to investigate the proton radioactivity based on the CDFT~\cite{Vretenar1999PRL,Yao2008PRC,Ferreira2011PLB,Zhao2014PRC,Lim2016PRC}. Because of its self-consistent treatment of deformation, pairing correlation, and continuum, it is natural to apply the DRHBc theory to study the proton radioactivity to understand the physics beyond drip line.

%----------------------------------------------------------------------------------------
\begin{figure}[htbp]
  \centering
  \includegraphics[scale=0.5,angle=0]{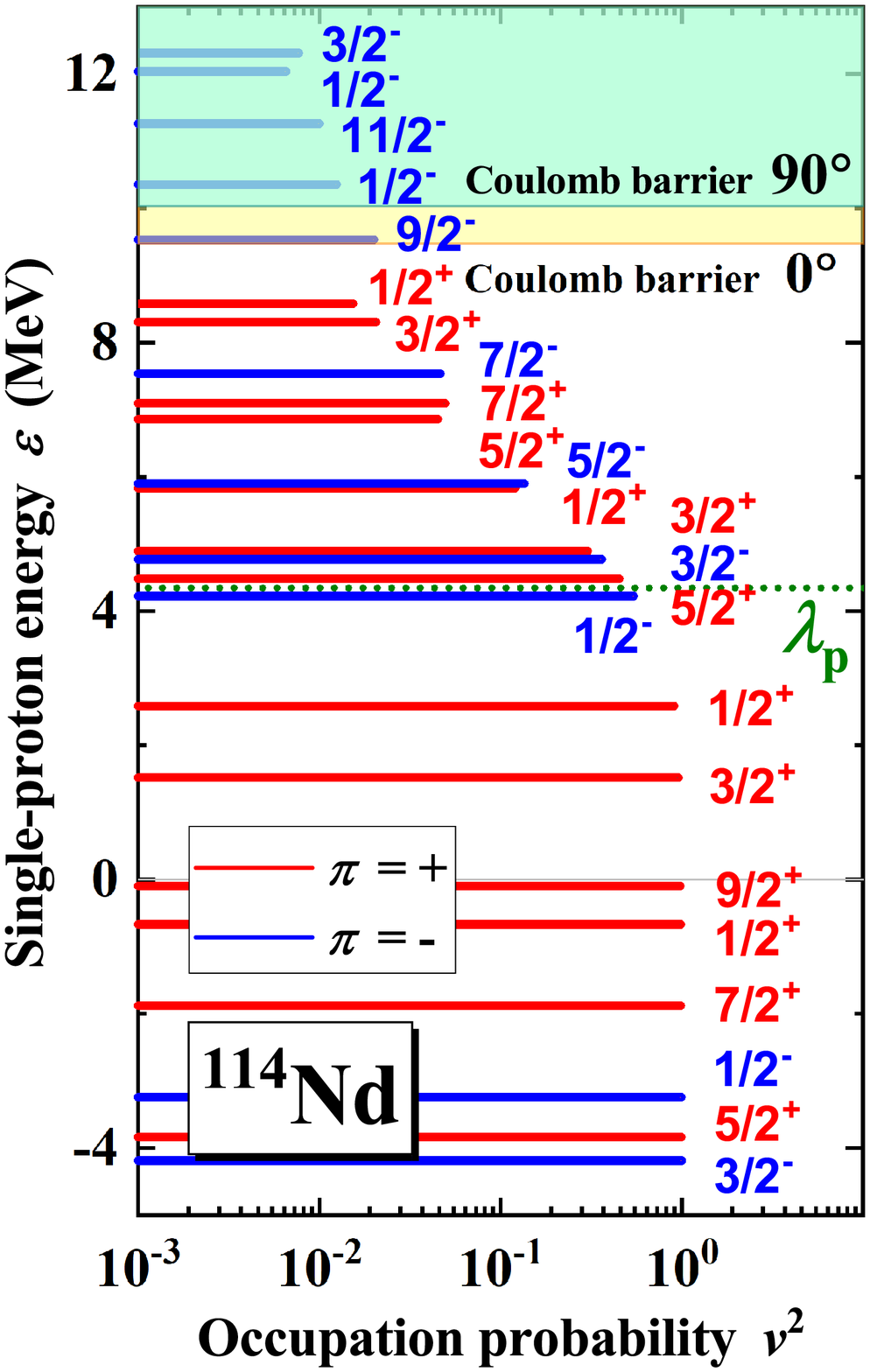}
  \caption{Single-proton levels around the Fermi energy in the canonical basis for $^{114}$Nd versus the occupation probability $v^2$. Each level is labeled by the quantum numbers $m^\pi$. The Fermi energy $\lambda_\mathrm{p}$ is shown as a dotted line. The continuum threshold is represented by a thin solid line. The shaded areas represent respectively the regions above the Coulomb barrier at $0^\circ$ (along the symmetry axis) and $90^\circ$ (perpendicular to the symmetry axis).}
\label{fig18}
\end{figure}
%----------------------------------------------------------------------------------------

As shown in Fig.~\ref{fig10}(b), the proton drip-line nucleus for Nd isotopes in the DRHBc theory is $^{120}$Nd. The proton Fermi energies for $^{118}$Nd and other lighter even-even Nd isotopes are positive, and they might be unstable against the proton emission. However, due to the existence of Coulomb barrier, some of them may become quasi-bound proton emitters with certain half-lives. To explore such exotic phenomena, the single-proton spectrum around the proton Fermi energy $\lambda_p$ in the canonical basis for $^{114}$Nd is shown in Fig.~\ref{fig18} as an example. The heights of Coulomb barrier along and perpendicular to the symmetry axis are given. $^{114}$Nd is prolate deformed with $\beta_2=0.248$ and $\beta_{\mathrm{p},2}=0.269$. Accordingly, the height of Coulomb barrier at $\theta=0^\circ$ is $9.12~\mathrm{MeV}$ and at $\theta=90^\circ$ is $10.05~\mathrm{MeV}$. For $^{114}$Nd, the proton Fermi energy $\lambda_p=4.35~\mathrm{MeV}$ is below the Coulomb barrier at either $\theta=0^\circ$ or $90^\circ$. Therefore the protons ($\approx 8$) above the continuum threshold are still quasi-bound by the Coulomb barrier. These protons may undergo quantum tunneling. By comparing the calculated binding energies of $^{114}$Nd and its two-proton emission daughter nucleus $^{112}$Ce, one can find the decay energy $Q_{2p}=-S_{2p}=8.53~\mathrm{MeV}$ is positive and thus the two-proton radioactivity is energetically allowed. Similarly, the decay energies $Q_{4p}=14.95~\mathrm{MeV}$, $Q_{6p}=19.03~\mathrm{MeV}$, and $Q_{8p}=21.17~\mathrm{MeV}$ for $^{114}$Nd are obtained by comparing its binding energy with those of its corresponding daughter nuclei, which suggests the possibility of multi-proton radioactivity in $^{114}$Nd.
Further calculations indicate that $^{116}$Nd and $^{118}$Nd are also candidates for two-proton and even multi-proton radioactivity. Systematical investigation of the proton radioactivity including not only even-even nuclei but also odd mass nuclei and odd-odd nuclei as well as the decay half-lives is highly demanded.

%%%%%%%%%%%%%%%%%%%%%%%%%%%%%%%%%%%%%%%%%%%%%%%%%%%%%%%%%%
%                    begin  summary
%%%%%%%%%%%%%%%%%%%%%%%%%%%%%%%%%%%%%%%%%%%%%%%%%%%%%%%%%%

%----------------------------------------------------------------------------------------
\section{Summary}\label{summary}
%----------------------------------------------------------------------------------------

In summary, the DRHBc theory based on the point-coupling density functionals including both the deformation and continuum effects is developed. Numerical details towards constructing the DRHBc mass table have been examined. The DRHBc calculation previously accessible only for light nuclei up to magnesium isotopes has been extended for all even-even nuclei in the nuclear chart. Taking even-even neodymium isotopes from the proton drip line to the neutron drip line as examples, the ground-state properties and exotic structures are investigated.

The numerical details towards constructing the DRHBc mass table for even-even nuclei with satisfactory accuracy have been examined. For the Dirac Woods-Saxon basis, the box size $R_{\mathrm{box}}=20$ fm, the mesh size $\Delta r=0.1$ fm, the energy cutoff $E^+_{\mathrm{cut}}=300$ MeV, and the angular momentum cutoff $J_{\max}=23/2~\hbar$ are recommended. For the pairing channel, the pairing strength $V_0=-325.0~\mathrm{MeV~fm}^3$ and the pairing window of $100$ MeV are suggested. For the Legendre expansion of deformed densities and potentials, the expansion truncation $\lambda_{\max}=6$ is suggested for $Z\le 80$, and $\lambda_{\max}=8$ is suggested for $Z>80$.

Taking even-even neodymium isotopes from the neutron drip line to the proton drip line as examples, the DRHBc calculations with the density functional PC-PK1 are systematically performed. The strategy to locate the ground states is suggested and confirmed by constrained calculations. The ground-state properties for even-even neodymium isotopes thus obtained are compared with available data and the results in the spherical RCHB mass table~\cite{Xia2018ADNDT}.

The experimental binding energies for even-even neodymium isotopes are reproduced by the DRHBc calculations with a rms deviation of $0.958$ MeV with the rotational correction and $2.668$ MeV without the rotational correction, in comparison with $8.301$ MeV given by the spherical RCHB calculations. Accordingly, the two-neutron separation energies are better reproduced. The predicted proton and neutron drip-line nuclei are respectively $^{120}$Nd and $^{214}$Nd, in contrast with $^{126}$Nd and $^{228}$Nd in the RCHB theory.

The shapes and sizes for even-even neodymium isotopes are correctly reproduced by the DRHBc calculations. Good agreements with the observed quadrupole deformation and its evolution as well as the charge radius and its kink around the shell closure $N=82$ are obtained.

The neutron density distributions for neodymium isotopic chain are examined. It is found that their spherical components increase with the mass number monotonically. The density of a prolate deformed nucleus is more elongated along the symmetry axis, and an oblate deformed one is more elongated perpendicular to the symmetry axis.

For the most neutron-rich neodymium isotope $^{214}$Nd, its two-neutron separation energy is smaller than $0.1~\mathrm{MeV}$, its neutron skin thickness is around $0.9~\mathrm{fm}$, and there are more than $4$ neutrons in continuum. By decomposing the neutron density of $^{214}$Nd, an interesting decoupling between the oblate shape $\beta_2=-0.273$ contributed by bound states and the nearly spherical one $\beta_2=0.047$ contributed by continuum is found. Contributions of different single-particle states to the total neutron density show that, the neutron density in the region of large $r$ is mainly contributed by the deeply bound low-$m$ states with $\epsilon<-2.6~\mathrm{MeV}$. Therefore, the exotic character in $^{214}$Nd is concluded as neutron skin instead of halo.

For the proton-rich side, by examining the proton single-particle energies, the Fermi energy, and the Coulomb barrier for $^{114}$Nd beyond the proton drip line, possible two-proton and even multi-proton emissions are predicted. Further calculations show that $^{116}$Nd and $^{118}$Nd are also candidates for two-proton and even multi-proton radioactivity. Future investigation of the proton radioactivity including not only even-even nuclei but also odd mass nuclei and odd-odd nuclei as well as the decay half-lives is highly demanded.

\begin{acknowledgments}

The authors thanks P. W. Zhao for helpful discussion and careful reading of the manuscript. This work was partly supported by the National Science Foundation of China (NSFC) under Grants No.~11935003, No.~11875070, No.~11875075, No.~11875225, No.~11621131001, No.~11975031, No.~11525524, No.~11947302, No.~11775276, No.~11961141004, No.~11711540016, No.~11735003, No.~11975041, and No.~11775014, the National Key R\&D Program of China (Contracts No.~2018YFA0404400, No.~2018YFA0404402, and No.~2017YFE0116700), the State Key Laboratory of Nuclear Physics and Technology, Peking University (No.~NPT2020ZZ01), the CAS Key Research Program of Frontier Sciences (No.~QYZDB-SSWSYS013), the CAS Key Research Program (No.~XDPB09-02), the National Research Foundation of Korea (NRF) grants funded by the Korea government (No.~2016R1A5A1013277 and No.~2018R1D1A1B07048599), and the Rare Isotope Science Project of Institute for Basic Science funded by Ministry of Science and ICT and
National Research Foundation of Korea (2013M7A1A1075764).

\end{acknowledgments}

\appendix
\section{Tabulation of ground-state properties}\label{appendix}
\newpage

\setlength{\tabcolsep}{1.2\tabcolsep}

%\renewcommand{\arraystretch}{1.0}
%\begin{center}
\begin{landscape}
\pagestyle{empty}
\begin{longtable}{cccccccccccccccc}
%\setcaptionwidth{10cm}
\caption{Ground-state properties of Nd isotopes calculated by the DRHBc theory, in comparison with the available data of masses and charge radii. In addition, the data labeled with underline means the nucleus is unbound.}\\		
\hline
\hline
\multirow{2}*{$A$}	&	\multirow{2}*{$N$}	&		$E_{\mathrm{b}}^{\mathrm{Cal.}}$	&	$E_{\mathrm{b}}^{\mathrm{Exp.}}$	&	$S_{2n}$	&	$E_{\mathrm{rot.}}$	&	$R_\mathrm{n}$	&	$R_\mathrm{p}$	&	$R_\mathrm{m}$	&	$R_\mathrm{C}^{\mathrm{Cal.}}$	&	$R_{\mathrm{C}}^{\mathrm{Exp.}}$	&	\multirow{2}*{$\beta_\mathrm{n}$}	&	\multirow{2}*{$\beta_\mathrm{p}$}	&	\multirow{2}*{$\beta_{\mathrm{tot}}$}	&	$\lambda_\mathrm{n}$	&	$\lambda_\mathrm{p}$	\\
			&	&	(MeV)	&	(MeV)	&	(MeV)	&	(MeV)	&	(fm)	&	(fm)	&	(fm)	&	(fm)	&	(fm)	&		&		&		&	(MeV)	&	(MeV)	\\
\hline \endfirsthead

\hline
\multirow{2}*{$A$}	&	\multirow{2}*{$N$}	&		$E_{\mathrm{b}}^{\mathrm{Cal.}}$	&	$E_{\mathrm{b}}^{\mathrm{Exp.}}$	&	$S_{2n}$	&	$E_{\mathrm{rot.}}$	&	$R_\mathrm{n}$	&	$R_\mathrm{p}$	&	$R_\mathrm{m}$	&	$R_\mathrm{C}^{\mathrm{Cal.}}$	&	$R_{\mathrm{C}}^{\mathrm{Exp.}}$	&	\multirow{2}*{$\beta_\mathrm{n}$}	&	\multirow{2}*{$\beta_\mathrm{p}$}	&	\multirow{2}*{$\beta_{\mathrm{tot}}$}	&	$\lambda_\mathrm{n}$	&	$\lambda_\mathrm{p}$	\\
			&	&	(MeV)	&	(MeV)	&	(MeV)	&	(MeV)	&	(fm)	&	(fm)	&	(fm)	&	(fm)	&	(fm)	&		&		&		&	(MeV)	&	(MeV)	\\
\hline \endhead

\multicolumn{8}{l}{$Z=60$ (Nd)} &\\
118	&	58	&	914.566 	&		&		&	2.534 	&	4.707 	&	4.826 	&	4.768 	&	4.892 	&		&	0.395 	&	0.427 	&	0.411 	&	-15.622 	&	\underline{0.752} 	\\
120	&	60	&	945.402 	&		&	30.837 	&	2.424 	&	4.745 	&	4.833 	&	4.789 	&	4.899 	&		&	0.411 	&	0.433 	&	0.422 	&	-15.915 	&	-0.320 	\\
122	&	62	&	971.808 	&		&	26.406 	&	2.664 	&	4.784 	&	4.842 	&	4.812 	&	4.907 	&		&	0.418 	&	0.431 	&	0.425 	&	-12.925 	&	-0.804 	\\
124	&	64	&	996.962 	&		&	25.154 	&	2.510 	&	4.813 	&	4.848 	&	4.830 	&	4.913 	&		&	0.404 	&	0.418 	&	0.411 	&	-12.409 	&	-1.552 	\\
126	&	66	&	1021.140 	&		&	24.178 	&	2.398 	&	4.841 	&	4.854 	&	4.847 	&	4.919 	&		&	0.389 	&	0.405 	&	0.396 	&	-11.836 	&	-1.051 	\\
128	&	68	&	1043.878 	&		&	22.739 	&	2.571 	&	4.874 	&	4.863 	&	4.869 	&	4.928 	&		&	0.375 	&	0.391 	&	0.383 	&	-11.182 	&	-1.718 	\\
130	&	70	&	1065.647 	&	1068.93 	&	21.768 	&	2.789 	&	4.942 	&	4.906 	&	4.926 	&	4.971 	&		&	0.429 	&	0.437 	&	0.433 	&	-10.801 	&	-2.361 	\\
132	&	72	&	1086.839 	&	1089.90 	&	21.192 	&	2.750 	&	4.991 	&	4.935 	&	4.965 	&	4.999 	&	4.917 	&	0.451 	&	0.459 	&	0.455 	&	-11.057 	&	-3.034 	\\
134	&	74	&	1106.479 	&	1110.26 	&	19.640 	&	2.464 	&	4.923 	&	4.845 	&	4.888 	&	4.911 	&	4.911 	&	0.218 	&	0.233 	&	0.224 	&	-10.210 	&	-3.275 	\\
136	&	76	&	1126.347 	&	1129.96 	&	19.869 	&	2.466 	&	4.945 	&	4.846 	&	4.902 	&	4.911 	&	4.911 	&	0.174 	&	0.193 	&	0.182 	&	-9.918 	&	-3.769 	\\
138	&	78	&	1145.921 	&	1148.92 	&	19.574 	&	2.250 	&	4.968 	&	4.847 	&	4.916 	&	4.913 	&	4.912 	&	0.126 	&	0.148 	&	0.136 	&	-9.867 	&	-4.320 	\\
140	&	80	&	1165.771 	&	1167.30 	&	19.850 	&	0.000 	&	4.988 	&	4.846 	&	4.928 	&	4.912 	&	4.910 	&	0.000 	&	0.000 	&	0.000 	&	-10.367 	&	-4.957 	\\
142	&	82	&	1186.396 	&	1185.14 	&	20.625 	&	0.000 	&	5.014 	&	4.854 	&	4.947 	&	4.920 	&	4.912 	&	0.000 	&	0.000 	&	0.000 	&	-11.276 	&	-5.564 	\\
144	&	84	&	1197.425 	&	1199.08 	&	11.029 	&	0.000 	&	5.060 	&	4.879 	&	4.985 	&	4.944 	&	4.942 	&	0.000 	&	0.000 	&	0.000 	&	-5.586 	&	-6.181 	\\
146	&	86	&	1209.470 	&	1212.40 	&	12.044 	&	1.858 	&	5.116 	&	4.915 	&	5.034 	&	4.979 	&	4.970 	&	0.152 	&	0.157 	&	0.154 	&	-6.464 	&	-6.880 	\\
148	&	88	&	1222.446 	&	1225.02 	&	12.976 	&	1.795 	&	5.167 	&	4.949 	&	5.080 	&	5.013 	&	5.000 	&	0.210 	&	0.218 	&	0.213 	&	-6.357 	&	-7.536 	\\
150	&	90	&	1235.217 	&	1237.44 	&	12.771 	&	2.306 	&	5.264 	&	5.034 	&	5.173 	&	5.098 	&	5.040 	&	0.365 	&	0.380 	&	0.371 	&	-6.791 	&	-8.666 	\\
152	&	92	&	1248.387 	&	1250.05 	&	13.170 	&	2.136 	&	5.289 	&	5.046 	&	5.194 	&	5.109 	&		&	0.353 	&	0.370 	&	0.360 	&	-6.040 	&	-9.198 	\\
154	&	94	&	1259.341 	&	1261.73 	&	10.954 	&	2.333 	&	5.329 	&	5.063 	&	5.227 	&	5.126 	&		&	0.362 	&	0.374 	&	0.367 	&	-5.440 	&	-11.073 	\\
156	&	96	&	1269.848 	&	1272.66 	&	10.507 	&	2.318 	&	5.368 	&	5.083 	&	5.260 	&	5.145 	&		&	0.371 	&	0.377 	&	0.373 	&	-5.227 	&	-11.636 	\\
158	&	98	&	1280.006 	&		&	10.158 	&	2.173 	&	5.406 	&	5.102 	&	5.293 	&	5.164 	&		&	0.379 	&	0.380 	&	0.379 	&	-5.027 	&	-12.190 	\\
160	&	100	&	1289.755 	&		&	9.750 	&	0.000 	&	5.445 	&	5.120 	&	5.325 	&	5.182 	&		&	0.385 	&	0.381 	&	0.383 	&	-5.321 	&	-11.700 	\\
162	&	102	&	1297.790 	&		&	8.034 	&	2.329 	&	5.493 	&	5.145 	&	5.367 	&	5.207 	&		&	0.404 	&	0.393 	&	0.400 	&	-3.976 	&	-12.153 	\\
164	&	104	&	1305.500 	&		&	7.710 	&	2.420 	&	5.551 	&	5.178 	&	5.417 	&	5.240 	&		&	0.441 	&	0.418 	&	0.433 	&	-3.771 	&	-12.553 	\\
166	&	106	&	1312.218 	&		&	6.718 	&	2.459 	&	5.534 	&	5.150 	&	5.398 	&	5.211 	&		&	0.332 	&	0.325 	&	0.329 	&	-3.686 	&	-12.993 	\\
168	&	108	&	1319.405 	&		&	7.186 	&	2.353 	&	5.562 	&	5.156 	&	5.420 	&	5.218 	&		&	0.305 	&	0.297 	&	0.302 	&	-3.623 	&	-13.316 	\\
170	&	110	&	1326.380 	&		&	6.975 	&	2.248 	&	5.593 	&	5.166 	&	5.446 	&	5.227 	&		&	0.288 	&	0.280 	&	0.285 	&	-3.405 	&	-13.639 	\\
172	&	112	&	1332.717 	&		&	6.337 	&	2.363 	&	5.619 	&	5.174 	&	5.468 	&	5.235 	&		&	0.264 	&	0.260 	&	0.262 	&	-3.102 	&	-13.944 	\\
174	&	114	&	1338.609 	&		&	5.892 	&	2.337 	&	5.641 	&	5.177 	&	5.486 	&	5.239 	&		&	0.226 	&	0.231 	&	0.228 	&	-3.036 	&	-14.169 	\\
176	&	116	&	1344.536 	&		&	5.927 	&	2.185 	&	5.663 	&	5.180 	&	5.503 	&	5.242 	&		&	0.178 	&	0.190 	&	0.182 	&	-3.169 	&	-14.390 	\\
178	&	118	&	1350.839 	&		&	6.303 	&	2.006 	&	5.687 	&	5.179 	&	5.521 	&	5.240 	&		&	0.108 	&	0.119 	&	0.112 	&	-3.431 	&	-14.672 	\\
180	&	120	&	1357.637 	&		&	6.798 	&	0.000 	&	5.713 	&	5.186 	&	5.543 	&	5.247 	&		&	0.041 	&	0.046 	&	0.043 	&	-3.509 	&	-14.925 	\\
182	&	122	&	1364.528 	&		&	6.891 	&	0.000 	&	5.739 	&	5.200 	&	5.567 	&	5.261 	&		&	0.000 	&	0.000 	&	0.000 	&	-3.442 	&	-15.282 	\\
184	&	124	&	1371.289 	&		&	6.761 	&	0.000 	&	5.765 	&	5.216 	&	5.592 	&	5.277 	&		&	0.000 	&	0.000 	&	0.000 	&	-3.327 	&	-15.689 	\\
186	&	126	&	1377.904 	&		&	6.615 	&	0.000 	&	5.790 	&	5.232 	&	5.616 	&	5.293 	&		&	0.000 	&	0.000 	&	0.000 	&	-4.294 	&	-16.104 	\\
188	&	128	&	1378.449 	&		&	0.545 	&	0.000 	&	5.835 	&	5.246 	&	5.654 	&	5.306 	&		&	0.000 	&	0.000 	&	0.000 	&	-0.357 	&	-16.377 	\\
190	&	130	&	1378.950 	&		&	0.510 	&	0.000 	&	5.880 	&	5.259 	&	5.691 	&	5.320 	&		&	0.000 	&	0.000 	&	0.000 	&	-0.339 	&	-16.650 	\\
192	&	132	&	1379.474 	&		&	0.524 	&	1.318 	&	5.937 	&	5.284 	&	5.740 	&	5.344 	&		&	0.141 	&	0.100 	&	0.128 	&	-0.744 	&	-17.208 	\\
194	&	134	&	1381.344 	&		&	1.870 	&	1.423 	&	5.981 	&	5.305 	&	5.781 	&	5.365 	&		&	0.174 	&	0.129 	&	0.160 	&	-0.740 	&	-17.564 	\\
196	&	136	&	1382.679 	&		&	1.335 	&	1.540 	&	6.026 	&	5.325 	&	5.820 	&	5.385 	&		&	0.200 	&	0.153 	&	0.186 	&	-0.738 	&	-17.835 	\\
198	&	138	&	1384.006 	&		&	1.327 	&	1.562 	&	6.070 	&	5.345 	&	5.860 	&	5.405 	&		&	0.222 	&	0.175 	&	0.208 	&	-0.698 	&	-18.072 	\\
200	&	140	&	1385.306 	&		&	1.300 	&	1.802 	&	6.144 	&	5.388 	&	5.928 	&	5.447 	&		&	-0.255 	&	-0.238 	&	-0.250 	&	-0.940 	&	-18.574 	\\
202	&	142	&	1386.801 	&		&	1.495 	&	1.928 	&	6.180 	&	5.406 	&	5.961 	&	5.465 	&		&	-0.260 	&	-0.242 	&	-0.255 	&	-0.720 	&	-18.871 	\\
204	&	144	&	1387.977 	&		&	1.176 	&	2.014 	&	6.216 	&	5.422 	&	5.993 	&	5.481 	&		&	-0.263 	&	-0.242 	&	-0.257 	&	-0.632 	&	-19.141 	\\
206	&	146	&	1389.044 	&		&	1.066 	&	2.008 	&	6.251 	&	5.437 	&	6.025 	&	5.496 	&		&	-0.266 	&	-0.242 	&	-0.259 	&	-0.567 	&	-19.405 	\\
208	&	148	&	1389.992 	&		&	0.948 	&	1.919 	&	6.287 	&	5.452 	&	6.058 	&	5.510 	&		&	-0.269 	&	-0.242 	&	-0.261 	&	-0.485 	&	-19.667 	\\
210	&	150	&	1390.745 	&		&	0.753 	&	1.803 	&	6.322 	&	5.468 	&	6.090 	&	5.526 	&		&	-0.271 	&	-0.243 	&	-0.263 	&	-0.350 	&	-19.935 	\\
212	&	152	&	1391.163 	&		&	0.418 	&	1.795 	&	6.357 	&	5.484 	&	6.122 	&	5.542 	&		&	-0.271 	&	-0.243 	&	-0.263 	&	-0.178 	&	-20.204 	\\
214	&	154	&	1391.261 	&		&	0.097 	&	1.870 	&	6.390 	&	5.495 	&	6.152 	&	5.553 	&		&	-0.264 	&	-0.237 	&	-0.257 	&	-0.071 	&	-20.428 	\\
216	&	156	&	1391.204 	&		&	\underline{-0.057} 	&	1.889 	&	6.421 	&	5.503 	&	6.179 	&	5.561 	&		&	-0.252 	&	-0.225 	&	-0.244 	&	-0.025 	&	-20.623 	\\
\hline
\hline
\end{longtable}
\newpage
\end{landscape}

%======================================================================================%

%\end{CJK*}

\begin{thebibliography}{134}
\expandafter\ifx\csname natexlab\endcsname\relax\def\natexlab#1{#1}\fi
\expandafter\ifx\csname bibnamefont\endcsname\relax
  \def\bibnamefont#1{#1}\fi
\expandafter\ifx\csname bibfnamefont\endcsname\relax
  \def\bibfnamefont#1{#1}\fi
\expandafter\ifx\csname citenamefont\endcsname\relax
  \def\citenamefont#1{#1}\fi
\expandafter\ifx\csname url\endcsname\relax
  \def\url#1{\texttt{#1}}\fi
\expandafter\ifx\csname urlprefix\endcsname\relax\def\urlprefix{URL }\fi
\providecommand{\bibinfo}[2]{#2}
\providecommand{\eprint}[2][]{\url{#2}}

\bibitem[{\citenamefont{Meng}(2016)}]{Meng2016Book}
\bibinfo{author}{\bibfnamefont{J.}~\bibnamefont{Meng}},
  \emph{\bibinfo{title}{Relativistic Density Functional for Nuclear Structure}}
  (\bibinfo{publisher}{World Scientific}, \bibinfo{year}{2016}).

\bibitem[{\citenamefont{Zhan et~al.}(2010)\citenamefont{Zhan, Xu, Xiao, Xia,
  Zhao, and Yuan}}]{ZHAN2010694c}
\bibinfo{author}{\bibfnamefont{W.}~\bibnamefont{Zhan}},
  \bibinfo{author}{\bibfnamefont{H.}~\bibnamefont{Xu}},
  \bibinfo{author}{\bibfnamefont{G.}~\bibnamefont{Xiao}},
  \bibinfo{author}{\bibfnamefont{J.}~\bibnamefont{Xia}},
  \bibinfo{author}{\bibfnamefont{H.}~\bibnamefont{Zhao}}, \bibnamefont{and}
  \bibinfo{author}{\bibfnamefont{Y.}~\bibnamefont{Yuan}},
  \bibinfo{journal}{Nucl. Phys. A} \textbf{\bibinfo{volume}{834}},
  \bibinfo{pages}{694c } (\bibinfo{year}{2010}).

\bibitem[{\citenamefont{Motobayashi}(2010)}]{MOTOBAYASHI2010707c}
\bibinfo{author}{\bibfnamefont{T.}~\bibnamefont{Motobayashi}},
  \bibinfo{journal}{Nucl. Phys. A} \textbf{\bibinfo{volume}{834}},
  \bibinfo{pages}{707c } (\bibinfo{year}{2010}), ISSN
  \bibinfo{issn}{0375-9474}, \bibinfo{note}{the 10th International Conference
  on Nucleus-Nucleus Collisions (NN2009)}.

\bibitem[{\citenamefont{Tshoo et~al.}(2013)\citenamefont{Tshoo, Kim, Kwon, Woo,
  Kim, Kim, Kang, Park, Park, Yoon et~al.}}]{TSHOO2013242}
\bibinfo{author}{\bibfnamefont{K.}~\bibnamefont{Tshoo}},
  \bibinfo{author}{\bibfnamefont{Y.}~\bibnamefont{Kim}},
  \bibinfo{author}{\bibfnamefont{Y.}~\bibnamefont{Kwon}},
  \bibinfo{author}{\bibfnamefont{H.}~\bibnamefont{Woo}},
  \bibinfo{author}{\bibfnamefont{G.}~\bibnamefont{Kim}},
  \bibinfo{author}{\bibfnamefont{Y.}~\bibnamefont{Kim}},
  \bibinfo{author}{\bibfnamefont{B.}~\bibnamefont{Kang}},
  \bibinfo{author}{\bibfnamefont{S.}~\bibnamefont{Park}},
  \bibinfo{author}{\bibfnamefont{Y.-H.} \bibnamefont{Park}},
  \bibinfo{author}{\bibfnamefont{J.}~\bibnamefont{Yoon}}, \bibnamefont{et~al.},
  \bibinfo{journal}{Nucl. Instrum. Methods Phys. Res. A}
  \textbf{\bibinfo{volume}{317}}, \bibinfo{pages}{242} (\bibinfo{year}{2013}).

\bibitem[{\citenamefont{Sturm et~al.}(2010)\citenamefont{Sturm, Sharkov, and
  Stcker}}]{STURM2010682c}
\bibinfo{author}{\bibfnamefont{C.}~\bibnamefont{Sturm}},
  \bibinfo{author}{\bibfnamefont{B.}~\bibnamefont{Sharkov}}, \bibnamefont{and}
  \bibinfo{author}{\bibfnamefont{H.}~\bibnamefont{Stcker}},
  \bibinfo{journal}{Nucl. Phys. A} \textbf{\bibinfo{volume}{834}},
  \bibinfo{pages}{682c } (\bibinfo{year}{2010}), ISSN
  \bibinfo{issn}{0375-9474}, \bibinfo{note}{the 10th International Conference
  on Nucleus-Nucleus Collisions (NN2009)}.

\bibitem[{\citenamefont{Gales}(2010)}]{GALES2010717c}
\bibinfo{author}{\bibfnamefont{S.}~\bibnamefont{Gales}},
  \bibinfo{journal}{Nucl. Phys. A} \textbf{\bibinfo{volume}{834}},
  \bibinfo{pages}{717c } (\bibinfo{year}{2010}).

\bibitem[{\citenamefont{Thoennessen}(2010)}]{THOENNESSEN2010688c}
\bibinfo{author}{\bibfnamefont{M.}~\bibnamefont{Thoennessen}},
  \bibinfo{journal}{Nucl. Phys. A} \textbf{\bibinfo{volume}{834}},
  \bibinfo{pages}{688c } (\bibinfo{year}{2010}).

\bibitem[{\citenamefont{Lunney et~al.}(2003)\citenamefont{Lunney, Pearson, and
  Thibault}}]{Lunney20031021}
\bibinfo{author}{\bibfnamefont{D.}~\bibnamefont{Lunney}},
  \bibinfo{author}{\bibfnamefont{J.}~\bibnamefont{Pearson}}, \bibnamefont{and}
  \bibinfo{author}{\bibfnamefont{C.}~\bibnamefont{Thibault}},
  \bibinfo{journal}{Rev. Mod. Phys.} \textbf{\bibinfo{volume}{75}},
  \bibinfo{pages}{1021} (\bibinfo{year}{2003}).

\bibitem[{\citenamefont{Blaum}(2006)}]{BLAUM20061}
\bibinfo{author}{\bibfnamefont{K.}~\bibnamefont{Blaum}},
  \bibinfo{journal}{Phys. Rep.} \textbf{\bibinfo{volume}{425}},
  \bibinfo{pages}{1 } (\bibinfo{year}{2006}).

\bibitem[{\citenamefont{Erler et~al.}(2012)\citenamefont{Erler, Birge,
  Kortelainen, Nazarewicz, Olsen, Perhac, and Stoitsov}}]{Erler2012}
\bibinfo{author}{\bibfnamefont{J.}~\bibnamefont{Erler}},
  \bibinfo{author}{\bibfnamefont{N.}~\bibnamefont{Birge}},
  \bibinfo{author}{\bibfnamefont{M.}~\bibnamefont{Kortelainen}},
  \bibinfo{author}{\bibfnamefont{W.}~\bibnamefont{Nazarewicz}},
  \bibinfo{author}{\bibfnamefont{E.}~\bibnamefont{Olsen}},
  \bibinfo{author}{\bibfnamefont{A.~M.} \bibnamefont{Perhac}},
  \bibnamefont{and} \bibinfo{author}{\bibfnamefont{M.}~\bibnamefont{Stoitsov}},
  \bibinfo{journal}{Nature} \textbf{\bibinfo{volume}{486}},
  \bibinfo{pages}{509} (\bibinfo{year}{2012}).

\bibitem[{\citenamefont{Thoennessen}(2013)}]{Thoennessen2013}
\bibinfo{author}{\bibfnamefont{M.}~\bibnamefont{Thoennessen}},
  \bibinfo{journal}{Rep. Progr. Phys.} \textbf{\bibinfo{volume}{76}},
  \bibinfo{pages}{056301} (\bibinfo{year}{2013}).

\bibitem[{\citenamefont{Xia et~al.}(2018)\citenamefont{Xia, Lim, Zhao, Liang,
  Qu, Chen, Liu, Zhang, Zhang, Kim et~al.}}]{Xia2018ADNDT}
\bibinfo{author}{\bibfnamefont{X.~W.} \bibnamefont{Xia}},
  \bibinfo{author}{\bibfnamefont{Y.}~\bibnamefont{Lim}},
  \bibinfo{author}{\bibfnamefont{P.~W.} \bibnamefont{Zhao}},
  \bibinfo{author}{\bibfnamefont{H.~Z.} \bibnamefont{Liang}},
  \bibinfo{author}{\bibfnamefont{X.~Y.} \bibnamefont{Qu}},
  \bibinfo{author}{\bibfnamefont{Y.}~\bibnamefont{Chen}},
  \bibinfo{author}{\bibfnamefont{H.}~\bibnamefont{Liu}},
  \bibinfo{author}{\bibfnamefont{L.~F.} \bibnamefont{Zhang}},
  \bibinfo{author}{\bibfnamefont{S.~Q.} \bibnamefont{Zhang}},
  \bibinfo{author}{\bibfnamefont{Y.}~\bibnamefont{Kim}}, \bibnamefont{et~al.},
  \bibinfo{journal}{Atom. Data Nucl. Data Tabl.}
  \textbf{\bibinfo{volume}{121-122}}, \bibinfo{pages}{1}
  (\bibinfo{year}{2018}).

\bibitem[{\citenamefont{{National Nuclear Data Center (NNDC)}}()}]{NNDC}
\bibinfo{author}{\bibnamefont{{National Nuclear Data Center (NNDC)}}},
  \bibinfo{howpublished}{\url{http://www.nndc.bnl.gov/}}.

\bibitem[{\citenamefont{Huang et~al.}(2017)\citenamefont{Huang, Audi, Wang,
  Kondev, Naimi, and Xu}}]{AME20161}
\bibinfo{author}{\bibfnamefont{W.~J.} \bibnamefont{Huang}},
  \bibinfo{author}{\bibfnamefont{G.}~\bibnamefont{Audi}},
  \bibinfo{author}{\bibfnamefont{M.}~\bibnamefont{Wang}},
  \bibinfo{author}{\bibfnamefont{F.~G.} \bibnamefont{Kondev}},
  \bibinfo{author}{\bibfnamefont{S.}~\bibnamefont{Naimi}}, \bibnamefont{and}
  \bibinfo{author}{\bibfnamefont{X.}~\bibnamefont{Xu}}, \bibinfo{journal}{Chin.
  Phys. C} \textbf{\bibinfo{volume}{41}}, \bibinfo{pages}{030002}
  (\bibinfo{year}{2017}).

\bibitem[{\citenamefont{Wang et~al.}(2017)\citenamefont{Wang, Audi, Kondev,
  Huang, Naimi, and Xu}}]{AME2016}
\bibinfo{author}{\bibfnamefont{M.}~\bibnamefont{Wang}},
  \bibinfo{author}{\bibfnamefont{G.}~\bibnamefont{Audi}},
  \bibinfo{author}{\bibfnamefont{F.~G.} \bibnamefont{Kondev}},
  \bibinfo{author}{\bibfnamefont{W.~J.} \bibnamefont{Huang}},
  \bibinfo{author}{\bibfnamefont{S.}~\bibnamefont{Naimi}}, \bibnamefont{and}
  \bibinfo{author}{\bibfnamefont{X.}~\bibnamefont{Xu}}, \bibinfo{journal}{Chin.
  Phys. C} \textbf{\bibinfo{volume}{41}}, \bibinfo{pages}{030003}
  (\bibinfo{year}{2017}).

\bibitem[{\citenamefont{Zhang et~al.}(2019{\natexlab{a}})\citenamefont{Zhang,
  Gan, Yang, Ma, Huang, Yang, Zhang, Tian, Wang, Sun
  et~al.}}]{PhysRevLett.122.192503}
\bibinfo{author}{\bibfnamefont{Z.~Y.} \bibnamefont{Zhang}},
  \bibinfo{author}{\bibfnamefont{Z.~G.} \bibnamefont{Gan}},
  \bibinfo{author}{\bibfnamefont{H.~B.} \bibnamefont{Yang}},
  \bibinfo{author}{\bibfnamefont{L.}~\bibnamefont{Ma}},
  \bibinfo{author}{\bibfnamefont{M.~H.} \bibnamefont{Huang}},
  \bibinfo{author}{\bibfnamefont{C.~L.} \bibnamefont{Yang}},
  \bibinfo{author}{\bibfnamefont{M.~M.} \bibnamefont{Zhang}},
  \bibinfo{author}{\bibfnamefont{Y.~L.} \bibnamefont{Tian}},
  \bibinfo{author}{\bibfnamefont{Y.~S.} \bibnamefont{Wang}},
  \bibinfo{author}{\bibfnamefont{M.~D.} \bibnamefont{Sun}},
  \bibnamefont{et~al.}, \bibinfo{journal}{Phys. Rev. Lett.}
  \textbf{\bibinfo{volume}{122}}, \bibinfo{pages}{192503}
  (\bibinfo{year}{2019}{\natexlab{a}}).

\bibitem[{\citenamefont{M\"oller et~al.}(2016)\citenamefont{M\"oller, Sierk,
  Ichikawa, and Sagawa}}]{Moller2016ADNDT}
\bibinfo{author}{\bibfnamefont{P.}~\bibnamefont{M\"oller}},
  \bibinfo{author}{\bibfnamefont{A.}~\bibnamefont{Sierk}},
  \bibinfo{author}{\bibfnamefont{T.}~\bibnamefont{Ichikawa}}, \bibnamefont{and}
  \bibinfo{author}{\bibfnamefont{H.}~\bibnamefont{Sagawa}},
  \bibinfo{journal}{Atom. Data Nucl. Data Tabl.}
  \textbf{\bibinfo{volume}{109-110}}, \bibinfo{pages}{1 }
  (\bibinfo{year}{2016}).

\bibitem[{\citenamefont{Aboussir et~al.}(1995)\citenamefont{Aboussir, Pearson,
  Dutta, and Tondeur}}]{Aboussir1995ADNDT}
\bibinfo{author}{\bibfnamefont{Y.}~\bibnamefont{Aboussir}},
  \bibinfo{author}{\bibfnamefont{J.}~\bibnamefont{Pearson}},
  \bibinfo{author}{\bibfnamefont{A.}~\bibnamefont{Dutta}}, \bibnamefont{and}
  \bibinfo{author}{\bibfnamefont{F.}~\bibnamefont{Tondeur}},
  \bibinfo{journal}{Atom. Data Nucl. Data Tabl.} \textbf{\bibinfo{volume}{61}},
  \bibinfo{pages}{127 } (\bibinfo{year}{1995}).

\bibitem[{\citenamefont{Wang et~al.}(2014)\citenamefont{Wang, Liu, Wu, and
  Meng}}]{Wang2014PLB}
\bibinfo{author}{\bibfnamefont{N.}~\bibnamefont{Wang}},
  \bibinfo{author}{\bibfnamefont{M.}~\bibnamefont{Liu}},
  \bibinfo{author}{\bibfnamefont{X.}~\bibnamefont{Wu}}, \bibnamefont{and}
  \bibinfo{author}{\bibfnamefont{J.}~\bibnamefont{Meng}},
  \bibinfo{journal}{Phys. Lett. B} \textbf{\bibinfo{volume}{734}},
  \bibinfo{pages}{215 } (\bibinfo{year}{2014}).

\bibitem[{\citenamefont{Zhang et~al.}(2014{\natexlab{a}})\citenamefont{Zhang,
  Dong, Ma, Royer, Li, and Zhang}}]{Zhang2014NPA}
\bibinfo{author}{\bibfnamefont{H.}~\bibnamefont{Zhang}},
  \bibinfo{author}{\bibfnamefont{J.}~\bibnamefont{Dong}},
  \bibinfo{author}{\bibfnamefont{N.}~\bibnamefont{Ma}},
  \bibinfo{author}{\bibfnamefont{G.}~\bibnamefont{Royer}},
  \bibinfo{author}{\bibfnamefont{J.}~\bibnamefont{Li}}, \bibnamefont{and}
  \bibinfo{author}{\bibfnamefont{H.}~\bibnamefont{Zhang}},
  \bibinfo{journal}{Nucl. Phys. A} \textbf{\bibinfo{volume}{929}},
  \bibinfo{pages}{38 } (\bibinfo{year}{2014}{\natexlab{a}}).

\bibitem[{\citenamefont{Samyn et~al.}(2002)\citenamefont{Samyn, Goriely,
  Heenen, Pearson, and Tondeur}}]{Samyn2002NPA}
\bibinfo{author}{\bibfnamefont{M.}~\bibnamefont{Samyn}},
  \bibinfo{author}{\bibfnamefont{S.}~\bibnamefont{Goriely}},
  \bibinfo{author}{\bibfnamefont{P.-H.} \bibnamefont{Heenen}},
  \bibinfo{author}{\bibfnamefont{J.}~\bibnamefont{Pearson}}, \bibnamefont{and}
  \bibinfo{author}{\bibfnamefont{F.}~\bibnamefont{Tondeur}},
  \bibinfo{journal}{Nucl. Phys. A} \textbf{\bibinfo{volume}{700}},
  \bibinfo{pages}{142 } (\bibinfo{year}{2002}).

\bibitem[{\citenamefont{Stoitsov et~al.}(2003)\citenamefont{Stoitsov,
  Dobaczewski, Nazarewicz, Pittel, and Dean}}]{Stoitsov2003PRC}
\bibinfo{author}{\bibfnamefont{M.~V.} \bibnamefont{Stoitsov}},
  \bibinfo{author}{\bibfnamefont{J.}~\bibnamefont{Dobaczewski}},
  \bibinfo{author}{\bibfnamefont{W.}~\bibnamefont{Nazarewicz}},
  \bibinfo{author}{\bibfnamefont{S.}~\bibnamefont{Pittel}}, \bibnamefont{and}
  \bibinfo{author}{\bibfnamefont{D.~J.} \bibnamefont{Dean}},
  \bibinfo{journal}{Phys. Rev. C} \textbf{\bibinfo{volume}{68}},
  \bibinfo{pages}{054312} (\bibinfo{year}{2003}).

\bibitem[{\citenamefont{Goriely
  et~al.}(2009{\natexlab{a}})\citenamefont{Goriely, Chamel, and
  Pearson}}]{Goriely2009PRLSkyrme}
\bibinfo{author}{\bibfnamefont{S.}~\bibnamefont{Goriely}},
  \bibinfo{author}{\bibfnamefont{N.}~\bibnamefont{Chamel}}, \bibnamefont{and}
  \bibinfo{author}{\bibfnamefont{J.~M.} \bibnamefont{Pearson}},
  \bibinfo{journal}{Phys. Rev. Lett.} \textbf{\bibinfo{volume}{102}},
  \bibinfo{pages}{152503} (\bibinfo{year}{2009}{\natexlab{a}}).

\bibitem[{\citenamefont{Goriely et~al.}(2013)\citenamefont{Goriely, Chamel, and
  Pearson}}]{Goriely2013PRC}
\bibinfo{author}{\bibfnamefont{S.}~\bibnamefont{Goriely}},
  \bibinfo{author}{\bibfnamefont{N.}~\bibnamefont{Chamel}}, \bibnamefont{and}
  \bibinfo{author}{\bibfnamefont{J.~M.} \bibnamefont{Pearson}},
  \bibinfo{journal}{Phys. Rev. C} \textbf{\bibinfo{volume}{88}},
  \bibinfo{pages}{024308} (\bibinfo{year}{2013}).

\bibitem[{\citenamefont{Hilaire and Girod}(2007)}]{Hilaire2007EPJA}
\bibinfo{author}{\bibfnamefont{S.}~\bibnamefont{Hilaire}} \bibnamefont{and}
  \bibinfo{author}{\bibfnamefont{M.}~\bibnamefont{Girod}},
  \bibinfo{journal}{Eur. Phys. J. A} \textbf{\bibinfo{volume}{33}},
  \bibinfo{pages}{237} (\bibinfo{year}{2007}).

\bibitem[{\citenamefont{Goriely
  et~al.}(2009{\natexlab{b}})\citenamefont{Goriely, Hilaire, Girod, and
  P\'eru}}]{Goriely2009PRLGogny}
\bibinfo{author}{\bibfnamefont{S.}~\bibnamefont{Goriely}},
  \bibinfo{author}{\bibfnamefont{S.}~\bibnamefont{Hilaire}},
  \bibinfo{author}{\bibfnamefont{M.}~\bibnamefont{Girod}}, \bibnamefont{and}
  \bibinfo{author}{\bibfnamefont{S.}~\bibnamefont{P\'eru}},
  \bibinfo{journal}{Phys. Rev. Lett.} \textbf{\bibinfo{volume}{102}},
  \bibinfo{pages}{242501} (\bibinfo{year}{2009}{\natexlab{b}}).

\bibitem[{\citenamefont{Delaroche et~al.}(2010)\citenamefont{Delaroche, Girod,
  Libert, Goutte, Hilaire, P\'eru, Pillet, and Bertsch}}]{Delaroche2010PRC}
\bibinfo{author}{\bibfnamefont{J.~P.} \bibnamefont{Delaroche}},
  \bibinfo{author}{\bibfnamefont{M.}~\bibnamefont{Girod}},
  \bibinfo{author}{\bibfnamefont{J.}~\bibnamefont{Libert}},
  \bibinfo{author}{\bibfnamefont{H.}~\bibnamefont{Goutte}},
  \bibinfo{author}{\bibfnamefont{S.}~\bibnamefont{Hilaire}},
  \bibinfo{author}{\bibfnamefont{S.}~\bibnamefont{P\'eru}},
  \bibinfo{author}{\bibfnamefont{N.}~\bibnamefont{Pillet}}, \bibnamefont{and}
  \bibinfo{author}{\bibfnamefont{G.~F.} \bibnamefont{Bertsch}},
  \bibinfo{journal}{Phys. Rev. C} \textbf{\bibinfo{volume}{81}},
  \bibinfo{pages}{014303} (\bibinfo{year}{2010}).

\bibitem[{\citenamefont{Lalazissis et~al.}(1999)\citenamefont{Lalazissis,
  Raman, and Ring}}]{Lalazissis1999ADNDT}
\bibinfo{author}{\bibfnamefont{G.}~\bibnamefont{Lalazissis}},
  \bibinfo{author}{\bibfnamefont{S.}~\bibnamefont{Raman}}, \bibnamefont{and}
  \bibinfo{author}{\bibfnamefont{P.}~\bibnamefont{Ring}},
  \bibinfo{journal}{Atom. Data Nucl. Data Tabl.} \textbf{\bibinfo{volume}{71}},
  \bibinfo{pages}{1 } (\bibinfo{year}{1999}).

\bibitem[{\citenamefont{Geng et~al.}(2005)\citenamefont{Geng, Toki, and
  Meng}}]{Geng2005Prog.Theor.Phys}
\bibinfo{author}{\bibfnamefont{L.-S.} \bibnamefont{Geng}},
  \bibinfo{author}{\bibfnamefont{H.}~\bibnamefont{Toki}}, \bibnamefont{and}
  \bibinfo{author}{\bibfnamefont{J.}~\bibnamefont{Meng}},
  \bibinfo{journal}{Prog. Theor. Phys.} \textbf{\bibinfo{volume}{113}},
  \bibinfo{pages}{785} (\bibinfo{year}{2005}).

\bibitem[{\citenamefont{Meng et~al.}(2013)\citenamefont{Meng, Peng, Zhang, and
  Zhao}}]{Meng2013Front}
\bibinfo{author}{\bibfnamefont{J.}~\bibnamefont{Meng}},
  \bibinfo{author}{\bibfnamefont{J.}~\bibnamefont{Peng}},
  \bibinfo{author}{\bibfnamefont{S.~Q.} \bibnamefont{Zhang}}, \bibnamefont{and}
  \bibinfo{author}{\bibfnamefont{P.~W.} \bibnamefont{Zhao}},
  \bibinfo{journal}{Front. Phys.} \textbf{\bibinfo{volume}{8}},
  \bibinfo{pages}{55} (\bibinfo{year}{2013}).

\bibitem[{\citenamefont{Zhang et~al.}(2014{\natexlab{b}})\citenamefont{Zhang,
  Niu, Li, Yao, and Meng}}]{Zhang2014Front.Phys.529}
\bibinfo{author}{\bibfnamefont{Q.~S.} \bibnamefont{Zhang}},
  \bibinfo{author}{\bibfnamefont{Z.~M.} \bibnamefont{Niu}},
  \bibinfo{author}{\bibfnamefont{Z.~P.} \bibnamefont{Li}},
  \bibinfo{author}{\bibfnamefont{J.~M.} \bibnamefont{Yao}}, \bibnamefont{and}
  \bibinfo{author}{\bibfnamefont{J.}~\bibnamefont{Meng}},
  \bibinfo{journal}{Front. Phys.} \textbf{\bibinfo{volume}{9}},
  \bibinfo{pages}{529} (\bibinfo{year}{2014}{\natexlab{b}}).

\bibitem[{\citenamefont{Agbemava et~al.}(2014)\citenamefont{Agbemava,
  Afanasjev, Ray, and Ring}}]{Agbemava2014PRC}
\bibinfo{author}{\bibfnamefont{S.~E.} \bibnamefont{Agbemava}},
  \bibinfo{author}{\bibfnamefont{A.~V.} \bibnamefont{Afanasjev}},
  \bibinfo{author}{\bibfnamefont{D.}~\bibnamefont{Ray}}, \bibnamefont{and}
  \bibinfo{author}{\bibfnamefont{P.}~\bibnamefont{Ring}},
  \bibinfo{journal}{Phys. Rev. C} \textbf{\bibinfo{volume}{89}},
  \bibinfo{pages}{054320} (\bibinfo{year}{2014}).

\bibitem[{\citenamefont{Afanasjev et~al.}(2015)\citenamefont{Afanasjev,
  Agbemava, Ray, and Ring}}]{Afanasjev2015PRC}
\bibinfo{author}{\bibfnamefont{A.~V.} \bibnamefont{Afanasjev}},
  \bibinfo{author}{\bibfnamefont{S.~E.} \bibnamefont{Agbemava}},
  \bibinfo{author}{\bibfnamefont{D.}~\bibnamefont{Ray}}, \bibnamefont{and}
  \bibinfo{author}{\bibfnamefont{P.}~\bibnamefont{Ring}},
  \bibinfo{journal}{Phys. Rev. C} \textbf{\bibinfo{volume}{91}},
  \bibinfo{pages}{014324} (\bibinfo{year}{2015}).

\bibitem[{\citenamefont{Lu et~al.}(2015)\citenamefont{Lu, Li, Li, Yao, and
  Meng}}]{Lu2015Phys.Rev.C27304}
\bibinfo{author}{\bibfnamefont{K.~Q.} \bibnamefont{Lu}},
  \bibinfo{author}{\bibfnamefont{Z.~X.} \bibnamefont{Li}},
  \bibinfo{author}{\bibfnamefont{Z.~P.} \bibnamefont{Li}},
  \bibinfo{author}{\bibfnamefont{J.~M.} \bibnamefont{Yao}}, \bibnamefont{and}
  \bibinfo{author}{\bibfnamefont{J.}~\bibnamefont{Meng}},
  \bibinfo{journal}{Phys. Rev. C} \textbf{\bibinfo{volume}{91}},
  \bibinfo{pages}{027304} (\bibinfo{year}{2015}).

\bibitem[{\citenamefont{Pe{\~{n}}a-Arteaga
  et~al.}(2016)\citenamefont{Pe{\~{n}}a-Arteaga, Goriely, and
  Chamel}}]{Pena-Arteaga2016EPJA}
\bibinfo{author}{\bibfnamefont{D.}~\bibnamefont{Pe{\~{n}}a-Arteaga}},
  \bibinfo{author}{\bibfnamefont{S.}~\bibnamefont{Goriely}}, \bibnamefont{and}
  \bibinfo{author}{\bibfnamefont{N.}~\bibnamefont{Chamel}},
  \bibinfo{journal}{Eur. Phys. J. A} \textbf{\bibinfo{volume}{52}},
  \bibinfo{pages}{320} (\bibinfo{year}{2016}).

\bibitem[{\citenamefont{Ring}(1996)}]{Ring1996}
\bibinfo{author}{\bibfnamefont{P.}~\bibnamefont{Ring}}, \bibinfo{journal}{Prog.
  Part. Nucl. Phys.} \textbf{\bibinfo{volume}{37}}, \bibinfo{pages}{193 }
  (\bibinfo{year}{1996}).

\bibitem[{\citenamefont{Vretenar et~al.}(2005)\citenamefont{Vretenar,
  Afanasjev, Lalazissis, and Ring}}]{Vretenar2005PhysRep}
\bibinfo{author}{\bibfnamefont{D.}~\bibnamefont{Vretenar}},
  \bibinfo{author}{\bibfnamefont{A.~V.} \bibnamefont{Afanasjev}},
  \bibinfo{author}{\bibfnamefont{G.~A.} \bibnamefont{Lalazissis}},
  \bibnamefont{and} \bibinfo{author}{\bibfnamefont{P.}~\bibnamefont{Ring}},
  \bibinfo{journal}{Phys. Rep.} \textbf{\bibinfo{volume}{409}},
  \bibinfo{pages}{101 } (\bibinfo{year}{2005}).

\bibitem[{\citenamefont{Meng et~al.}(2006{\natexlab{a}})\citenamefont{Meng,
  Toki, Zhou, Zhang, Long, and Geng}}]{Meng2006PPNP}
\bibinfo{author}{\bibfnamefont{J.}~\bibnamefont{Meng}},
  \bibinfo{author}{\bibfnamefont{H.}~\bibnamefont{Toki}},
  \bibinfo{author}{\bibfnamefont{S.~G.} \bibnamefont{Zhou}},
  \bibinfo{author}{\bibfnamefont{S.~Q.} \bibnamefont{Zhang}},
  \bibinfo{author}{\bibfnamefont{W.~H.} \bibnamefont{Long}}, \bibnamefont{and}
  \bibinfo{author}{\bibfnamefont{L.~S.} \bibnamefont{Geng}},
  \bibinfo{journal}{Prog. Part. Nucl. Phys.} \textbf{\bibinfo{volume}{57}},
  \bibinfo{pages}{470} (\bibinfo{year}{2006}{\natexlab{a}}).

\bibitem[{\citenamefont{Niksic et~al.}(2011)\citenamefont{Niksic, Vretenar, and
  Ring}}]{Niksic2011PPNP}
\bibinfo{author}{\bibfnamefont{T.}~\bibnamefont{Niksic}},
  \bibinfo{author}{\bibfnamefont{D.}~\bibnamefont{Vretenar}}, \bibnamefont{and}
  \bibinfo{author}{\bibfnamefont{P.}~\bibnamefont{Ring}},
  \bibinfo{journal}{Prog. Part. Nucl. Phys.} \textbf{\bibinfo{volume}{66}},
  \bibinfo{pages}{519 } (\bibinfo{year}{2011}).

\bibitem[{\citenamefont{Meng and Zhou}(2015)}]{Meng2015JPhysG}
\bibinfo{author}{\bibfnamefont{J.}~\bibnamefont{Meng}} \bibnamefont{and}
  \bibinfo{author}{\bibfnamefont{S.~G.} \bibnamefont{Zhou}},
  \bibinfo{journal}{J. Phys. G} \textbf{\bibinfo{volume}{42}},
  \bibinfo{pages}{093101} (\bibinfo{year}{2015}).

\bibitem[{\citenamefont{Zhou}(2016)}]{Zhou2016PhysScr}
\bibinfo{author}{\bibfnamefont{S.-G.} \bibnamefont{Zhou}},
  \bibinfo{journal}{Phys. Scr.} \textbf{\bibinfo{volume}{91}},
  \bibinfo{pages}{063008} (\bibinfo{year}{2016}).

\bibitem[{\citenamefont{Shen et~al.}(2019)\citenamefont{Shen, Liang, Long,
  Meng, and Ring}}]{Shen2019PPNP}
\bibinfo{author}{\bibfnamefont{S.}~\bibnamefont{Shen}},
  \bibinfo{author}{\bibfnamefont{H.}~\bibnamefont{Liang}},
  \bibinfo{author}{\bibfnamefont{W.~H.} \bibnamefont{Long}},
  \bibinfo{author}{\bibfnamefont{J.}~\bibnamefont{Meng}}, \bibnamefont{and}
  \bibinfo{author}{\bibfnamefont{P.}~\bibnamefont{Ring}},
  \bibinfo{journal}{Prog. Part. Nucl. Phys.} \textbf{\bibinfo{volume}{109}},
  \bibinfo{pages}{103713} (\bibinfo{year}{2019}).

\bibitem[{\citenamefont{Ginocchio}(1997)}]{PhysRevLett.78.436}
\bibinfo{author}{\bibfnamefont{J.~N.} \bibnamefont{Ginocchio}},
  \bibinfo{journal}{Phys. Rev. Lett.} \textbf{\bibinfo{volume}{78}},
  \bibinfo{pages}{436} (\bibinfo{year}{1997}).

\bibitem[{\citenamefont{Meng et~al.}(1998{\natexlab{a}})\citenamefont{Meng,
  Sugawara-Tanabe, Yamaji, Ring, and Arima}}]{Meng1998Phys.Rev.C628}
\bibinfo{author}{\bibfnamefont{J.}~\bibnamefont{Meng}},
  \bibinfo{author}{\bibfnamefont{K.}~\bibnamefont{Sugawara-Tanabe}},
  \bibinfo{author}{\bibfnamefont{S.}~\bibnamefont{Yamaji}},
  \bibinfo{author}{\bibfnamefont{P.}~\bibnamefont{Ring}}, \bibnamefont{and}
  \bibinfo{author}{\bibfnamefont{A.}~\bibnamefont{Arima}},
  \bibinfo{journal}{Phys. Rev. C} \textbf{\bibinfo{volume}{58}},
  \bibinfo{pages}{R628} (\bibinfo{year}{1998}{\natexlab{a}}).

\bibitem[{\citenamefont{Meng et~al.}(1999)\citenamefont{Meng, Sugawara-Tanabe,
  Yamaji, and Arima}}]{Meng1999Phys.Rev.C154}
\bibinfo{author}{\bibfnamefont{J.}~\bibnamefont{Meng}},
  \bibinfo{author}{\bibfnamefont{K.}~\bibnamefont{Sugawara-Tanabe}},
  \bibinfo{author}{\bibfnamefont{S.}~\bibnamefont{Yamaji}}, \bibnamefont{and}
  \bibinfo{author}{\bibfnamefont{A.}~\bibnamefont{Arima}},
  \bibinfo{journal}{Phys. Rev. C} \textbf{\bibinfo{volume}{59}},
  \bibinfo{pages}{154} (\bibinfo{year}{1999}).

\bibitem[{\citenamefont{Chen et~al.}(2003)\citenamefont{Chen, Lu, Meng, Zhang,
  and Zhou}}]{Chen2003Chin.Phys.Lett.358}
\bibinfo{author}{\bibfnamefont{T.~S.} \bibnamefont{Chen}},
  \bibinfo{author}{\bibfnamefont{H.~F.} \bibnamefont{Lu}},
  \bibinfo{author}{\bibfnamefont{J.}~\bibnamefont{Meng}},
  \bibinfo{author}{\bibfnamefont{S.~Q.} \bibnamefont{Zhang}}, \bibnamefont{and}
  \bibinfo{author}{\bibfnamefont{S.~G.} \bibnamefont{Zhou}},
  \bibinfo{journal}{Chin. Phys. Lett.} \textbf{\bibinfo{volume}{20}},
  \bibinfo{pages}{358} (\bibinfo{year}{2003}).

\bibitem[{\citenamefont{Ginocchio}(2005)}]{Ginocchio2005PhysRep}
\bibinfo{author}{\bibfnamefont{J.~N.} \bibnamefont{Ginocchio}},
  \bibinfo{journal}{Phys. Rep.} \textbf{\bibinfo{volume}{414}},
  \bibinfo{pages}{165 } (\bibinfo{year}{2005}).

\bibitem[{\citenamefont{Liang et~al.}(2015)\citenamefont{Liang, Meng, and
  Zhou}}]{Liang2015Phys.Rept.1}
\bibinfo{author}{\bibfnamefont{H.}~\bibnamefont{Liang}},
  \bibinfo{author}{\bibfnamefont{J.}~\bibnamefont{Meng}}, \bibnamefont{and}
  \bibinfo{author}{\bibfnamefont{S.-G.} \bibnamefont{Zhou}},
  \bibinfo{journal}{Phys. Rep.} \textbf{\bibinfo{volume}{570}},
  \bibinfo{pages}{1} (\bibinfo{year}{2015}).

\bibitem[{\citenamefont{Zhou et~al.}(2003{\natexlab{a}})\citenamefont{Zhou,
  Meng, and Ring}}]{Zhou2003Phys.Rev.Lett.262501}
\bibinfo{author}{\bibfnamefont{S.-G.} \bibnamefont{Zhou}},
  \bibinfo{author}{\bibfnamefont{J.}~\bibnamefont{Meng}}, \bibnamefont{and}
  \bibinfo{author}{\bibfnamefont{P.}~\bibnamefont{Ring}},
  \bibinfo{journal}{Phys. Rev. Lett.} \textbf{\bibinfo{volume}{91}},
  \bibinfo{pages}{262501} (\bibinfo{year}{2003}{\natexlab{a}}).

\bibitem[{\citenamefont{He et~al.}(2006)\citenamefont{He, Zhou, Meng, Zhao, and
  Scheid}}]{He2006Eur.Phys.J.A265}
\bibinfo{author}{\bibfnamefont{X.~T.} \bibnamefont{He}},
  \bibinfo{author}{\bibfnamefont{S.~G.} \bibnamefont{Zhou}},
  \bibinfo{author}{\bibfnamefont{J.}~\bibnamefont{Meng}},
  \bibinfo{author}{\bibfnamefont{E.~G.} \bibnamefont{Zhao}}, \bibnamefont{and}
  \bibinfo{author}{\bibfnamefont{W.}~\bibnamefont{Scheid}},
  \bibinfo{journal}{Eur. Phys. J. A} \textbf{\bibinfo{volume}{28}},
  \bibinfo{pages}{265} (\bibinfo{year}{2006}).

\bibitem[{\citenamefont{Koepf and Ring}(1989)}]{Koepf1989NPA}
\bibinfo{author}{\bibfnamefont{W.}~\bibnamefont{Koepf}} \bibnamefont{and}
  \bibinfo{author}{\bibfnamefont{P.}~\bibnamefont{Ring}},
  \bibinfo{journal}{Nucl. Phys. A} \textbf{\bibinfo{volume}{493}},
  \bibinfo{pages}{61 } (\bibinfo{year}{1989}).

\bibitem[{\citenamefont{Yao et~al.}(2006)\citenamefont{Yao, Chen, and
  Meng}}]{Yao2006Phys.Rev.C24307}
\bibinfo{author}{\bibfnamefont{J.~M.} \bibnamefont{Yao}},
  \bibinfo{author}{\bibfnamefont{H.}~\bibnamefont{Chen}}, \bibnamefont{and}
  \bibinfo{author}{\bibfnamefont{J.}~\bibnamefont{Meng}},
  \bibinfo{journal}{Phys. Rev. C} \textbf{\bibinfo{volume}{74}},
  \bibinfo{pages}{024307} (\bibinfo{year}{2006}).

\bibitem[{\citenamefont{Arima}(2011)}]{Arima2011}
\bibinfo{author}{\bibfnamefont{A.}~\bibnamefont{Arima}}, \bibinfo{journal}{Sci.
  China Phys. Mech. Astron.} \textbf{\bibinfo{volume}{54}},
  \bibinfo{pages}{188} (\bibinfo{year}{2011}).

\bibitem[{\citenamefont{Li et~al.}(2011{\natexlab{a}})\citenamefont{Li, Meng,
  Ring, Yao, and Arima}}]{Li2011Sci.ChinaPhys.Mech.Astron.204}
\bibinfo{author}{\bibfnamefont{J.}~\bibnamefont{Li}},
  \bibinfo{author}{\bibfnamefont{J.}~\bibnamefont{Meng}},
  \bibinfo{author}{\bibfnamefont{P.}~\bibnamefont{Ring}},
  \bibinfo{author}{\bibfnamefont{J.~M.} \bibnamefont{Yao}}, \bibnamefont{and}
  \bibinfo{author}{\bibfnamefont{A.}~\bibnamefont{Arima}},
  \bibinfo{journal}{Sci. China Phys. Mech. Astron.}
  \textbf{\bibinfo{volume}{54}}, \bibinfo{pages}{204}
  (\bibinfo{year}{2011}{\natexlab{a}}).

\bibitem[{\citenamefont{Li et~al.}(2011{\natexlab{b}})\citenamefont{Li, Yao,
  Meng, and Arima}}]{Li2011Prog.Theor.Phys.1185}
\bibinfo{author}{\bibfnamefont{J.}~\bibnamefont{Li}},
  \bibinfo{author}{\bibfnamefont{J.~M.} \bibnamefont{Yao}},
  \bibinfo{author}{\bibfnamefont{J.}~\bibnamefont{Meng}}, \bibnamefont{and}
  \bibinfo{author}{\bibfnamefont{A.}~\bibnamefont{Arima}},
  \bibinfo{journal}{Prog. Theor. Phys.} \textbf{\bibinfo{volume}{125}},
  \bibinfo{pages}{1185} (\bibinfo{year}{2011}{\natexlab{b}}).

\bibitem[{\citenamefont{Li and Meng}(2018)}]{Li2018Front.Phys.Beijing132109}
\bibinfo{author}{\bibfnamefont{J.}~\bibnamefont{Li}} \bibnamefont{and}
  \bibinfo{author}{\bibfnamefont{J.}~\bibnamefont{Meng}},
  \bibinfo{journal}{Front. Phys.} \textbf{\bibinfo{volume}{13}},
  \bibinfo{pages}{132109} (\bibinfo{year}{2018}).

\bibitem[{\citenamefont{K\"onig and Ring}(1993)}]{PhysRevLett.71.3079}
\bibinfo{author}{\bibfnamefont{J.}~\bibnamefont{K\"onig}} \bibnamefont{and}
  \bibinfo{author}{\bibfnamefont{P.}~\bibnamefont{Ring}},
  \bibinfo{journal}{Phys. Rev. Lett.} \textbf{\bibinfo{volume}{71}},
  \bibinfo{pages}{3079} (\bibinfo{year}{1993}).

\bibitem[{\citenamefont{Afanasjev et~al.}(2000)\citenamefont{Afanasjev, Ring,
  and Konig}}]{Afanasjev2000NPA}
\bibinfo{author}{\bibfnamefont{A.~V.} \bibnamefont{Afanasjev}},
  \bibinfo{author}{\bibfnamefont{P.}~\bibnamefont{Ring}}, \bibnamefont{and}
  \bibinfo{author}{\bibfnamefont{J.}~\bibnamefont{Konig}},
  \bibinfo{journal}{Nucl. Phys. A} \textbf{\bibinfo{volume}{676}},
  \bibinfo{pages}{196 } (\bibinfo{year}{2000}).

\bibitem[{\citenamefont{Afanasjev and Ring}(2000)}]{PhysRevC.62.031302}
\bibinfo{author}{\bibfnamefont{A.~V.} \bibnamefont{Afanasjev}}
  \bibnamefont{and} \bibinfo{author}{\bibfnamefont{P.}~\bibnamefont{Ring}},
  \bibinfo{journal}{Phys. Rev. C} \textbf{\bibinfo{volume}{62}},
  \bibinfo{pages}{031302(R)} (\bibinfo{year}{2000}).

\bibitem[{\citenamefont{Afanasjev and Abusara}(2010)}]{PhysRevC.82.034329}
\bibinfo{author}{\bibfnamefont{A.~V.} \bibnamefont{Afanasjev}}
  \bibnamefont{and} \bibinfo{author}{\bibfnamefont{H.}~\bibnamefont{Abusara}},
  \bibinfo{journal}{Phys. Rev. C} \textbf{\bibinfo{volume}{82}},
  \bibinfo{pages}{034329} (\bibinfo{year}{2010}).

\bibitem[{\citenamefont{Zhao et~al.}(2011{\natexlab{a}})\citenamefont{Zhao,
  Peng, Liang, Ring, and Meng}}]{Zhao2011Phys.Rev.Lett.122501}
\bibinfo{author}{\bibfnamefont{P.~W.} \bibnamefont{Zhao}},
  \bibinfo{author}{\bibfnamefont{J.}~\bibnamefont{Peng}},
  \bibinfo{author}{\bibfnamefont{H.~Z.} \bibnamefont{Liang}},
  \bibinfo{author}{\bibfnamefont{P.}~\bibnamefont{Ring}}, \bibnamefont{and}
  \bibinfo{author}{\bibfnamefont{J.}~\bibnamefont{Meng}},
  \bibinfo{journal}{Phys. Rev. Lett.} \textbf{\bibinfo{volume}{107}},
  \bibinfo{pages}{122501} (\bibinfo{year}{2011}{\natexlab{a}}).

\bibitem[{\citenamefont{Zhao et~al.}(2011{\natexlab{b}})\citenamefont{Zhao,
  Zhang, Peng, Liang, Ring, and Meng}}]{Zhao2011Phys.Lett.B181}
\bibinfo{author}{\bibfnamefont{P.~W.} \bibnamefont{Zhao}},
  \bibinfo{author}{\bibfnamefont{S.~Q.} \bibnamefont{Zhang}},
  \bibinfo{author}{\bibfnamefont{J.}~\bibnamefont{Peng}},
  \bibinfo{author}{\bibfnamefont{H.~Z.} \bibnamefont{Liang}},
  \bibinfo{author}{\bibfnamefont{P.}~\bibnamefont{Ring}}, \bibnamefont{and}
  \bibinfo{author}{\bibfnamefont{J.}~\bibnamefont{Meng}},
  \bibinfo{journal}{Phys. Lett. B} \textbf{\bibinfo{volume}{699}},
  \bibinfo{pages}{181} (\bibinfo{year}{2011}{\natexlab{b}}).

\bibitem[{\citenamefont{Zhao et~al.}(2012{\natexlab{a}})\citenamefont{Zhao,
  Peng, Liang, Ring, and Meng}}]{Zhao2012Phys.Rev.C54310}
\bibinfo{author}{\bibfnamefont{P.~W.} \bibnamefont{Zhao}},
  \bibinfo{author}{\bibfnamefont{J.}~\bibnamefont{Peng}},
  \bibinfo{author}{\bibfnamefont{H.~Z.} \bibnamefont{Liang}},
  \bibinfo{author}{\bibfnamefont{P.}~\bibnamefont{Ring}}, \bibnamefont{and}
  \bibinfo{author}{\bibfnamefont{J.}~\bibnamefont{Meng}},
  \bibinfo{journal}{Phys. Rev. C} \textbf{\bibinfo{volume}{85}},
  \bibinfo{pages}{054310} (\bibinfo{year}{2012}{\natexlab{a}}).

\bibitem[{\citenamefont{Zhao et~al.}(2015)\citenamefont{Zhao, Itagaki, and
  Meng}}]{Zhao2015PRL}
\bibinfo{author}{\bibfnamefont{P.~W.} \bibnamefont{Zhao}},
  \bibinfo{author}{\bibfnamefont{N.}~\bibnamefont{Itagaki}}, \bibnamefont{and}
  \bibinfo{author}{\bibfnamefont{J.}~\bibnamefont{Meng}},
  \bibinfo{journal}{Phys. Rev. Lett.} \textbf{\bibinfo{volume}{115}},
  \bibinfo{pages}{022501} (\bibinfo{year}{2015}).

\bibitem[{\citenamefont{Wang}(2017)}]{Wang2017Phys.Rev.C54324}
\bibinfo{author}{\bibfnamefont{Y.~K.} \bibnamefont{Wang}},
  \bibinfo{journal}{Phys. Rev. C} \textbf{\bibinfo{volume}{96}},
  \bibinfo{pages}{054324} (\bibinfo{year}{2017}).

\bibitem[{\citenamefont{Wang}(2018)}]{Wang2018Phys.Rev.64321}
\bibinfo{author}{\bibfnamefont{Y.~K.} \bibnamefont{Wang}},
  \bibinfo{journal}{Phys. Rev. C} \textbf{\bibinfo{volume}{97}},
  \bibinfo{pages}{064321} (\bibinfo{year}{2018}).

\bibitem[{\citenamefont{Ren et~al.}(2019)\citenamefont{Ren, Zhang, Zhao,
  Itagaki, Maruhn, and Meng}}]{Ren2018Sci}
\bibinfo{author}{\bibfnamefont{Z.~X.} \bibnamefont{Ren}},
  \bibinfo{author}{\bibfnamefont{S.~Q.} \bibnamefont{Zhang}},
  \bibinfo{author}{\bibfnamefont{P.~W.} \bibnamefont{Zhao}},
  \bibinfo{author}{\bibfnamefont{N.}~\bibnamefont{Itagaki}},
  \bibinfo{author}{\bibfnamefont{J.~A.} \bibnamefont{Maruhn}},
  \bibnamefont{and} \bibinfo{author}{\bibfnamefont{J.}~\bibnamefont{Meng}},
  \bibinfo{journal}{Sci. China Phys. Mech. Astron.}
  \textbf{\bibinfo{volume}{62}}, \bibinfo{pages}{112026}
  (\bibinfo{year}{2019}).

\bibitem[{\citenamefont{Meng and Ring}(1996)}]{Meng1996PRL}
\bibinfo{author}{\bibfnamefont{J.}~\bibnamefont{Meng}} \bibnamefont{and}
  \bibinfo{author}{\bibfnamefont{P.}~\bibnamefont{Ring}},
  \bibinfo{journal}{Phys. Rev. Lett.} \textbf{\bibinfo{volume}{77}},
  \bibinfo{pages}{3963} (\bibinfo{year}{1996}).

\bibitem[{\citenamefont{Meng}(1998)}]{Meng1998NPA}
\bibinfo{author}{\bibfnamefont{J.}~\bibnamefont{Meng}}, \bibinfo{journal}{Nucl.
  Phys. A} \textbf{\bibinfo{volume}{635}}, \bibinfo{pages}{3}
  (\bibinfo{year}{1998}).

\bibitem[{\citenamefont{Meng and Ring}(1998)}]{Meng1998PRL}
\bibinfo{author}{\bibfnamefont{J.}~\bibnamefont{Meng}} \bibnamefont{and}
  \bibinfo{author}{\bibfnamefont{P.}~\bibnamefont{Ring}},
  \bibinfo{journal}{Phys. Rev. Lett.} \textbf{\bibinfo{volume}{80}},
  \bibinfo{pages}{460} (\bibinfo{year}{1998}).

\bibitem[{\citenamefont{Meng et~al.}(2002{\natexlab{a}})\citenamefont{Meng,
  Toki, Zeng, Zhang, and Zhou}}]{Meng2002Phys.Rev.C41302}
\bibinfo{author}{\bibfnamefont{J.}~\bibnamefont{Meng}},
  \bibinfo{author}{\bibfnamefont{H.}~\bibnamefont{Toki}},
  \bibinfo{author}{\bibfnamefont{J.~Y.} \bibnamefont{Zeng}},
  \bibinfo{author}{\bibfnamefont{S.~Q.} \bibnamefont{Zhang}}, \bibnamefont{and}
  \bibinfo{author}{\bibfnamefont{S.-G.} \bibnamefont{Zhou}},
  \bibinfo{journal}{Phys. Rev. C} \textbf{\bibinfo{volume}{65}},
  \bibinfo{pages}{041302(R)} (\bibinfo{year}{2002}{\natexlab{a}}).

\bibitem[{\citenamefont{Zhang et~al.}(2002)\citenamefont{Zhang, Meng, Zhou, and
  Zeng}}]{Zhang2002Chin.Phys.Lett.312}
\bibinfo{author}{\bibfnamefont{S.~Q.} \bibnamefont{Zhang}},
  \bibinfo{author}{\bibfnamefont{J.}~\bibnamefont{Meng}},
  \bibinfo{author}{\bibfnamefont{S.~G.} \bibnamefont{Zhou}}, \bibnamefont{and}
  \bibinfo{author}{\bibfnamefont{J.~Y.} \bibnamefont{Zeng}},
  \bibinfo{journal}{Chin. Phys. Lett.} \textbf{\bibinfo{volume}{19}},
  \bibinfo{pages}{312} (\bibinfo{year}{2002}).

\bibitem[{\citenamefont{Meng et~al.}(1998{\natexlab{b}})\citenamefont{Meng,
  Tanihata, and Yamaji}}]{Meng1998Phys.Lett.B1}
\bibinfo{author}{\bibfnamefont{J.}~\bibnamefont{Meng}},
  \bibinfo{author}{\bibfnamefont{I.}~\bibnamefont{Tanihata}}, \bibnamefont{and}
  \bibinfo{author}{\bibfnamefont{S.}~\bibnamefont{Yamaji}},
  \bibinfo{journal}{Phys. Lett. B} \textbf{\bibinfo{volume}{419}},
  \bibinfo{pages}{1} (\bibinfo{year}{1998}{\natexlab{b}}).

\bibitem[{\citenamefont{Meng et~al.}(2002{\natexlab{b}})\citenamefont{Meng,
  Zhou, and Tanihata}}]{Meng2002Phys.Lett.B209}
\bibinfo{author}{\bibfnamefont{J.}~\bibnamefont{Meng}},
  \bibinfo{author}{\bibfnamefont{S.~G.} \bibnamefont{Zhou}}, \bibnamefont{and}
  \bibinfo{author}{\bibfnamefont{I.}~\bibnamefont{Tanihata}},
  \bibinfo{journal}{Phys. Lett. B} \textbf{\bibinfo{volume}{532}},
  \bibinfo{pages}{209} (\bibinfo{year}{2002}{\natexlab{b}}).

\bibitem[{\citenamefont{Zhang et~al.}(2005)\citenamefont{Zhang, Meng, Zhang,
  Geng, and Toki}}]{Zhang2005Nucl.Phys.A106}
\bibinfo{author}{\bibfnamefont{W.}~\bibnamefont{Zhang}},
  \bibinfo{author}{\bibfnamefont{J.}~\bibnamefont{Meng}},
  \bibinfo{author}{\bibfnamefont{S.~Q.} \bibnamefont{Zhang}},
  \bibinfo{author}{\bibfnamefont{L.~S.} \bibnamefont{Geng}}, \bibnamefont{and}
  \bibinfo{author}{\bibfnamefont{H.}~\bibnamefont{Toki}},
  \bibinfo{journal}{Nucl. Phys. A} \textbf{\bibinfo{volume}{753}},
  \bibinfo{pages}{106} (\bibinfo{year}{2005}).

\bibitem[{\citenamefont{Lu et~al.}(2003)\citenamefont{Lu, Meng, Zhang, and
  Zhou}}]{Lu2003Eur.Phys.J.A19}
\bibinfo{author}{\bibfnamefont{H.~F.} \bibnamefont{Lu}},
  \bibinfo{author}{\bibfnamefont{J.}~\bibnamefont{Meng}},
  \bibinfo{author}{\bibfnamefont{S.~Q.} \bibnamefont{Zhang}}, \bibnamefont{and}
  \bibinfo{author}{\bibfnamefont{S.~G.} \bibnamefont{Zhou}},
  \bibinfo{journal}{Eur. Phys. J. A} \textbf{\bibinfo{volume}{17}},
  \bibinfo{pages}{19} (\bibinfo{year}{2003}).

\bibitem[{\citenamefont{Zhao et~al.}(2010)\citenamefont{Zhao, Li, Yao, and
  Meng}}]{Zhao2010Phys.Rev.C54319}
\bibinfo{author}{\bibfnamefont{P.~W.} \bibnamefont{Zhao}},
  \bibinfo{author}{\bibfnamefont{Z.~P.} \bibnamefont{Li}},
  \bibinfo{author}{\bibfnamefont{J.~M.} \bibnamefont{Yao}}, \bibnamefont{and}
  \bibinfo{author}{\bibfnamefont{J.}~\bibnamefont{Meng}},
  \bibinfo{journal}{Phys. Rev. C} \textbf{\bibinfo{volume}{82}},
  \bibinfo{pages}{054319} (\bibinfo{year}{2010}).

\bibitem[{\citenamefont{Zhang and Xia}(2016)}]{Zhang2016CPC}
\bibinfo{author}{\bibfnamefont{L.-F.} \bibnamefont{Zhang}} \bibnamefont{and}
  \bibinfo{author}{\bibfnamefont{X.-W.} \bibnamefont{Xia}},
  \bibinfo{journal}{Chin. Phys. C} \textbf{\bibinfo{volume}{40}},
  \bibinfo{pages}{054102} (\bibinfo{year}{2016}).

\bibitem[{\citenamefont{Lim et~al.}(2016)\citenamefont{Lim, Xia, and
  Kim}}]{Lim2016PRC}
\bibinfo{author}{\bibfnamefont{Y.}~\bibnamefont{Lim}},
  \bibinfo{author}{\bibfnamefont{X.}~\bibnamefont{Xia}}, \bibnamefont{and}
  \bibinfo{author}{\bibfnamefont{Y.}~\bibnamefont{Kim}},
  \bibinfo{journal}{Phys. Rev. C} \textbf{\bibinfo{volume}{93}},
  \bibinfo{pages}{014314} (\bibinfo{year}{2016}).

\bibitem[{\citenamefont{Zhou et~al.}(2000)\citenamefont{Zhou, Meng, Yamaji, and
  Yang}}]{Zhou2000CPL}
\bibinfo{author}{\bibfnamefont{S.-G.} \bibnamefont{Zhou}},
  \bibinfo{author}{\bibfnamefont{J.}~\bibnamefont{Meng}},
  \bibinfo{author}{\bibfnamefont{S.}~\bibnamefont{Yamaji}}, \bibnamefont{and}
  \bibinfo{author}{\bibfnamefont{S.-C.} \bibnamefont{Yang}},
  \bibinfo{journal}{Chin. Phys. Lett.} \textbf{\bibinfo{volume}{17}},
  \bibinfo{pages}{717} (\bibinfo{year}{2000}).

\bibitem[{\citenamefont{Zhou et~al.}(2010)\citenamefont{Zhou, Meng, Ring, and
  Zhao}}]{Zhou2010PRC}
\bibinfo{author}{\bibfnamefont{S.-G.} \bibnamefont{Zhou}},
  \bibinfo{author}{\bibfnamefont{J.}~\bibnamefont{Meng}},
  \bibinfo{author}{\bibfnamefont{P.}~\bibnamefont{Ring}}, \bibnamefont{and}
  \bibinfo{author}{\bibfnamefont{E.-G.} \bibnamefont{Zhao}},
  \bibinfo{journal}{Phys. Rev. C} \textbf{\bibinfo{volume}{82}},
  \bibinfo{pages}{011301(R)} (\bibinfo{year}{2010}).

\bibitem[{\citenamefont{Li et~al.}(2012{\natexlab{a}})\citenamefont{Li, Meng,
  Ring, Zhao, and Zhou}}]{Li2012PRC}
\bibinfo{author}{\bibfnamefont{L.}~\bibnamefont{Li}},
  \bibinfo{author}{\bibfnamefont{J.}~\bibnamefont{Meng}},
  \bibinfo{author}{\bibfnamefont{P.}~\bibnamefont{Ring}},
  \bibinfo{author}{\bibfnamefont{E.-G.} \bibnamefont{Zhao}}, \bibnamefont{and}
  \bibinfo{author}{\bibfnamefont{S.-G.} \bibnamefont{Zhou}},
  \bibinfo{journal}{Phys. Rev. C} \textbf{\bibinfo{volume}{85}},
  \bibinfo{pages}{024312} (\bibinfo{year}{2012}{\natexlab{a}}).

\bibitem[{\citenamefont{Zhou et~al.}(2003{\natexlab{b}})\citenamefont{Zhou,
  Meng, and Ring}}]{Zhou2003PRC}
\bibinfo{author}{\bibfnamefont{S.-G.} \bibnamefont{Zhou}},
  \bibinfo{author}{\bibfnamefont{J.}~\bibnamefont{Meng}}, \bibnamefont{and}
  \bibinfo{author}{\bibfnamefont{P.}~\bibnamefont{Ring}},
  \bibinfo{journal}{Phys. Rev. C} \textbf{\bibinfo{volume}{68}},
  \bibinfo{pages}{034323} (\bibinfo{year}{2003}{\natexlab{b}}).

\bibitem[{\citenamefont{Chen et~al.}(2012)\citenamefont{Chen, Li, Liang, and
  Meng}}]{Chen2012Phys.Rev.C67301}
\bibinfo{author}{\bibfnamefont{Y.}~\bibnamefont{Chen}},
  \bibinfo{author}{\bibfnamefont{L.}~\bibnamefont{Li}},
  \bibinfo{author}{\bibfnamefont{H.}~\bibnamefont{Liang}}, \bibnamefont{and}
  \bibinfo{author}{\bibfnamefont{J.}~\bibnamefont{Meng}},
  \bibinfo{journal}{Phys. Rev. C} \textbf{\bibinfo{volume}{85}},
  \bibinfo{pages}{067301} (\bibinfo{year}{2012}).

\bibitem[{\citenamefont{Li et~al.}(2012{\natexlab{b}})\citenamefont{Li, Meng,
  Ring, Zhao, and Zhou}}]{Li2012CPL}
\bibinfo{author}{\bibfnamefont{L.}~\bibnamefont{Li}},
  \bibinfo{author}{\bibfnamefont{J.}~\bibnamefont{Meng}},
  \bibinfo{author}{\bibfnamefont{P.}~\bibnamefont{Ring}},
  \bibinfo{author}{\bibfnamefont{E.-G.} \bibnamefont{Zhao}}, \bibnamefont{and}
  \bibinfo{author}{\bibfnamefont{S.-G.} \bibnamefont{Zhou}},
  \bibinfo{journal}{Chin. Phys. Lett.} \textbf{\bibinfo{volume}{29}},
  \bibinfo{pages}{042101} (\bibinfo{year}{2012}{\natexlab{b}}).

\bibitem[{\citenamefont{Sun et~al.}(2018)\citenamefont{Sun, Zhao, and
  Zhou}}]{Sun2018PLB}
\bibinfo{author}{\bibfnamefont{X.-X.} \bibnamefont{Sun}},
  \bibinfo{author}{\bibfnamefont{J.}~\bibnamefont{Zhao}}, \bibnamefont{and}
  \bibinfo{author}{\bibfnamefont{S.-G.} \bibnamefont{Zhou}},
  \bibinfo{journal}{Phys. Lett. B} \textbf{\bibinfo{volume}{785}},
  \bibinfo{pages}{530 } (\bibinfo{year}{2018}).

\bibitem[{\citenamefont{Zhang et~al.}(2019{\natexlab{b}})\citenamefont{Zhang,
  Wang, and Zhang}}]{Zhang2019PRC}
\bibinfo{author}{\bibfnamefont{K.~Y.} \bibnamefont{Zhang}},
  \bibinfo{author}{\bibfnamefont{D.~Y.} \bibnamefont{Wang}}, \bibnamefont{and}
  \bibinfo{author}{\bibfnamefont{S.~Q.} \bibnamefont{Zhang}},
  \bibinfo{journal}{Phys. Rev. C} \textbf{\bibinfo{volume}{100}},
  \bibinfo{pages}{034312} (\bibinfo{year}{2019}{\natexlab{b}}).

\bibitem[{\citenamefont{Koepf and Ring}(1991)}]{Koepf1991}
\bibinfo{author}{\bibfnamefont{W.}~\bibnamefont{Koepf}} \bibnamefont{and}
  \bibinfo{author}{\bibfnamefont{P.}~\bibnamefont{Ring}}, \bibinfo{journal}{Z.
  Phys. A} \textbf{\bibinfo{volume}{339}}, \bibinfo{pages}{81}
  (\bibinfo{year}{1991}).

\bibitem[{\citenamefont{Ring and Schuck}(1980)}]{PeterBook}
\bibinfo{author}{\bibfnamefont{P.}~\bibnamefont{Ring}} \bibnamefont{and}
  \bibinfo{author}{\bibfnamefont{P.}~\bibnamefont{Schuck}},
  \emph{\bibinfo{title}{The Nuclear Many-body Problem}}
  (\bibinfo{publisher}{Springer-Verlag, Berlin}, \bibinfo{year}{1980}).

\bibitem[{\citenamefont{Pritychenko et~al.}(2016)\citenamefont{Pritychenko,
  Birch, Singh, and Horoi}}]{2016ADNDT}
\bibinfo{author}{\bibfnamefont{B.}~\bibnamefont{Pritychenko}},
  \bibinfo{author}{\bibfnamefont{M.}~\bibnamefont{Birch}},
  \bibinfo{author}{\bibfnamefont{B.}~\bibnamefont{Singh}}, \bibnamefont{and}
  \bibinfo{author}{\bibfnamefont{M.}~\bibnamefont{Horoi}},
  \bibinfo{journal}{Atom. Data Nucl. Data Tabl.}
  \textbf{\bibinfo{volume}{107}}, \bibinfo{pages}{1 } (\bibinfo{year}{2016}).

\bibitem[{\citenamefont{Angeli and Marinova}(2013)}]{Angeli2013ADNDT}
\bibinfo{author}{\bibfnamefont{I.}~\bibnamefont{Angeli}} \bibnamefont{and}
  \bibinfo{author}{\bibfnamefont{K.}~\bibnamefont{Marinova}},
  \bibinfo{journal}{Atom. Data Nucl. Data Tabl.} \textbf{\bibinfo{volume}{99}},
  \bibinfo{pages}{69 } (\bibinfo{year}{2013}).

\bibitem[{\citenamefont{B\"urvenich et~al.}(2002)\citenamefont{B\"urvenich,
  Madland, Maruhn, and Reinhard}}]{PhysRevC.65.044308}
\bibinfo{author}{\bibfnamefont{T.}~\bibnamefont{B\"urvenich}},
  \bibinfo{author}{\bibfnamefont{D.~G.} \bibnamefont{Madland}},
  \bibinfo{author}{\bibfnamefont{J.~A.} \bibnamefont{Maruhn}},
  \bibnamefont{and} \bibinfo{author}{\bibfnamefont{P.-G.}
  \bibnamefont{Reinhard}}, \bibinfo{journal}{Phys. Rev. C}
  \textbf{\bibinfo{volume}{65}}, \bibinfo{pages}{044308}
  (\bibinfo{year}{2002}).

\bibitem[{\citenamefont{Dobaczewski et~al.}(1984)\citenamefont{Dobaczewski,
  Flocard, and Treiner}}]{DOBACZEWSKI1984103}
\bibinfo{author}{\bibfnamefont{J.}~\bibnamefont{Dobaczewski}},
  \bibinfo{author}{\bibfnamefont{H.}~\bibnamefont{Flocard}}, \bibnamefont{and}
  \bibinfo{author}{\bibfnamefont{J.}~\bibnamefont{Treiner}},
  \bibinfo{journal}{Nucl. Phys. A} \textbf{\bibinfo{volume}{422}},
  \bibinfo{pages}{103 } (\bibinfo{year}{1984}).

\bibitem[{\citenamefont{Kucharek and Ring}(1991)}]{Kucharek1991}
\bibinfo{author}{\bibfnamefont{H.}~\bibnamefont{Kucharek}} \bibnamefont{and}
  \bibinfo{author}{\bibfnamefont{P.}~\bibnamefont{Ring}}, \bibinfo{journal}{Z.
  Phys. A} \textbf{\bibinfo{volume}{339}}, \bibinfo{pages}{23}
  (\bibinfo{year}{1991}).

\bibitem[{\citenamefont{Gonzalez-Llarena
  et~al.}(1996)\citenamefont{Gonzalez-Llarena, Egido, Lalazissis, and
  Ring}}]{GONZALEZLLARENA1996PLB}
\bibinfo{author}{\bibfnamefont{T.}~\bibnamefont{Gonzalez-Llarena}},
  \bibinfo{author}{\bibfnamefont{J.}~\bibnamefont{Egido}},
  \bibinfo{author}{\bibfnamefont{G.}~\bibnamefont{Lalazissis}},
  \bibnamefont{and} \bibinfo{author}{\bibfnamefont{P.}~\bibnamefont{Ring}},
  \bibinfo{journal}{Physics Letters B} \textbf{\bibinfo{volume}{379}},
  \bibinfo{pages}{13 } (\bibinfo{year}{1996}).

\bibitem[{\citenamefont{Serra and Ring}(2002)}]{Serra2002PRC}
\bibinfo{author}{\bibfnamefont{M.}~\bibnamefont{Serra}} \bibnamefont{and}
  \bibinfo{author}{\bibfnamefont{P.}~\bibnamefont{Ring}},
  \bibinfo{journal}{Phys. Rev. C} \textbf{\bibinfo{volume}{65}},
  \bibinfo{pages}{064324} (\bibinfo{year}{2002}).

\bibitem[{\citenamefont{Price and Walker}(1987)}]{PhysRevC.36.354}
\bibinfo{author}{\bibfnamefont{C.~E.} \bibnamefont{Price}} \bibnamefont{and}
  \bibinfo{author}{\bibfnamefont{G.~E.} \bibnamefont{Walker}},
  \bibinfo{journal}{Phys. Rev. C} \textbf{\bibinfo{volume}{36}},
  \bibinfo{pages}{354} (\bibinfo{year}{1987}).

\bibitem[{\citenamefont{Grasso et~al.}(2001)\citenamefont{Grasso, Sandulescu,
  Van~Giai, and Liotta}}]{Grasso2001PRC}
\bibinfo{author}{\bibfnamefont{M.}~\bibnamefont{Grasso}},
  \bibinfo{author}{\bibfnamefont{N.}~\bibnamefont{Sandulescu}},
  \bibinfo{author}{\bibfnamefont{N.}~\bibnamefont{Van~Giai}}, \bibnamefont{and}
  \bibinfo{author}{\bibfnamefont{R.~J.} \bibnamefont{Liotta}},
  \bibinfo{journal}{Phys. Rev. C} \textbf{\bibinfo{volume}{64}},
  \bibinfo{pages}{064321} (\bibinfo{year}{2001}).

\bibitem[{\citenamefont{Michel et~al.}(2008)\citenamefont{Michel, Matsuyanagi,
  and Stoitsov}}]{Michel2008PRC}
\bibinfo{author}{\bibfnamefont{N.}~\bibnamefont{Michel}},
  \bibinfo{author}{\bibfnamefont{K.}~\bibnamefont{Matsuyanagi}},
  \bibnamefont{and} \bibinfo{author}{\bibfnamefont{M.}~\bibnamefont{Stoitsov}},
  \bibinfo{journal}{Phys. Rev. C} \textbf{\bibinfo{volume}{78}},
  \bibinfo{pages}{044319} (\bibinfo{year}{2008}).

\bibitem[{\citenamefont{Pei et~al.}(2011)\citenamefont{Pei, Kruppa, and
  Nazarewicz}}]{Pei2011PRC}
\bibinfo{author}{\bibfnamefont{J.~C.} \bibnamefont{Pei}},
  \bibinfo{author}{\bibfnamefont{A.~T.} \bibnamefont{Kruppa}},
  \bibnamefont{and}
  \bibinfo{author}{\bibfnamefont{W.}~\bibnamefont{Nazarewicz}},
  \bibinfo{journal}{Phys. Rev. C} \textbf{\bibinfo{volume}{84}},
  \bibinfo{pages}{024311} (\bibinfo{year}{2011}).

\bibitem[{\citenamefont{Zhang et~al.}(2012)\citenamefont{Zhang, Matsuo, and
  Meng}}]{Zhang2012PRC}
\bibinfo{author}{\bibfnamefont{Y.}~\bibnamefont{Zhang}},
  \bibinfo{author}{\bibfnamefont{M.}~\bibnamefont{Matsuo}}, \bibnamefont{and}
  \bibinfo{author}{\bibfnamefont{J.}~\bibnamefont{Meng}},
  \bibinfo{journal}{Phys. Rev. C} \textbf{\bibinfo{volume}{86}},
  \bibinfo{pages}{054318} (\bibinfo{year}{2012}).

\bibitem[{\citenamefont{Bender et~al.}(2000)\citenamefont{Bender, Rutz,
  Reinhard, and Maruhn}}]{Bender2000}
\bibinfo{author}{\bibfnamefont{M.}~\bibnamefont{Bender}},
  \bibinfo{author}{\bibfnamefont{K.}~\bibnamefont{Rutz}},
  \bibinfo{author}{\bibfnamefont{P.-G.} \bibnamefont{Reinhard}},
  \bibnamefont{and} \bibinfo{author}{\bibfnamefont{J.}~\bibnamefont{Maruhn}},
  \bibinfo{journal}{Eur. Phys. J. A} \textbf{\bibinfo{volume}{7}},
  \bibinfo{pages}{467} (\bibinfo{year}{2000}).

\bibitem[{\citenamefont{Long et~al.}(2004)\citenamefont{Long, Meng, VanGiai,
  and Zhou}}]{Long2004Phys.Rev.C34319}
\bibinfo{author}{\bibfnamefont{W.}~\bibnamefont{Long}},
  \bibinfo{author}{\bibfnamefont{J.}~\bibnamefont{Meng}},
  \bibinfo{author}{\bibfnamefont{N.}~\bibnamefont{VanGiai}}, \bibnamefont{and}
  \bibinfo{author}{\bibfnamefont{S.-G.} \bibnamefont{Zhou}},
  \bibinfo{journal}{Phys. Rev. C} \textbf{\bibinfo{volume}{69}},
  \bibinfo{pages}{034319} (\bibinfo{year}{2004}).

\bibitem[{\citenamefont{Zhao et~al.}(2009)\citenamefont{Zhao, Sun, and
  Meng}}]{Zhao2009Chin.Phys.Lett.112102}
\bibinfo{author}{\bibfnamefont{P.~W.} \bibnamefont{Zhao}},
  \bibinfo{author}{\bibfnamefont{B.~Y.} \bibnamefont{Sun}}, \bibnamefont{and}
  \bibinfo{author}{\bibfnamefont{J.}~\bibnamefont{Meng}},
  \bibinfo{journal}{Chin. Phys. Lett.} \textbf{\bibinfo{volume}{26}},
  \bibinfo{pages}{112102} (\bibinfo{year}{2009}).

\bibitem[{\citenamefont{Zhao et~al.}(2012{\natexlab{b}})\citenamefont{Zhao,
  Song, Sun, Geissel, and Meng}}]{Zhao2012Phys.Rev.C64324}
\bibinfo{author}{\bibfnamefont{P.~W.} \bibnamefont{Zhao}},
  \bibinfo{author}{\bibfnamefont{L.~S.} \bibnamefont{Song}},
  \bibinfo{author}{\bibfnamefont{B.}~\bibnamefont{Sun}},
  \bibinfo{author}{\bibfnamefont{H.}~\bibnamefont{Geissel}}, \bibnamefont{and}
  \bibinfo{author}{\bibfnamefont{J.}~\bibnamefont{Meng}},
  \bibinfo{journal}{Phys. Rev. C} \textbf{\bibinfo{volume}{86}},
  \bibinfo{pages}{064324} (\bibinfo{year}{2012}{\natexlab{b}}).

\bibitem[{\citenamefont{Pan et~al.}(2019)\citenamefont{Pan, Zhang, and
  Zhang}}]{Pan2019IJMPE}
\bibinfo{author}{\bibfnamefont{C.}~\bibnamefont{Pan}},
  \bibinfo{author}{\bibfnamefont{K.}~\bibnamefont{Zhang}}, \bibnamefont{and}
  \bibinfo{author}{\bibfnamefont{S.}~\bibnamefont{Zhang}},
  \bibinfo{journal}{Int. J. Mod. Phys. E} \textbf{\bibinfo{volume}{28}},
  \bibinfo{pages}{1950082} (\bibinfo{year}{2019}).

\bibitem[{\citenamefont{Hofmann and Ring}(1988)}]{Hofmann1988PLB}
\bibinfo{author}{\bibfnamefont{U.}~\bibnamefont{Hofmann}} \bibnamefont{and}
  \bibinfo{author}{\bibfnamefont{P.}~\bibnamefont{Ring}},
  \bibinfo{journal}{Phys. Lett. B} \textbf{\bibinfo{volume}{214}},
  \bibinfo{pages}{307 } (\bibinfo{year}{1988}).

\bibitem[{\citenamefont{Rutz et~al.}(1998)\citenamefont{Rutz, Bender, Reinhard,
  Maruhn, and Greiner}}]{Rutz1998NPA}
\bibinfo{author}{\bibfnamefont{K.}~\bibnamefont{Rutz}},
  \bibinfo{author}{\bibfnamefont{M.}~\bibnamefont{Bender}},
  \bibinfo{author}{\bibfnamefont{P.-G.} \bibnamefont{Reinhard}},
  \bibinfo{author}{\bibfnamefont{J.}~\bibnamefont{Maruhn}}, \bibnamefont{and}
  \bibinfo{author}{\bibfnamefont{W.}~\bibnamefont{Greiner}},
  \bibinfo{journal}{Nucl. Phys. A} \textbf{\bibinfo{volume}{634}},
  \bibinfo{pages}{67 } (\bibinfo{year}{1998}).

\bibitem[{\citenamefont{Rutz et~al.}(1999)\citenamefont{Rutz, Bender, Reinhard,
  and Maruhn}}]{Rutz1999PLB}
\bibinfo{author}{\bibfnamefont{K.}~\bibnamefont{Rutz}},
  \bibinfo{author}{\bibfnamefont{M.}~\bibnamefont{Bender}},
  \bibinfo{author}{\bibfnamefont{P.-G.} \bibnamefont{Reinhard}},
  \bibnamefont{and} \bibinfo{author}{\bibfnamefont{J.}~\bibnamefont{Maruhn}},
  \bibinfo{journal}{Phys. Lett. B} \textbf{\bibinfo{volume}{468}},
  \bibinfo{pages}{1 } (\bibinfo{year}{1999}).

\bibitem[{\citenamefont{Meng et~al.}(2006{\natexlab{b}})\citenamefont{Meng,
  Peng, Zhang, and Zhou}}]{Meng2006Phys.Rev.C37303}
\bibinfo{author}{\bibfnamefont{J.}~\bibnamefont{Meng}},
  \bibinfo{author}{\bibfnamefont{J.}~\bibnamefont{Peng}},
  \bibinfo{author}{\bibfnamefont{S.~Q.} \bibnamefont{Zhang}}, \bibnamefont{and}
  \bibinfo{author}{\bibfnamefont{S.-G.} \bibnamefont{Zhou}},
  \bibinfo{journal}{Phys. Rev. C} \textbf{\bibinfo{volume}{73}},
  \bibinfo{pages}{037303} (\bibinfo{year}{2006}{\natexlab{b}}).

\bibitem[{\citenamefont{Lu et~al.}(2007)\citenamefont{Lu, Geng, and
  Meng}}]{Lu2007Eur.Phys.J.A273}
\bibinfo{author}{\bibfnamefont{H.~F.} \bibnamefont{Lu}},
  \bibinfo{author}{\bibfnamefont{L.~S.} \bibnamefont{Geng}}, \bibnamefont{and}
  \bibinfo{author}{\bibfnamefont{J.}~\bibnamefont{Meng}},
  \bibinfo{journal}{Eur. Phys. J. A} \textbf{\bibinfo{volume}{31}},
  \bibinfo{pages}{273} (\bibinfo{year}{2007}).

\bibitem[{\citenamefont{Sun and Li}(2008)}]{Sun2008Chin.Phys.C882}
\bibinfo{author}{\bibfnamefont{B.-H.} \bibnamefont{Sun}} \bibnamefont{and}
  \bibinfo{author}{\bibfnamefont{J.}~\bibnamefont{Li}}, \bibinfo{journal}{Chin.
  Phys. C} \textbf{\bibinfo{volume}{32}}, \bibinfo{pages}{882}
  (\bibinfo{year}{2008}).

\bibitem[{\citenamefont{Li et~al.}(2009{\natexlab{a}})\citenamefont{Li, Yao,
  and Meng}}]{Li2009Chin.Phys.C98}
\bibinfo{author}{\bibfnamefont{J.}~\bibnamefont{Li}},
  \bibinfo{author}{\bibfnamefont{J.-M.} \bibnamefont{Yao}}, \bibnamefont{and}
  \bibinfo{author}{\bibfnamefont{J.}~\bibnamefont{Meng}},
  \bibinfo{journal}{Chin. Phys. C} \textbf{\bibinfo{volume}{33}},
  \bibinfo{pages}{98} (\bibinfo{year}{2009}{\natexlab{a}}).

\bibitem[{\citenamefont{Zhang et~al.}(2009)\citenamefont{Zhang, Peng, and
  Zhang}}]{Zhang2009Chin.Phys.Lett.52101}
\bibinfo{author}{\bibfnamefont{W.}~\bibnamefont{Zhang}},
  \bibinfo{author}{\bibfnamefont{J.}~\bibnamefont{Peng}}, \bibnamefont{and}
  \bibinfo{author}{\bibfnamefont{S.-Q.} \bibnamefont{Zhang}},
  \bibinfo{journal}{Chin. Phys. Lett.} \textbf{\bibinfo{volume}{26}},
  \bibinfo{pages}{052101} (\bibinfo{year}{2009}).

\bibitem[{\citenamefont{Staszczak et~al.}(2010)\citenamefont{Staszczak,
  Stoitsov, Baran, and Nazarewicz}}]{Staszczak2010}
\bibinfo{author}{\bibfnamefont{A.}~\bibnamefont{Staszczak}},
  \bibinfo{author}{\bibfnamefont{M.}~\bibnamefont{Stoitsov}},
  \bibinfo{author}{\bibfnamefont{A.}~\bibnamefont{Baran}}, \bibnamefont{and}
  \bibinfo{author}{\bibfnamefont{W.}~\bibnamefont{Nazarewicz}},
  \bibinfo{journal}{Eur. Phys. J. A} \textbf{\bibinfo{volume}{46}},
  \bibinfo{pages}{85} (\bibinfo{year}{2010}).

\bibitem[{\citenamefont{Li et~al.}(2012{\natexlab{c}})\citenamefont{Li,
  Nik\ifmmode \check{s}\else \v{s}\fi{}i\ifmmode~\acute{c}\else \'{c}\fi{},
  Ring, Vretenar, Yao, and Meng}}]{Li2012Phys.Rev.C34334}
\bibinfo{author}{\bibfnamefont{Z.~P.} \bibnamefont{Li}},
  \bibinfo{author}{\bibfnamefont{T.}~\bibnamefont{Nik\ifmmode \check{s}\else
  \v{s}\fi{}i\ifmmode~\acute{c}\else \'{c}\fi{}}},
  \bibinfo{author}{\bibfnamefont{P.}~\bibnamefont{Ring}},
  \bibinfo{author}{\bibfnamefont{D.}~\bibnamefont{Vretenar}},
  \bibinfo{author}{\bibfnamefont{J.~M.} \bibnamefont{Yao}}, \bibnamefont{and}
  \bibinfo{author}{\bibfnamefont{J.}~\bibnamefont{Meng}},
  \bibinfo{journal}{Phys. Rev. C} \textbf{\bibinfo{volume}{86}},
  \bibinfo{pages}{034334} (\bibinfo{year}{2012}{\natexlab{c}}).

\bibitem[{\citenamefont{Libert et~al.}(1999)\citenamefont{Libert, Girod, and
  Delaroche}}]{Libert1999PRC}
\bibinfo{author}{\bibfnamefont{J.}~\bibnamefont{Libert}},
  \bibinfo{author}{\bibfnamefont{M.}~\bibnamefont{Girod}}, \bibnamefont{and}
  \bibinfo{author}{\bibfnamefont{J.-P.} \bibnamefont{Delaroche}},
  \bibinfo{journal}{Phys. Rev. C} \textbf{\bibinfo{volume}{60}},
  \bibinfo{pages}{054301} (\bibinfo{year}{1999}).

\bibitem[{\citenamefont{Bender et~al.}(2006)\citenamefont{Bender, Bertsch, and
  Heenen}}]{Bender2006PRC}
\bibinfo{author}{\bibfnamefont{M.}~\bibnamefont{Bender}},
  \bibinfo{author}{\bibfnamefont{G.~F.} \bibnamefont{Bertsch}},
  \bibnamefont{and} \bibinfo{author}{\bibfnamefont{P.-H.}
  \bibnamefont{Heenen}}, \bibinfo{journal}{Phys. Rev. C}
  \textbf{\bibinfo{volume}{73}}, \bibinfo{pages}{034322}
  (\bibinfo{year}{2006}).

\bibitem[{\citenamefont{Bender et~al.}(2008)\citenamefont{Bender, Bertsch, and
  Heenen}}]{Bender2008PRC}
\bibinfo{author}{\bibfnamefont{M.}~\bibnamefont{Bender}},
  \bibinfo{author}{\bibfnamefont{G.~F.} \bibnamefont{Bertsch}},
  \bibnamefont{and} \bibinfo{author}{\bibfnamefont{P.-H.}
  \bibnamefont{Heenen}}, \bibinfo{journal}{Phys. Rev. C}
  \textbf{\bibinfo{volume}{78}}, \bibinfo{pages}{054312}
  (\bibinfo{year}{2008}).

\bibitem[{\citenamefont{Xiang et~al.}(2018)\citenamefont{Xiang, Li, Long,
  Nik\ifmmode \check{s}\else \v{s}\fi{}i\ifmmode~\acute{c}\else \'{c}\fi{}, and
  Vretenar}}]{Xiang2018PRC}
\bibinfo{author}{\bibfnamefont{J.}~\bibnamefont{Xiang}},
  \bibinfo{author}{\bibfnamefont{Z.~P.} \bibnamefont{Li}},
  \bibinfo{author}{\bibfnamefont{W.~H.} \bibnamefont{Long}},
  \bibinfo{author}{\bibfnamefont{T.}~\bibnamefont{Nik\ifmmode \check{s}\else
  \v{s}\fi{}i\ifmmode~\acute{c}\else \'{c}\fi{}}}, \bibnamefont{and}
  \bibinfo{author}{\bibfnamefont{D.}~\bibnamefont{Vretenar}},
  \bibinfo{journal}{Phys. Rev. C} \textbf{\bibinfo{volume}{98}},
  \bibinfo{pages}{054308} (\bibinfo{year}{2018}).

\bibitem[{\citenamefont{Li et~al.}(2009{\natexlab{b}})\citenamefont{Li,
  Nik\ifmmode \check{s}\else \v{s}\fi{}i\ifmmode~\acute{c}\else \'{c}\fi{},
  Vretenar, Meng, Lalazissis, and Ring}}]{Li2009PRC}
\bibinfo{author}{\bibfnamefont{Z.~P.} \bibnamefont{Li}},
  \bibinfo{author}{\bibfnamefont{T.}~\bibnamefont{Nik\ifmmode \check{s}\else
  \v{s}\fi{}i\ifmmode~\acute{c}\else \'{c}\fi{}}},
  \bibinfo{author}{\bibfnamefont{D.}~\bibnamefont{Vretenar}},
  \bibinfo{author}{\bibfnamefont{J.}~\bibnamefont{Meng}},
  \bibinfo{author}{\bibfnamefont{G.~A.} \bibnamefont{Lalazissis}},
  \bibnamefont{and} \bibinfo{author}{\bibfnamefont{P.}~\bibnamefont{Ring}},
  \bibinfo{journal}{Phys. Rev. C} \textbf{\bibinfo{volume}{79}},
  \bibinfo{pages}{054301} (\bibinfo{year}{2009}{\natexlab{b}}).

\bibitem[{\citenamefont{Scamps et~al.}(2013)\citenamefont{Scamps, Lacroix,
  Adamian, and Antonenko}}]{Scamps2013PRC}
\bibinfo{author}{\bibfnamefont{G.}~\bibnamefont{Scamps}},
  \bibinfo{author}{\bibfnamefont{D.}~\bibnamefont{Lacroix}},
  \bibinfo{author}{\bibfnamefont{G.~G.} \bibnamefont{Adamian}},
  \bibnamefont{and} \bibinfo{author}{\bibfnamefont{N.~V.}
  \bibnamefont{Antonenko}}, \bibinfo{journal}{Phys. Rev. C}
  \textbf{\bibinfo{volume}{88}}, \bibinfo{pages}{064327}
  (\bibinfo{year}{2013}).

\bibitem[{\citenamefont{Dobaczewski et~al.}(1996)\citenamefont{Dobaczewski,
  Nazarewicz, Werner, Berger, Chinn, and Decharg\'e}}]{Dobaczewski1996PRC}
\bibinfo{author}{\bibfnamefont{J.}~\bibnamefont{Dobaczewski}},
  \bibinfo{author}{\bibfnamefont{W.}~\bibnamefont{Nazarewicz}},
  \bibinfo{author}{\bibfnamefont{T.~R.} \bibnamefont{Werner}},
  \bibinfo{author}{\bibfnamefont{J.~F.} \bibnamefont{Berger}},
  \bibinfo{author}{\bibfnamefont{C.~R.} \bibnamefont{Chinn}}, \bibnamefont{and}
  \bibinfo{author}{\bibfnamefont{J.}~\bibnamefont{Decharg\'e}},
  \bibinfo{journal}{Phys. Rev. C} \textbf{\bibinfo{volume}{53}},
  \bibinfo{pages}{2809} (\bibinfo{year}{1996}).

\bibitem[{\citenamefont{Terasaki et~al.}(2006)\citenamefont{Terasaki, Zhang,
  Zhou, and Meng}}]{Terasaki2006Phys.Rev.C54318}
\bibinfo{author}{\bibfnamefont{J.}~\bibnamefont{Terasaki}},
  \bibinfo{author}{\bibfnamefont{S.~Q.} \bibnamefont{Zhang}},
  \bibinfo{author}{\bibfnamefont{S.~G.} \bibnamefont{Zhou}}, \bibnamefont{and}
  \bibinfo{author}{\bibfnamefont{J.}~\bibnamefont{Meng}},
  \bibinfo{journal}{Phys. Rev. C} \textbf{\bibinfo{volume}{74}},
  \bibinfo{pages}{054318} (\bibinfo{year}{2006}).

\bibitem[{\citenamefont{Rotival and Duguet}(2009)}]{Rotival2009PRC1}
\bibinfo{author}{\bibfnamefont{V.}~\bibnamefont{Rotival}} \bibnamefont{and}
  \bibinfo{author}{\bibfnamefont{T.}~\bibnamefont{Duguet}},
  \bibinfo{journal}{Phys. Rev. C} \textbf{\bibinfo{volume}{79}},
  \bibinfo{pages}{054308} (\bibinfo{year}{2009}).

\bibitem[{\citenamefont{Rotival et~al.}(2009)\citenamefont{Rotival, Bennaceur,
  and Duguet}}]{Rotival2009PRC2}
\bibinfo{author}{\bibfnamefont{V.}~\bibnamefont{Rotival}},
  \bibinfo{author}{\bibfnamefont{K.}~\bibnamefont{Bennaceur}},
  \bibnamefont{and} \bibinfo{author}{\bibfnamefont{T.}~\bibnamefont{Duguet}},
  \bibinfo{journal}{Phys. Rev. C} \textbf{\bibinfo{volume}{79}},
  \bibinfo{pages}{054309} (\bibinfo{year}{2009}).

\bibitem[{\citenamefont{Im and Meng}(2000)}]{Im2000PRC}
\bibinfo{author}{\bibfnamefont{S.}~\bibnamefont{Im}} \bibnamefont{and}
  \bibinfo{author}{\bibfnamefont{J.}~\bibnamefont{Meng}},
  \bibinfo{journal}{Phys. Rev. C} \textbf{\bibinfo{volume}{61}},
  \bibinfo{pages}{047302} (\bibinfo{year}{2000}).

\bibitem[{\citenamefont{Mizutori et~al.}(2000)\citenamefont{Mizutori,
  Dobaczewski, Lalazissis, Nazarewicz, and Reinhard}}]{Mizutori2000PRC}
\bibinfo{author}{\bibfnamefont{S.}~\bibnamefont{Mizutori}},
  \bibinfo{author}{\bibfnamefont{J.}~\bibnamefont{Dobaczewski}},
  \bibinfo{author}{\bibfnamefont{G.~A.} \bibnamefont{Lalazissis}},
  \bibinfo{author}{\bibfnamefont{W.}~\bibnamefont{Nazarewicz}},
  \bibnamefont{and} \bibinfo{author}{\bibfnamefont{P.-G.}
  \bibnamefont{Reinhard}}, \bibinfo{journal}{Phys. Rev. C}
  \textbf{\bibinfo{volume}{61}}, \bibinfo{pages}{044326}
  (\bibinfo{year}{2000}).

\bibitem[{\citenamefont{Blank and Borge}(2008)}]{Blank2008PPNP}
\bibinfo{author}{\bibfnamefont{B.}~\bibnamefont{Blank}} \bibnamefont{and}
  \bibinfo{author}{\bibfnamefont{M.}~\bibnamefont{Borge}},
  \bibinfo{journal}{Prog. Part. Nucl. Phys.} \textbf{\bibinfo{volume}{60}},
  \bibinfo{pages}{403 } (\bibinfo{year}{2008}).

\bibitem[{\citenamefont{Pf\"utzner et~al.}(2012)\citenamefont{Pf\"utzner,
  Karny, Grigorenko, and Riisager}}]{Pfutzner2012RMP}
\bibinfo{author}{\bibfnamefont{M.}~\bibnamefont{Pf\"utzner}},
  \bibinfo{author}{\bibfnamefont{M.}~\bibnamefont{Karny}},
  \bibinfo{author}{\bibfnamefont{L.~V.} \bibnamefont{Grigorenko}},
  \bibnamefont{and} \bibinfo{author}{\bibfnamefont{K.}~\bibnamefont{Riisager}},
  \bibinfo{journal}{Rev. Mod. Phys.} \textbf{\bibinfo{volume}{84}},
  \bibinfo{pages}{567} (\bibinfo{year}{2012}).

\bibitem[{\citenamefont{Vretenar et~al.}(1999)\citenamefont{Vretenar,
  Lalazissis, and Ring}}]{Vretenar1999PRL}
\bibinfo{author}{\bibfnamefont{D.}~\bibnamefont{Vretenar}},
  \bibinfo{author}{\bibfnamefont{G.~A.} \bibnamefont{Lalazissis}},
  \bibnamefont{and} \bibinfo{author}{\bibfnamefont{P.}~\bibnamefont{Ring}},
  \bibinfo{journal}{Phys. Rev. Lett.} \textbf{\bibinfo{volume}{82}},
  \bibinfo{pages}{4595} (\bibinfo{year}{1999}).

\bibitem[{\citenamefont{Yao et~al.}(2008)\citenamefont{Yao, Sun, Woods, and
  Meng}}]{Yao2008PRC}
\bibinfo{author}{\bibfnamefont{J.~M.} \bibnamefont{Yao}},
  \bibinfo{author}{\bibfnamefont{B.}~\bibnamefont{Sun}},
  \bibinfo{author}{\bibfnamefont{P.~J.} \bibnamefont{Woods}}, \bibnamefont{and}
  \bibinfo{author}{\bibfnamefont{J.}~\bibnamefont{Meng}},
  \bibinfo{journal}{Phys. Rev. C} \textbf{\bibinfo{volume}{77}},
  \bibinfo{pages}{024315} (\bibinfo{year}{2008}).

\bibitem[{\citenamefont{Ferreira et~al.}(2011)\citenamefont{Ferreira, Maglione,
  and Ring}}]{Ferreira2011PLB}
\bibinfo{author}{\bibfnamefont{L.}~\bibnamefont{Ferreira}},
  \bibinfo{author}{\bibfnamefont{E.}~\bibnamefont{Maglione}}, \bibnamefont{and}
  \bibinfo{author}{\bibfnamefont{P.}~\bibnamefont{Ring}},
  \bibinfo{journal}{Phys. Lett. B} \textbf{\bibinfo{volume}{701}},
  \bibinfo{pages}{508 } (\bibinfo{year}{2011}).

\bibitem[{\citenamefont{Zhao et~al.}(2014)\citenamefont{Zhao, Dong, Song, and
  Long}}]{Zhao2014PRC}
\bibinfo{author}{\bibfnamefont{Q.}~\bibnamefont{Zhao}},
  \bibinfo{author}{\bibfnamefont{J.~M.} \bibnamefont{Dong}},
  \bibinfo{author}{\bibfnamefont{J.~L.} \bibnamefont{Song}}, \bibnamefont{and}
  \bibinfo{author}{\bibfnamefont{W.~H.} \bibnamefont{Long}},
  \bibinfo{journal}{Phys. Rev. C} \textbf{\bibinfo{volume}{90}},
  \bibinfo{pages}{054326} (\bibinfo{year}{2014}).

\end{thebibliography}
\end{document}